\documentclass[
pra,
aps,
onecolumn,
superscriptaddress,
longbibliography,
floatfix,
notitlepage]{revtex4-2}

\usepackage[export]{adjustbox}

\usepackage[utf8]{inputenc}
\usepackage[T1]{fontenc}
\usepackage[english]{babel}
\usepackage{lmodern}
\usepackage{microtype}

\usepackage{graphicx}
\usepackage{epstopdf}

\usepackage{amsmath,amsfonts,amssymb,amsthm,bm,bbm,times,dcolumn}
\usepackage[normalem]{ulem}
\usepackage{braket}
\usepackage{physics}
\usepackage{gensymb}
\usepackage[title]{appendix}

\usepackage[colorlinks=true,citecolor=blue,filecolor=blue,linkcolor=blue,urlcolor=blue]{hyperref}
\usepackage[caption=false]{subfig}
\usepackage{xcolor}
\usepackage{changepage}
\newlength{\extralength}
\begin{document}
%\sloppy

\title{Quantum Information Flow in Microtubule Tryptophan Networks}

\author{Lea Gassab}
\email{lgassab@uwaterloo.ca}
\affiliation{Departments of Biology, Chemistry, Physics \& Astronomy, Waterloo Institute for Nanotechnology, University of Waterloo, Waterloo, ON, Canada}

\author{Onur Pusuluk}
\affiliation{Faculty of Engineering and Natural Sciences, Kadir Has University, 34083, Fatih, Istanbul, T\"{u}rkiye}

\author{Travis J.A.\ Craddock}
\email{travis.craddock@uwaterloo.ca}
\affiliation{Departments of Biology, Physics \& Astronomy, Waterloo Institute for Nanotechnology, University of Waterloo, Waterloo, ON, Canada}

\begin{abstract}
Networks of aromatic amino acid residues within microtubules, particularly those formed by tryptophan, may serve as pathways for optical information flow. Ultraviolet excitation dynamics in these networks are typically modeled with effective non-Hermitian Hamiltonians. By extending this approach to a Lindblad master equation that incorporates explicit site geometries and dipole orientations, we track how correlations are generated, routed, and dissipated, while capturing both energy dissipation and information propagation among coupled chromophores. We compare localized injections, fully delocalized preparations, and eigenmode-based initial states. To quantify the emerging quantum-informational structure, we evaluate the $L_1$ norm of coherence, the correlated coherence, and the logarithmic negativity within and between selected chromophore sub-networks. The results reveal a strong dependence of both the direction and persistence of information flow on the type of initial preparation. Superradiant components drive the rapid export of correlations to the environment, whereas subradiant components retain them and slow their leakage. Embedding single tubulin units into larger dimers and spirals reshapes pairwise correlation maps and enables site-selective routing. Scaling to larger ordered lattices strengthens both export and retention channels, whereas static energetic and structural disorder suppresses long-range transport and reduces overall correlation transfer. These findings provide a Lindbladian picture of information flow in cytoskeletal chromophore networks and identify structural and dynamical conditions that transiently preserve nonclassical correlations in microtubules.
\end{abstract}

\keywords{quantum biology; open quantum systems; non-Markovianity; spectral density; cooperative emission; excitonic transport; non-Hermitian dynamics; superradiance; subradiance}

\maketitle

\section{Introduction}

Neural systems display rhythmic synchronization and structured correlations that are often interpreted as signatures of directed information flow \cite{Fries2015,Bastos2015,Schreiber2000,Vicente2011}. The neuronal microtubule cytoskeleton, an intracellular network of protein polymers, has been proposed to play a significant role beyond mere structural support and axonal transport, acting as a potential substrate for propagating and processing information within the neuron. In contrast to the rapid turnover observed in many dividing cells, neuronal microtubules contain a substantial stable fraction as well as long-lived polymer domains. Experimentally, stable microtubule populations in neurons can persist for hours, coexisting with more labile populations that remodel on much shorter timescales \cite{Baas2016,LiBlack1996}. This relative stability is crucial for maintaining neuronal morphology and supporting long-distance transport, and it motivates exploring whether microtubule-embedded molecular networks could support structured correlation routing. Motivated by a general perspective on correlation and information routing, we turn to molecular networks in microtubules where optical interactions can be quantified with microscopic detail.

Microtubules contain dense arrays of aromatic amino acids, with tryptophan providing strong ultraviolet absorption and large transition dipoles that support collective optical effects. Recent experiments using tryptophan autofluorescence lifetimes have already demonstrated electronic energy migration over several nanometers along microtubules \cite{Kalra2023}. Complementary experimental and theoretical work reports ultraviolet superradiance in biological assemblies with extended tryptophan networks, suggesting that ordered cytoskeletal structures can sustain cooperative emission \cite{Craddock2014,kurian2017oxidative,babcock2024ultraviolet,patwa2024quantum}. For microtubules specifically, analyses based on effective non-Hermitian Hamiltonians predict superradiant and subradiant excitonic eigenstates that arise from radiative coupling to the electromagnetic field \cite{celardo2019existence,Rotter2009,Ashida2020,Celardo2012,Wang2023,Bergholtz2021}. These observations raise time-resolved questions with biological relevance. How are correlations generated and redistributed among chromophores on picosecond–nanosecond timescales, how rapidly are they exported radiatively, and how do structure, size, and disorder shape the balance between internal redistribution and emission?

Beyond these radiative and optical considerations, several studies have already treated microtubules as quantum information channels. Shirmovsky and collaborators modeled single–excitation migration along microtubule tryptophan chains and showed that the propagation speed of the excited state and the associated entanglement transfer can lie in the range of axonal conduction velocities, suggesting that tryptophan networks may mediate quantum‐assisted signaling \cite{ShirmovskyChizhov2023,Shirmovsky2024a,Shirmovsky2024b}. Related work on the quantum relaxation of tubulin dipole networks quantified decoherence times under dissipative dynamics and explored Markovian versus non-Markovian regimes in microtubules \cite{ShirmovskyShulga2021,ShirmovskyShulga2023,Saenko2025}. Other approaches include waveguide QED models in which tryptophans and surrounding water act as qubits coupled to guided modes in microtubules \cite{Nishiyama2025}, and proposals that employ error-correcting surface codes or mitotic spindle entanglement to link microtubules with objective reduction–type mechanisms \cite{Hameroff2004,AnChoi2025}. In contrast to these tunneling- and spin-based schemes, the present work focuses on radiative coupling to the electromagnetic field, larger ordered tryptophan assemblies, and a systematic comparison of distinct initial states to assess how microtubule geometry and disorder shape correlation generation, redistribution, and loss.

A more physically consistent open-system approach may provide a natural framework for addressing these questions. Accordingly, instead of using a non-Hermitian effective Hamiltonian, we describe radiative loss with a Markovian Lindblad master equation, so the dynamics remain completely positive while retaining the usual decay channel \cite{Lindblad1976,GKS1976,Ishizaki2009a,Ishizaki2009b,Huelga2013,Mohseni2008,Rebentrost2009,Engel2007,Panitchayangkoon2010,breuer2002theory}. Within this framework, we model networks of tryptophan chromophores using site-specific positions and dipole orientations, and we track information flow by following population redistribution along with correlation-oriented metrics. In particular, we quantify the information carried by coherent excitation delocalization by monitoring the $L_1$ norm of coherence~\cite{Baumgratz2014, Streltsov2017}. We further evaluate a basis-independent measure of coherence~\cite{2015_PRA_CoherenceAndDiscord}, called correlated coherence~\cite{2016_PRA_CorrelatedCoherenceAndDiscordAndEnt}, which captures how much of this information is shared between microtubule substructures, and we use the logarithmic negativity as an entanglement measure~\cite{Plenio2005,VidalWerner2002}.

Related open-system exciton-transport analyses in pigment--protein complexes, particularly the Fenna--Matthews--Olson (FMO) complex, provide a useful point of comparison for our tryptophan-network results: in both cases, the interplay of coherent couplings, dissipation, and disorder shapes how excitation delocalizes, how quickly coherences decay, and whether environmentally assisted dynamics enhance transport. Representative experimental and theoretical studies of coherence and open-system transport in FMO and related light-harvesting networks include Refs.~\cite{Engel2007,IshizakiFleming2009,Rebentrost2009,Chin2010,Scholes2017Nature}.

Our analysis probes initial conditions that represent distinct routes for correlation propagation. We consider preparations aligned with superradiant and subradiant sectors of the effective generator, a fully coherent uniform superposition across sites, a fully mixed uniform distribution, and site-localized injections that mimic single-photon absorption. We then examine how embedding a single tubulin into dimers and microtubule segments containing one or more spirals (where a spiral denotes one circumferential turn consisting of 13 tubulin dimers in our construction; see Appendix~\ref{sec:geometry_construction}) redirects correlation pathways, and how scaling to larger ordered assemblies or introducing static energetic disorder (random diagonal site-energy fluctuations) and structural disorder (MD-sampled geometric variability) modulates outward emission versus internal retention. We evolve the system with a trace-preserving Lindblad master equation with time-independent rates constructed from the collective radiative decay matrix, and we compare these dynamics with those generated by an effective non-Hermitian generator. We also quantify the degree of non-Markovianity using a dedicated measure of information backflow on reduced subsystems. This approach connects collective radiative physics to information flow in cytoskeletal chromophore networks, identifying the structural and dynamical settings under which microtubules transiently preserve nonclassical correlations.

In neurons, microtubules form long-lived cytoskeletal tracks and regulatory scaffolds whose properties and interactions are shaped by microtubule-associated proteins and by the tubulin code (isotypes and post-translational modifications) \cite{KapiteinHoogenraad2015,JankeMagiera2020}. In this cellular context, our model does not assume that ultraviolet excitations directly implement neuronal computation; rather, it shows that if localized excitation events occur (e.g., photon absorption or oxidative chemistry on aromatic residues), then the resulting dynamics are strongly initial-state dependent. This initial-state dependence can be interpreted as state-selective routing between fast bright channels (rapid export) and slower dark channels (transient retention), providing a concrete way to connect site specificity to testable downstream microtubule-dependent effects and timescales.

The article is organized as follows. Section~\ref{secmodel} describes the model and methods; Section~\ref{secresult} presents the results; and Section~\ref{secconc} discusses the findings and summarizes the study’s implications.

\section{Model and Methods}
\label{secmodel}

As a starting point for describing the dynamics of radiatively coupled dipoles, we first consider a non-Hermitian effective Hamiltonian formalism:
\begin{equation}
\label{non_her_eq}
H_{\mathrm{eff}} = H_0 + \Delta - \frac{i}{2} G.
\end{equation}
which has become widely used in recent literature~\citep{patwa2024quantum,celardo2019existence} because it captures both coherent interactions and radiative decay. Here, $H_0$ denotes on-site excitation energies,
\begin{equation}
H_0 = \sum_{n=0}^{N-1} \hbar \omega_0 |n\rangle \langle n|,\qquad\omega_0 = \frac{2\pi c}{\lambda_0},  \qquad\lambda_0 = 280\,\mathrm{nm}.
\end{equation}
We choose $\lambda_0=280$~nm to match the dominant near-UV absorption/excitation band used for tryptophan-rich proteins (A$_{280}$), so that $k_0=2\pi/\lambda_0$ corresponds to the radiative wavelength scale relevant for Trp transitions.
While $\Delta$ and $G$ represent the coherent dipole-dipole coupling and radiative decay matrices:
\begin{align}
\Delta &= \sum_{n \neq m} \Delta_{nm} |n\rangle \langle m|, \\
G &= \sum_{n} \gamma |n\rangle \langle n| + \sum_{n \neq m} G_{nm} |n\rangle \langle m|.
\end{align}

Diagonalizing $H_{\mathrm{eff}}$ yields complex eigenvalues:
\begin{equation}
E_j = \mathcal{E}_j - \frac{i}{2} \Gamma_j,
\end{equation}
where $\mathcal{E}_j$ is the mode energy and $\Gamma_j$ is its radiative decay rate. Superradiant and subradiant states correspond to modes with high and low $\Gamma_j$, respectively.

While insightful, this approach fails to preserve the trace of the density operator, i.e., it does not guarantee $\mathrm{Tr},\rho(t)=1$ once radiative decay to the ground state is incorporated, and is therefore inadequate for a consistent description of open quantum dynamics in the presence of decoherence. We therefore extract $\Delta_{nm}$ and $G_{nm}$ from the above framework and proceed with a Lindblad master equation description.

\subsection{Hamiltonian and Dipole Coupling Terms}

The full system Hamiltonian used in the Lindblad framework retains the same coherent interaction structure, expressed in the site basis using spin-1/2 excitation creation and excitation annihilation operators:
\begin{align}
H &= H_0 + \Delta, \\
H_0 &= \sum_{n=0}^{N-1} \hbar \omega_0\, \sigma_n^+ \sigma_n^-, \\
\Delta &= \sum_{n \neq m} \Delta_{nm} \left( \sigma_n^+ \sigma_m^- + \sigma_m^+ \sigma_n^- \right).
\end{align}
Here, $\sigma_n^+$ and $\sigma_n^-$ are the excitation creation and annihilation operators for site $n$ in the two-level (single-excitation) representation. The interaction strength $\Delta_{nm}$ between dipoles $n$ and $m$ is given by:
\begin{equation}
\begin{aligned}
&\Delta_{nm} = \frac{3\gamma}{4} \Bigg[ \left( -\frac{\cos \alpha_{nm}}{\alpha_{nm}} + \frac{\sin \alpha_{nm}}{\alpha_{nm}^2} + \frac{\cos \alpha_{nm}}{\alpha_{nm}^3} \right) \hat{\mu}_n \cdot \hat{\mu}_m \\
&\quad - \left( -\frac{\cos \alpha_{nm}}{\alpha_{nm}} + \frac{3\sin \alpha_{nm}}{\alpha_{nm}^2} + \frac{3\cos \alpha_{nm}}{\alpha_{nm}^3} \right) 
(\hat{\mu}_n \cdot \hat{r}_{nm})(\hat{\mu}_m \cdot \hat{r}_{nm}) \Bigg],
\end{aligned}
\end{equation}
where $\hat{\mu}_n$ is the transition dipole moment of site $n$, $\hat{r}_{nm}$ is the unit vector from $n$ to $m$, $r_{nm}=|\mathbf{r}_n-\mathbf{r}_m|$ is the center-to-center separation between dipoles, and $\alpha_{nm} = k_0 r_{nm}$ with $k_0 = \frac{2\pi}{\lambda_0}$.

\subsection{Decay Matrix and Lindblad Formalism}

The collective radiative decay matrix $G$ is similarly computed:
\begin{equation}
\begin{aligned}
&G_{nm} = \frac{3\gamma}{2} \Bigg[ \left( \frac{\sin \alpha_{nm}}{\alpha_{nm}} + \frac{\cos \alpha_{nm}}{\alpha_{nm}^2} - \frac{\sin \alpha_{nm}}{\alpha_{nm}^3} \right) \hat{\mu}_n \cdot \hat{\mu}_m \\
&\quad - \left( \frac{\sin \alpha_{nm}}{\alpha_{nm}} + \frac{3\cos \alpha_{nm}}{\alpha_{nm}^2} - \frac{3\sin \alpha_{nm}}{\alpha_{nm}^3} \right) 
(\hat{\mu}_n \cdot \hat{r}_{nm})(\hat{\mu}_m \cdot \hat{r}_{nm}) \Bigg].
\end{aligned}
\end{equation}

The open quantum system dynamics are then described using the Lindblad master equation:
\begin{equation}
\frac{d\rho(t)}{dt} = -i[H, \rho(t)] + \sum_j \left( L_j \rho(t) L_j^\dagger - \frac{1}{2} \{L_j^\dagger L_j, \rho(t)\} \right),
\end{equation}
where $\rho(t)$ is the system density matrix and $L_j$ are collapse (jump) operators encoding radiative losses; the term ``collapse'' reflects the quantum-trajectory picture in which application of $L_j$ corresponds to an emission event that conditionally updates the system state.

To define $L_j$, we diagonalize the decay matrix:
\begin{equation}
G = V \Lambda V^\dagger, \quad \Lambda = \text{diag}(\gamma_1, \ldots, \gamma_N),
\end{equation}
with $V$ containing the eigenvectors $v^{(j)}$ of $G$. The collapse operators are then constructed as:
\begin{equation}
L_j = \sqrt{\gamma_j} \sum_{n=0}^{N-1} v_n^{(j)} \sigma_n^-.
\end{equation}

This Lindblad-based treatment guarantees trace preservation and consistent thermodynamic behavior while incorporating the same dipole-mediated interactions as the effective non-Hermitian model. In practice, we enforce trace preservation by including an explicit ground (sink) state $|0\rangle$ and radiative jump operators that transfer population from the excitonic manifold to $|0\rangle$; thus emission reduces excited-state population without removing total probability from the enlarged state space. The transition to this formalism thus enables a more accurate and physically complete simulation of quantum dynamics in radiatively interacting systems.

\section{Results}
\label{secresult}

\begin{figure}[t]
  %  \centering
    \includegraphics[width=0.6\linewidth]{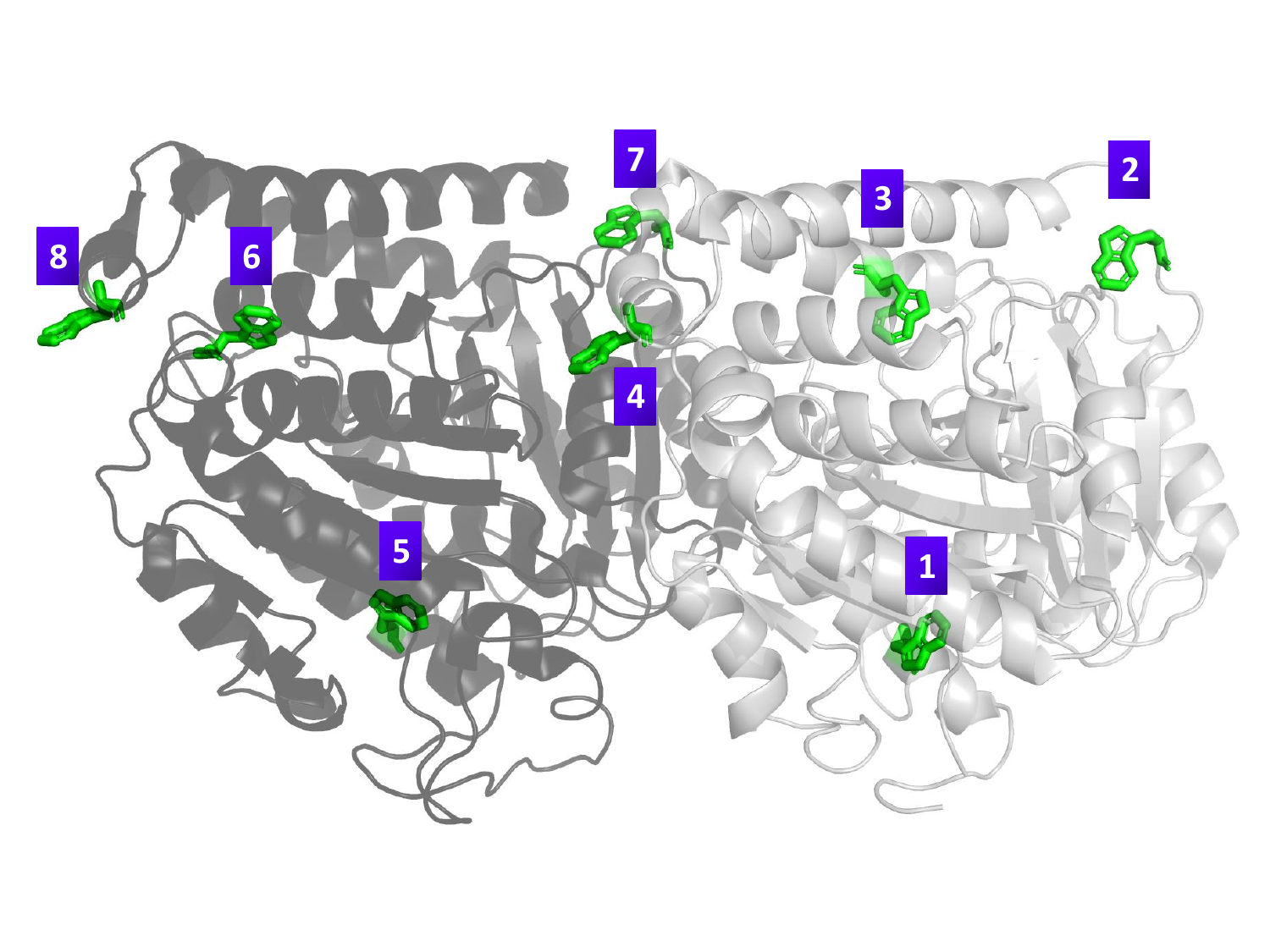}
    \caption{Structure of the tubulin dimer from Protein Data Bank (PDB) entry 1JFF, rendered using PyMOL (molecular visualization software) \cite{DeLano2002PyMOL}, highlighting the positions of the eight tryptophan residues. The $\alpha$-tubulin chain is shown in light gray and the $\beta$-tubulin chain in dark gray, with tryptophan residues displayed in green. Purple-labeled numbers correspond to specific tryptophan residues analyzed in the study: Trp1 ($\alpha$ 21), Trp2 ($\alpha$ 346), Trp3 ($\alpha$ 388), Trp4 ($\alpha$ 407), Trp5 ($\beta$ 21), Trp6 ($\beta$ 103), Trp7 ($\beta$ 346), and Trp8 ($\beta$ 407).}
    \label{fig:tub}
\end{figure}

In this section we report excitation dynamics in networks of eight tryptophan chromophores (see figure \ref{fig:tub}) embedded in a tubulin environment. The evolution is modeled with a Lindblad master equation built from an effective non-Hermitian generator that includes radiative loss, using site-specific positions and dipole orientations from structural data (see Appendix \ref{sec:geometry_construction}). Simulations are performed in QuTiP~\cite{johansson2012qutip}.

We consider five initial preparations that probe distinct routes for information flow: (i) superradiant eigenstates of \(H_{\mathrm{eff}}\); (ii) subradiant eigenstates of \(H_{\mathrm{eff}}\); (iii) a maximally coherent single excitation with uniform phase across the sites; (iv) a maximally mixed single excitation over the sites; and (v) a localized single site excitation. For the single-dimer 8-site network, the Lindblad simulations reported here span ps to tens of ns (e.g., 0--5~ns in Fig.~\ref{Figsup}, 0--80~ns in Fig.~\ref{Figsub}, and 0--15~ns in Figs.~\ref{Figfull}--\ref{Figmix}), as set by the time axes in each figure. We report site resolved populations together with quantum correlation measures, including the \(L_1\) norm of coherence, pairwise coherence among chromophores, and logarithmic negativity as an entanglement witness. When we refer to a chromophore pair \((i,j)\), the indices label the tryptophan site numbers Trp\(i\) and Trp\(j\) defined in Fig.~\ref{fig:tub}. Definitions and computational details for these measures are provided in Appendix~\ref{app:coh}.

Throughout the results, we track how correlations are generated, routed, and dissipated across the network, and how preparation and geometry shape the directionality and persistence of nonclassical information flow.

\subsection{Dynamics from the Superradiant Eigenstate}

\begin{figure}[t]
    \centering
    \subfloat[]{\includegraphics[width=0.40\linewidth]{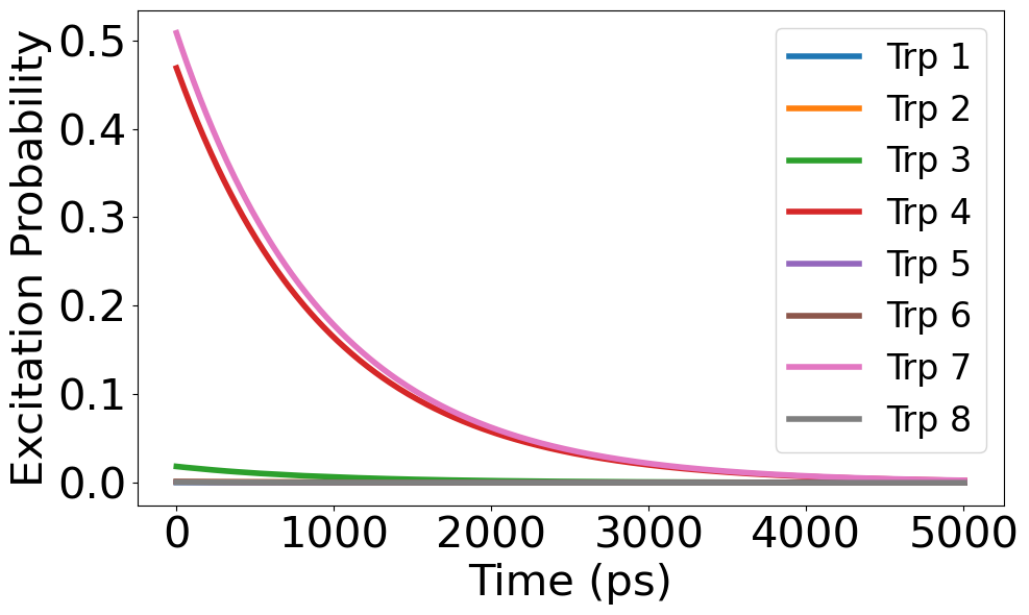} \label{Figsup:a}}
    \quad
    \subfloat[]{\includegraphics[width=0.40\linewidth]{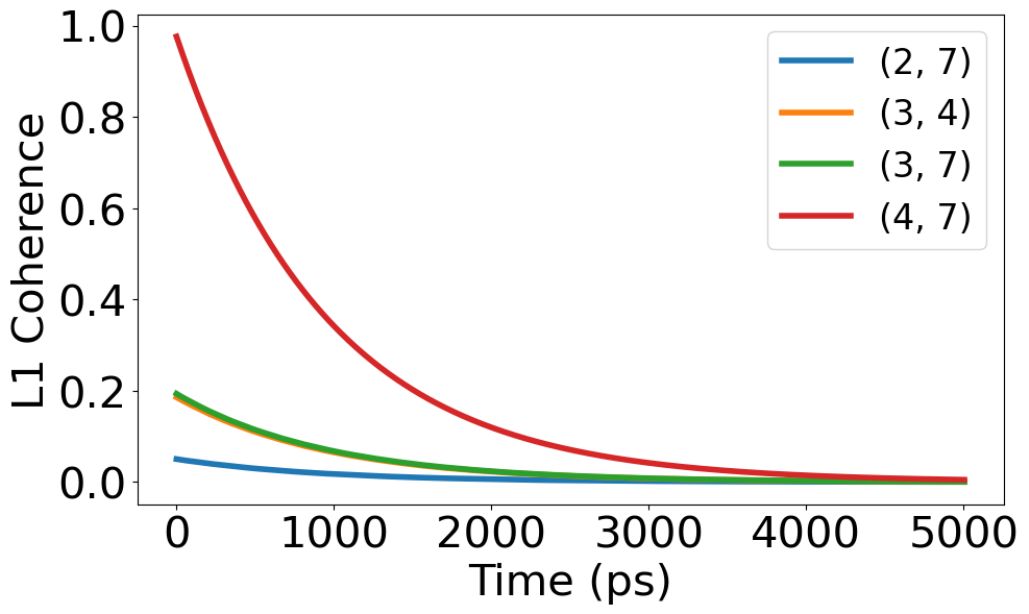} \label{Figsup:b}} \\
    \subfloat[]{\includegraphics[width=0.40\linewidth]{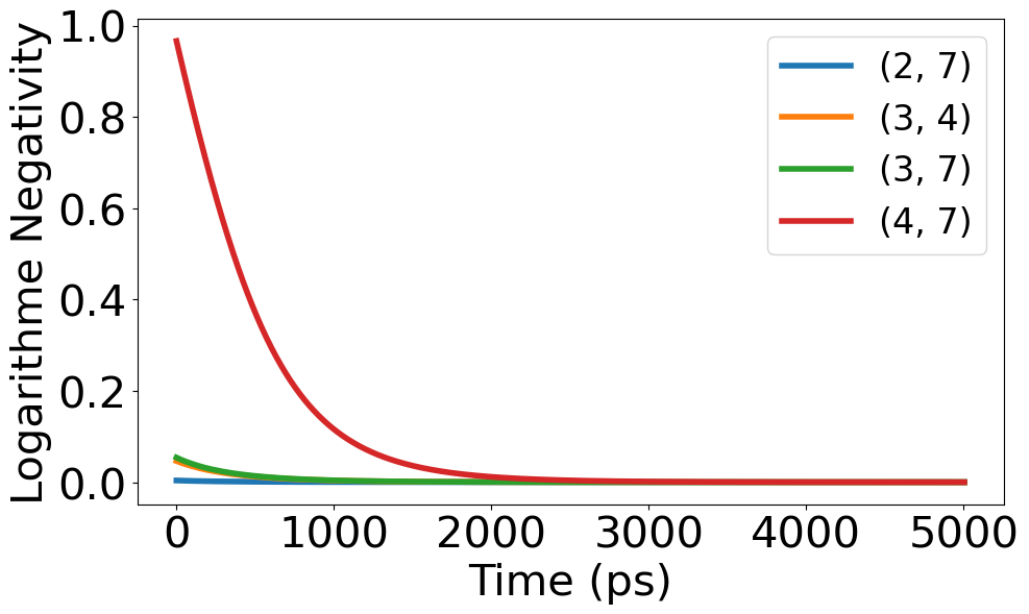} \label{Figsup:c}}
    \caption{Excitation dynamics for the initial state corresponding to the superradiant eigenstate of the non-Hermitian Hamiltonian. (a) shows rapid decay of excitation population across all tryptophan sites, characteristic of strong radiative coupling. (b) illustrates the rapid loss of coherence via the $L_1$ norm for the four most coherent chromophore pairs. (c) presents the logarithmic negativity, which peaks briefly before vanishing, indicating transient entanglement that dissipates alongside the excitation. Pairs $(i,j)$ denote tryptophan site indices (labels) defined in Fig.~\ref{fig:tub}.} Time is reported in picoseconds (ps) in all panels.
    \label{Figsup}
\end{figure}

When the system is initialized in the superradiant eigenstate of the non-Hermitian Hamiltonian, corresponding to the mode with the highest radiative decay rate, the excitation undergoes rapid collective dissipation. As shown in Figure~\ref{Figsup:a}, site populations decay almost synchronously, reflecting strong collective coupling to the radiation field. In this regime, the primary channel for information flow is outward: energy and correlations are efficiently transferred to the environment rather than redistributed within the network.

The evolution of correlations is illustrated in Figure~\ref{Figsup:b}, where the $L_1$ norm of coherence for the four most correlated chromophore pairs drops sharply within the first $\sim 10^3$~ps (about 1~ns) and is largely extinguished over the 0--5~ns window shown (0--5000~ps). The coherence carried by inter-site correlations is radiated away together with the excitation, indicating that the information flow is dominated by rapid leakage rather than internal exchange. The ns-scale lifetime of off-diagonal terms in the density matrix demonstrates that collective enhancement of emission occurs at the cost of retaining local or pairwise correlations.

Figure~\ref{Figsup:c} shows the corresponding logarithmic negativity. Entanglement appears briefly at early times as the excitation delocalizes, but it vanishes quickly as the system relaxes through the radiative channel. This transient entanglement marks a brief surge in correlation exchange before the information is lost to the environment. Overall, the superradiant eigenstate behaves as a fast exporting channel of quantum information, maximizing collective emission but minimizing internal retention of nonclassical correlations.

\subsection{Dynamics from the Subradiant Eigenstate}

\begin{figure}[t]
    \centering
    \subfloat[]{\includegraphics[width=0.40\linewidth]{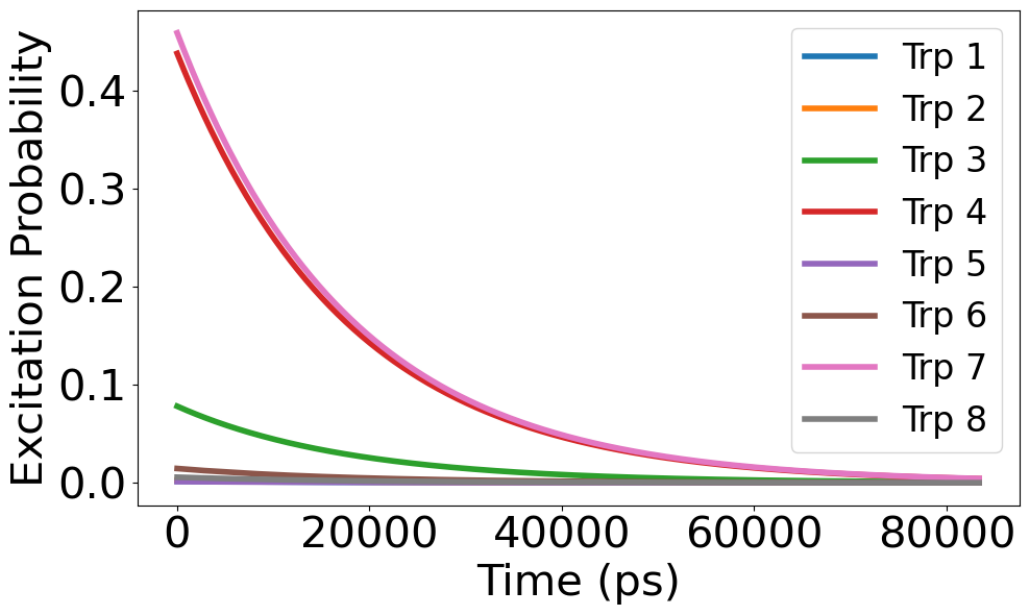} \label{Figsub:a}}
    \quad
    \subfloat[]{\includegraphics[width=0.40\linewidth]{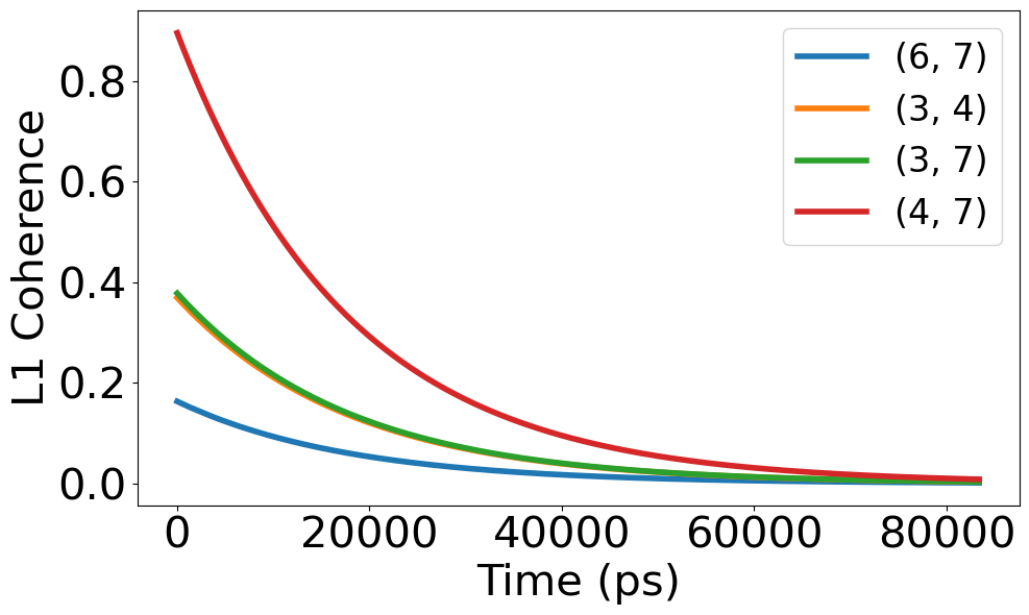} \label{Figsub:b}} \\
    \subfloat[]{\includegraphics[width=0.40\linewidth]{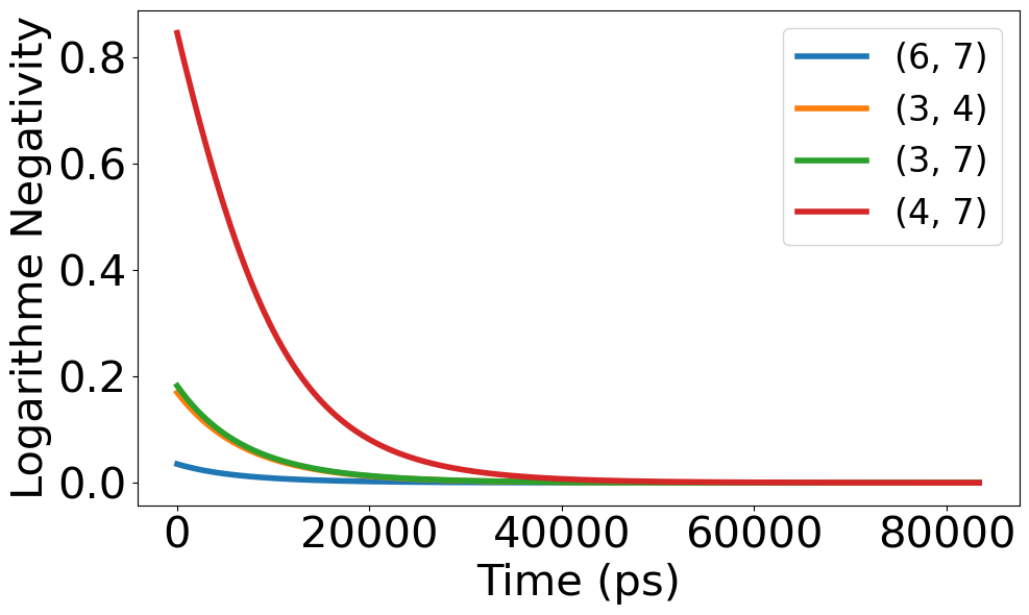} \label{Figsub:c}}
    \caption{Excitation dynamics for an initial state corresponding to the most subradiant eigenstate of the non-Hermitian Hamiltonian. (a) shows the site-resolved excitation population, which decays very slowly, indicating suppression of radiative losses. (b) illustrates the $L_1$ norm of coherence for the four most coherent chromophore pairs, remaining high throughout the evolution. (c) presents the logarithmic negativity between those pairs, showing sustained and robust bipartite entanglement over time. Pairs $(i,j)$ denote tryptophan site indices (labels) defined in Fig.~\ref{fig:tub}. Time is reported in picoseconds (ps) in all panels.}
    \label{Figsub}
\end{figure}

In contrast to the superradiant preparation, initializing the system in the most subradiant eigenstate of the non-Hermitian Hamiltonian yields a much slower release of excitation. As shown in Figure~\ref{Figsub:a}, populations remain within the network for tens of nanoseconds (0--80~ns, i.e., 0--$8\times10^4$~ps in the plotted window), which indicates strong suppression of radiative loss. In this setting the dominant direction of information flow is internal rather than outward, with excitations and correlations circulating among chromophores before any leakage to the environment.

Correlation dynamics follow the same pattern. In Figure~\ref{Figsub:b}, the $L_1$ norm for the four most strongly correlated chromophore pairs remains elevated throughout the full 0--80~ns window shown, demonstrating that off-diagonal terms in the density matrix persist even under weak decay. This persistence reflects sustained internal exchange of phase and amplitude information across the network.

Figure~\ref{Figsub:c} reports the logarithmic negativity for the same pairs. Entanglement remains stable with only gradual attenuation and clear oscillatory structure, consistent with recurrent redistribution of correlations within the subspace protected from radiation. Taken together, these trends identify the subradiant eigenstate as an internal retention channel for quantum information, where population, coherence, and entanglement are preserved and recirculated within the network rather than being quickly exported to the environment.

\subsection{Uniformly Shared Initial Excitation}

To explore how spatial distribution of the initial excitation shapes correlation dynamics and information flow, we compare two limiting cases: a fully coherent delocalized state and a completely incoherent mixed state, each uniformly involving all eight tryptophan chromophores.

\subsubsection{Fully Coherent Initial State}

We first consider a symmetric and fully coherent initial state where the excitation is delocalized equally across all sites,
\begin{equation}
    |\psi(0)\rangle = \frac{1}{\sqrt{8}} \sum_{j=1}^{8} |j\rangle,
\end{equation}
with $|j\rangle$ denoting an excitation localized on the $j$th chromophore.

\begin{figure}[t]
    \centering
    \subfloat[]{\includegraphics[width=0.40\linewidth]{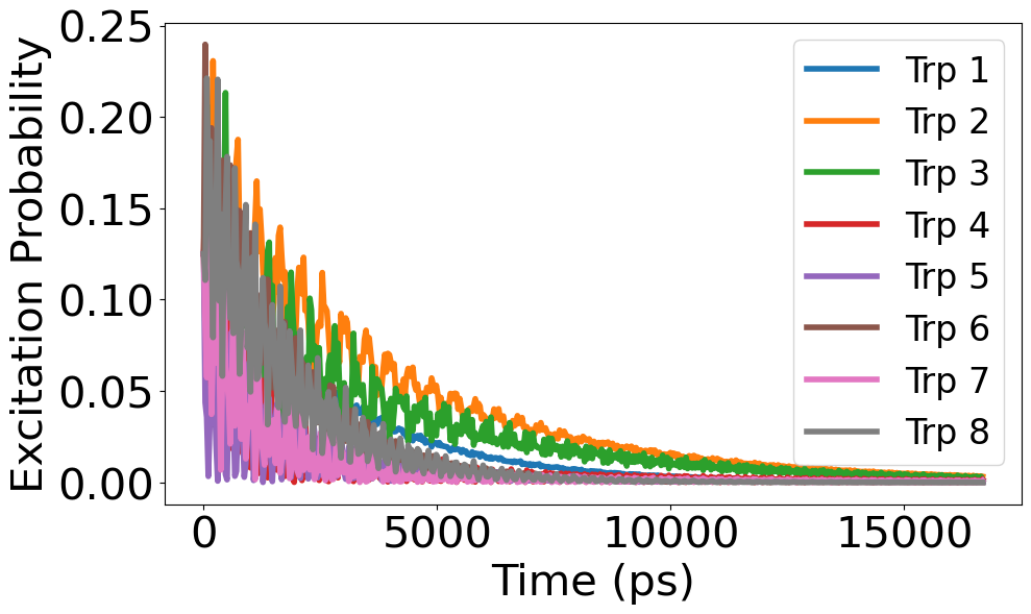} \label{Figfull:a}}
    \quad
    \subfloat[]{\includegraphics[width=0.40\linewidth]{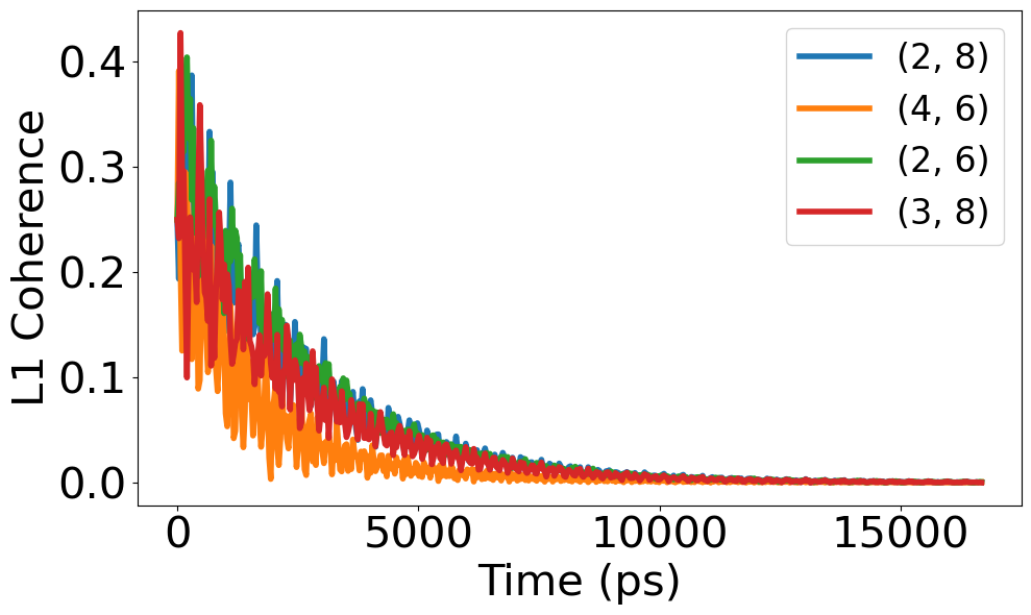} \label{Figfull:b}} \\
    \subfloat[]{\includegraphics[width=0.40\linewidth]{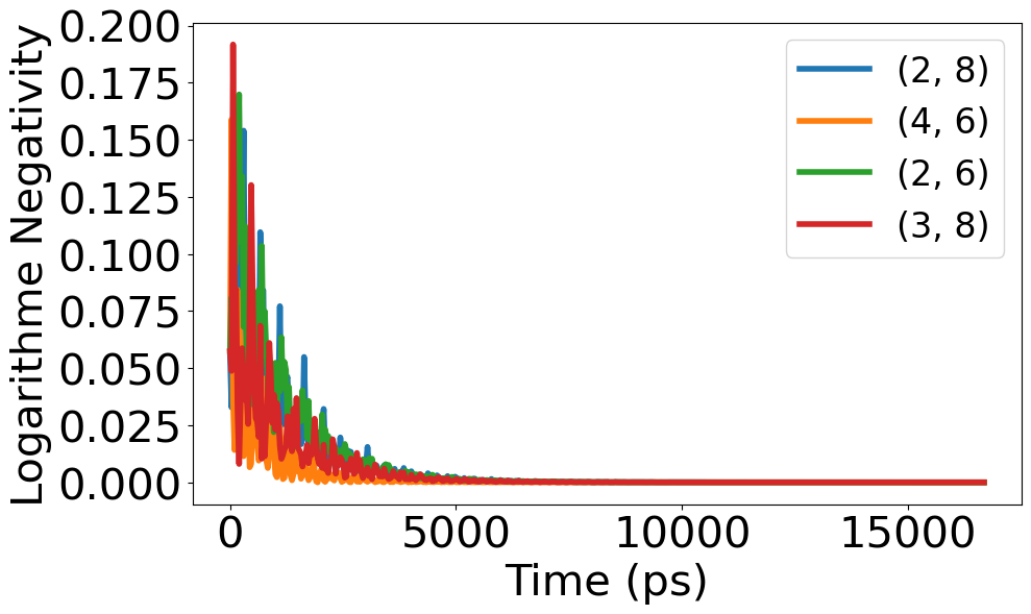} \label{Figfull:c}}
    \quad
    \subfloat[]{\includegraphics[width=0.40\linewidth]{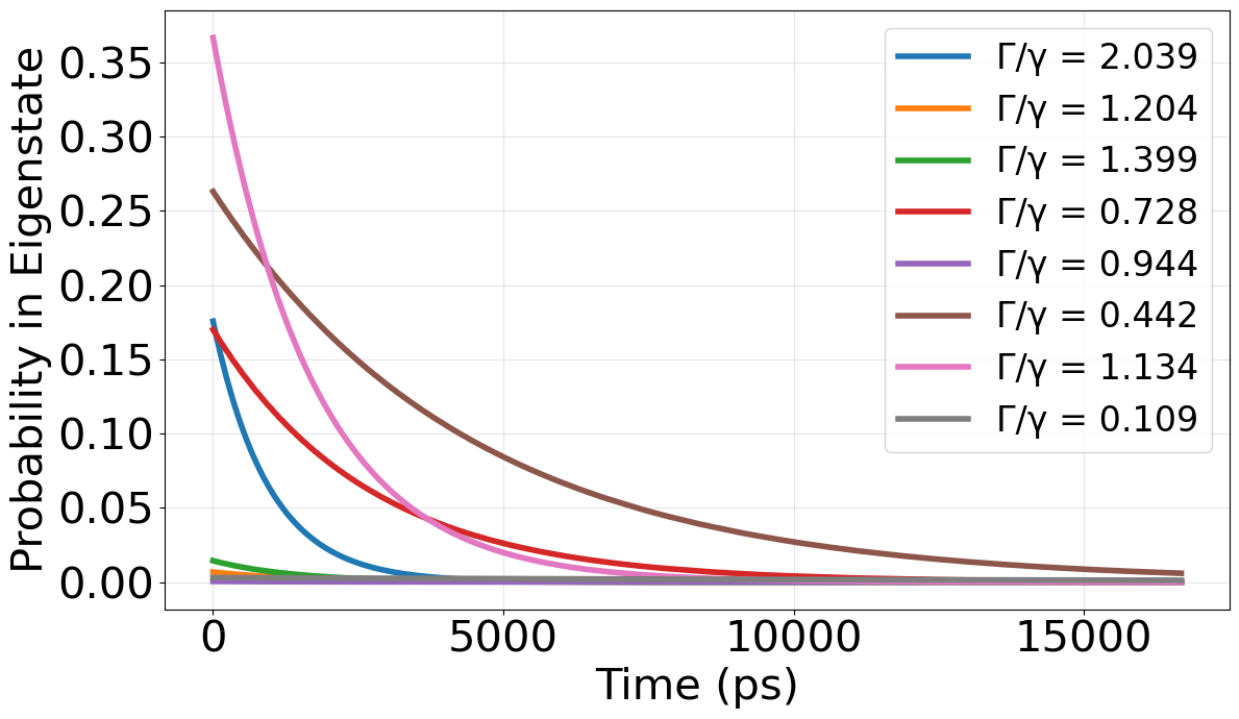} \label{Figfull:d}}
    \caption{Excitation dynamics for a fully coherent initial state delocalized across all eight tryptophan sites. (a) Site-resolved excitation populations over time. (b) $L_1$ norm of coherence for selected chromophore pairs. (c) Logarithmic negativity showing entanglement dynamics. (d) Projection onto the eigenmodes of the effective non-Hermitian Hamiltonian $H_{\mathrm{eff}}$, classified by their collective radiative rates $\Gamma_j$ relative to the single-site rate $\gamma$ (superradiant/bright: $\Gamma_j/\gamma>1$; subradiant/dark: $\Gamma_j/\gamma<1$). The bright-to-dark crossover is identified when the total projected weight in modes with $\Gamma_j/\gamma<1$ exceeds that in modes with $\Gamma_j/\gamma>1$. Pairs $(i,j)$ denote tryptophan site indices (labels) defined in Fig.~\ref{fig:tub}. Time is reported in picoseconds (ps) in all panels.}
    \label{Figfull}
\end{figure}

The resulting dynamics, shown in Figure~\ref{Figfull}, reveal that correlations propagate through the network as oscillatory information flow between sites. In Figure~\ref{Figfull:a}, population exchange displays clear interference patterns rather than simple exponential loss, indicating that coherent delocalization enables partial protection from radiative decay through destructive interference. 

The correlation measures in Figure~\ref{Figfull:b} show that the $L_1$ norm for selected chromophore pairs remains high over the full 0--15~ns window shown (0--15000~ps), confirming that internal information exchange persists even as total excitation diminishes. Entanglement, shown in Figure~\ref{Figfull:c} through logarithmic negativity, rises rapidly and then decays slowly over the same 0--15~ns window, indicating sustained quantum information sharing across multiple chromophore pairs.

Population projections onto the eigenstates of the non-Hermitian Hamiltonian (Figure~\ref{Figfull:d}) show that while both superradiant and subradiant components are initially populated, the dynamics naturally channel population toward subradiant sectors. This gradual self-selection redistributes information flow toward correlation preserving subspaces, where coherence and entanglement remain protected under radiative loss.

\subsubsection{Fully Incoherent Mixed State}

As a contrasting limit, we consider a maximally mixed initial state where excitation is equally distributed across all sites but carries no initial phase correlation,
\begin{equation}
    \rho(0) = \frac{1}{8} \sum_{j=1}^{8} |j\rangle\langle j|.
\end{equation}

\begin{figure}[t]
    \centering
    \subfloat[]{\includegraphics[width=0.40\linewidth]{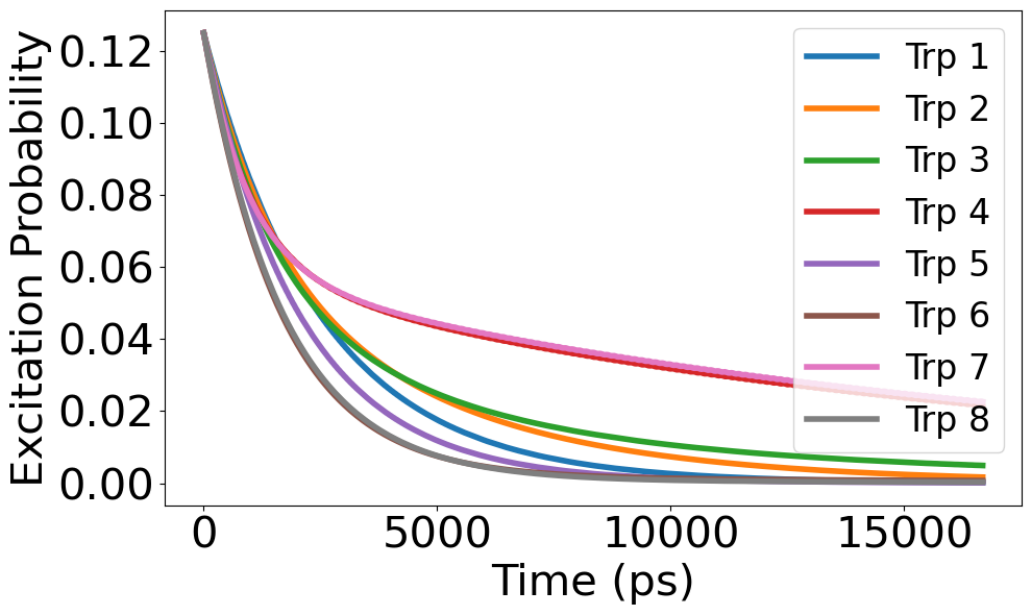} \label{Figmix:a}}
    \quad
    \subfloat[]{\includegraphics[width=0.40\linewidth]{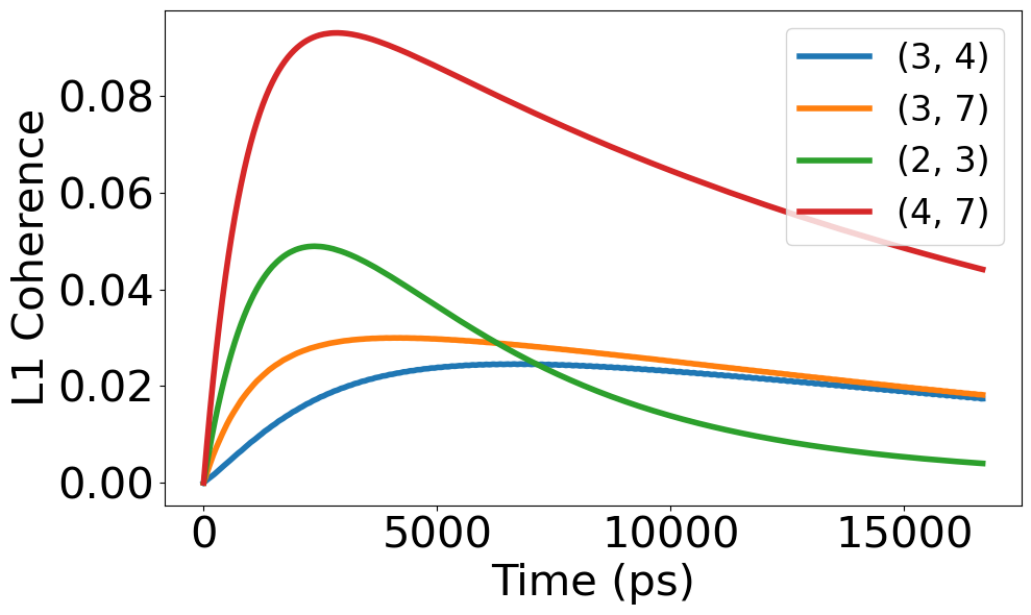} \label{Figmix:b}} \\
    \subfloat[]{\includegraphics[width=0.40\linewidth]{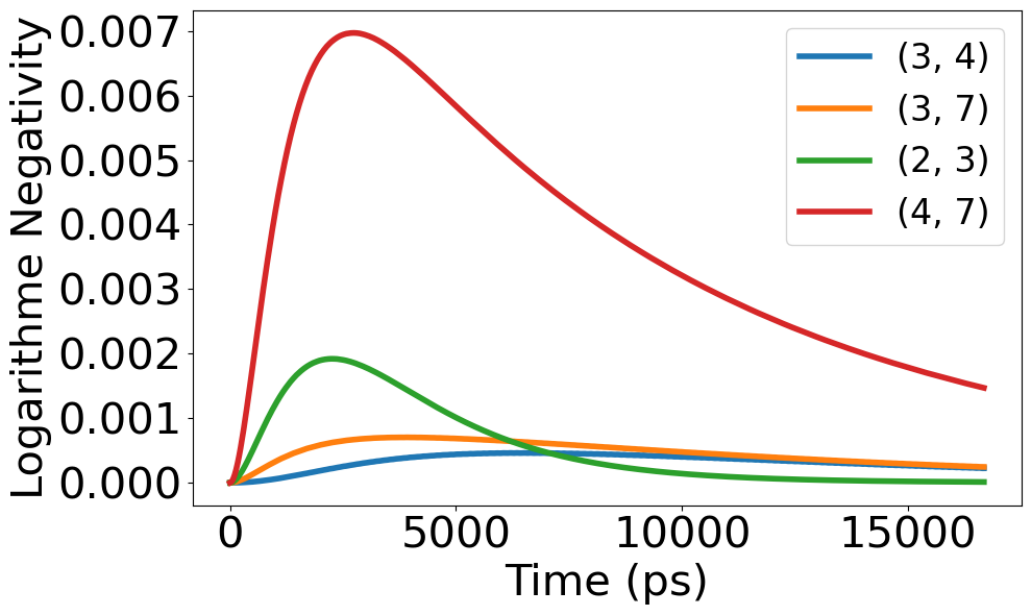} \label{Figmix:c}}
    \quad
    \subfloat[]{\includegraphics[width=0.40\linewidth]{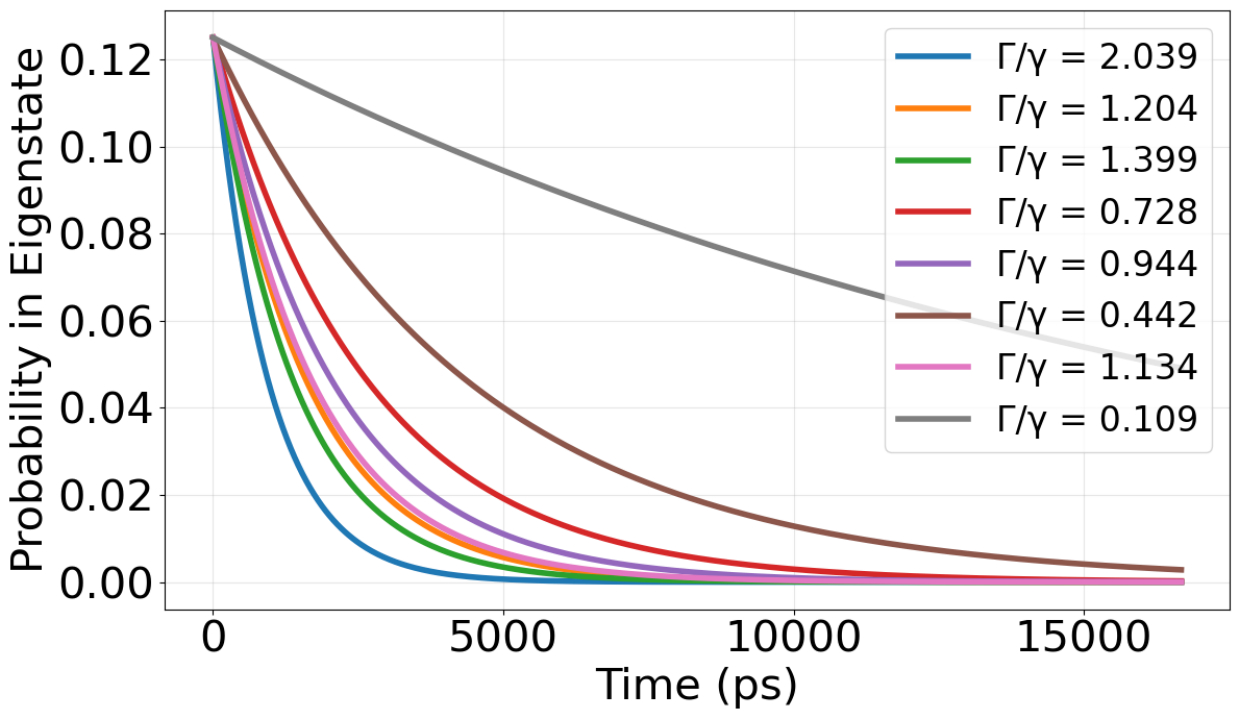} \label{Figmix:d}}
    \caption{Excitation dynamics for a fully incoherent mixed state uniformly distributed over all eight tryptophan sites. (a) Site-resolved excitation populations. (b) $L_1$ norm of coherence for selected chromophore pairs. (c) Logarithmic negativity reflecting lack of entanglement. (d) Projection onto eigenstates of the non-Hermitian Hamiltonian: the plotted weights indicate how population overlaps with eigenmodes $j$ of $H_{\mathrm{eff}}$, each characterized by a collective radiative rate $\Gamma_j$; modes with $\Gamma_j/\gamma>1$ are superradiant (bright) and those with $\Gamma_j/\gamma<1$ are subradiant (dark). For a fully mixed initial condition, the projection is broadly distributed and does not selectively target either sector, so the decay is effectively non-preferential across bright and dark channels. Pairs $(i,j)$ denote tryptophan site indices (labels) defined in Fig.~\ref{fig:tub}. Time is reported in picoseconds (ps) in all panels.}
    \label{Figmix}
\end{figure}

The corresponding dynamics, shown in Figure~\ref{Figmix}, demonstrate that the absence of initial coherence prevents any internal information circulation. In Figure~\ref{Figmix:a}, populations decay monotonically and nearly uniformly across all sites, with no evidence of interference mediated redistribution. The $L_1$ norm in Figure~\ref{Figmix:b} remains near zero, confirming that no new coherent correlations are generated during the evolution. Likewise, the logarithmic negativity in Figure~\ref{Figmix:c} stays negligible, indicating that entanglement does not emerge spontaneously in the absence of coherent phase relations.

The eigenstate population projections in Figure~\ref{Figmix:d} show that both superradiant and subradiant components are initially populated, but without phase coherence the dynamics do not preferentially channel excitation into the subradiant manifold. Information flow therefore proceeds primarily outward into the environment, resulting in rapid loss of population and vanishing internal correlations. This highlights that coherence is a prerequisite for sustained correlation transport and retention of quantum information within the microtubule network.

\subsection{Site Localized Initial Excitations}

To reflect biologically plausible conditions, we examine dynamics when the system is prepared with a single excitation localized on each of the eight tryptophan sites in turn. This mimics natural events in which a photon is absorbed by one chromophore and seeds a site-specific initial state, or in which reactive oxygen species generate localized electronic excitations on aromatic residues under oxidative stress conditions \cite{kurian2017oxidative}.

\begin{figure}[t]
    \centering
    \subfloat[]{\includegraphics[width=0.40\linewidth]{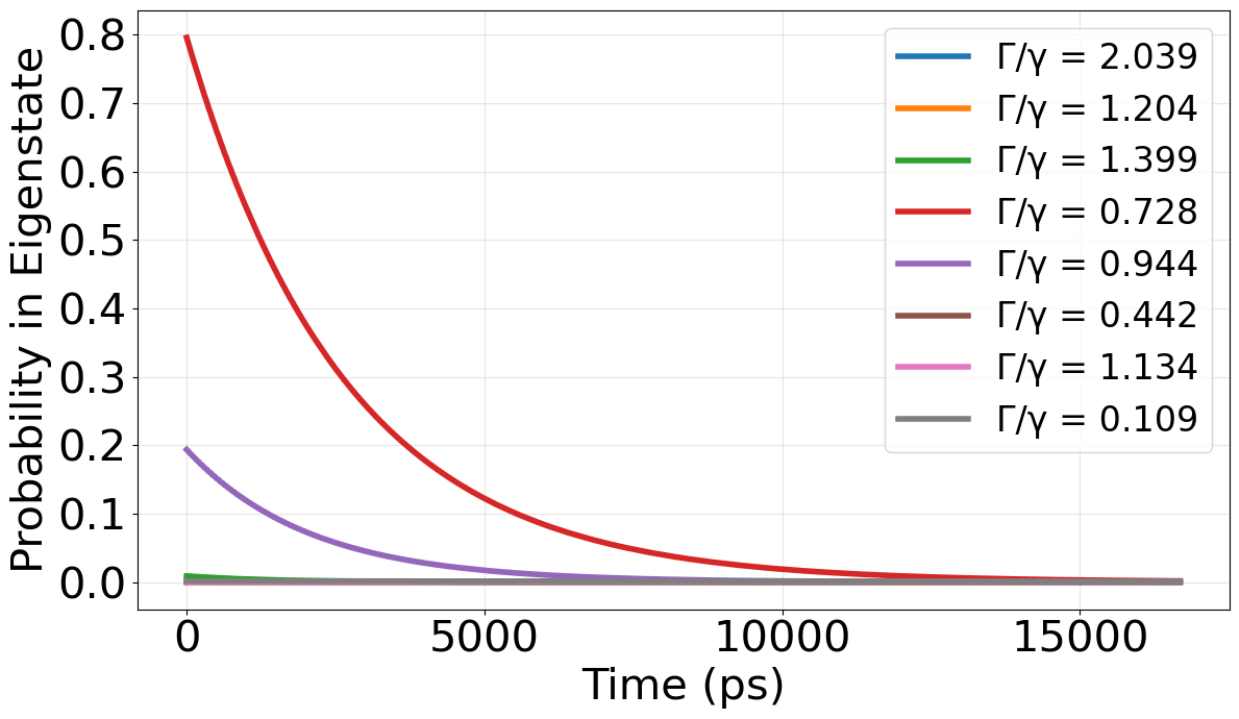} \label{poptrp:a}}
    \quad
    \subfloat[]{\includegraphics[width=0.40\linewidth]{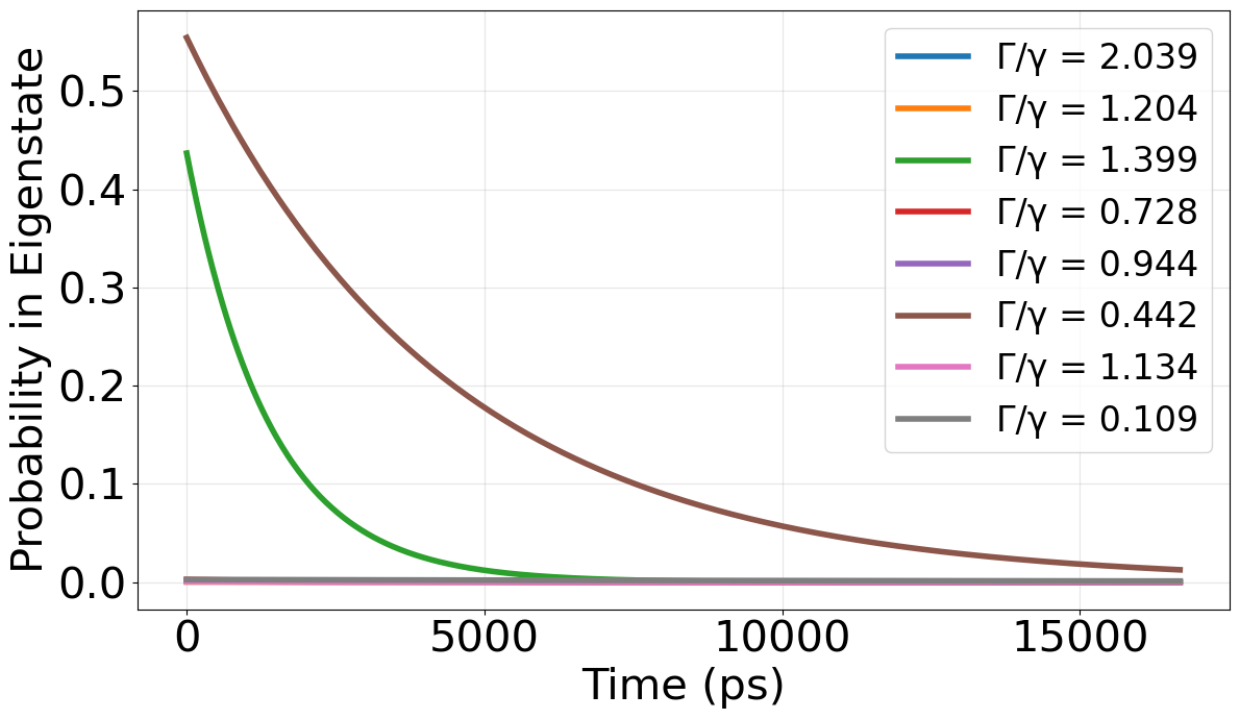} \label{poptrp:b}} \\
    \subfloat[]{\includegraphics[width=0.40\linewidth]{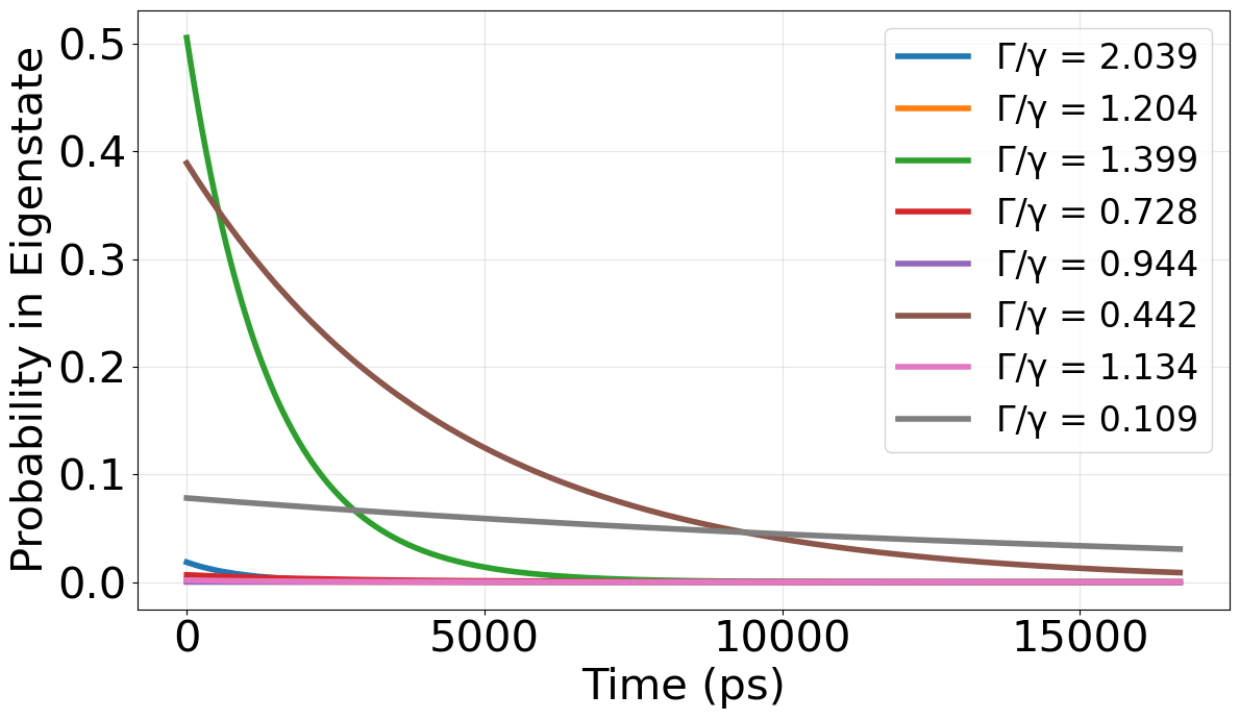} \label{poptrp:c}}
    \quad
    \subfloat[]{\includegraphics[width=0.40\linewidth]{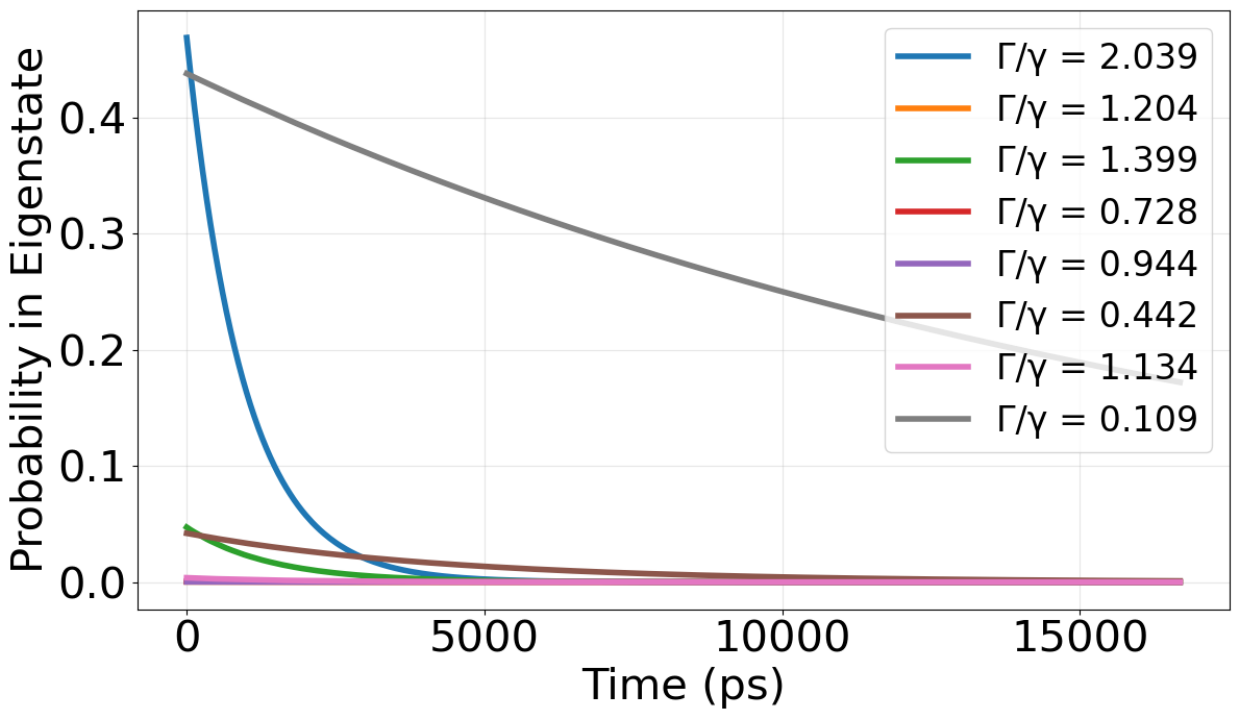} \label{poptrp:d}} \\
    \subfloat[]{\includegraphics[width=0.40\linewidth]{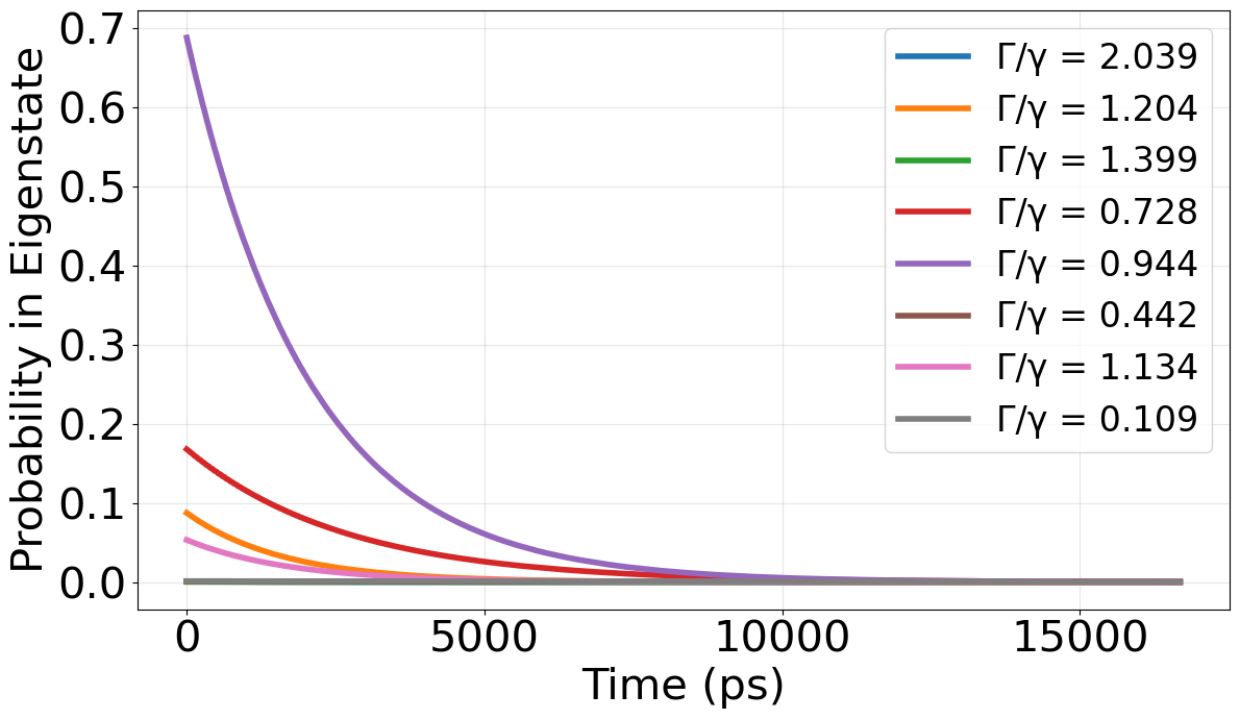} \label{poptrp:e}}
    \quad
    \subfloat[]{\includegraphics[width=0.40\linewidth]{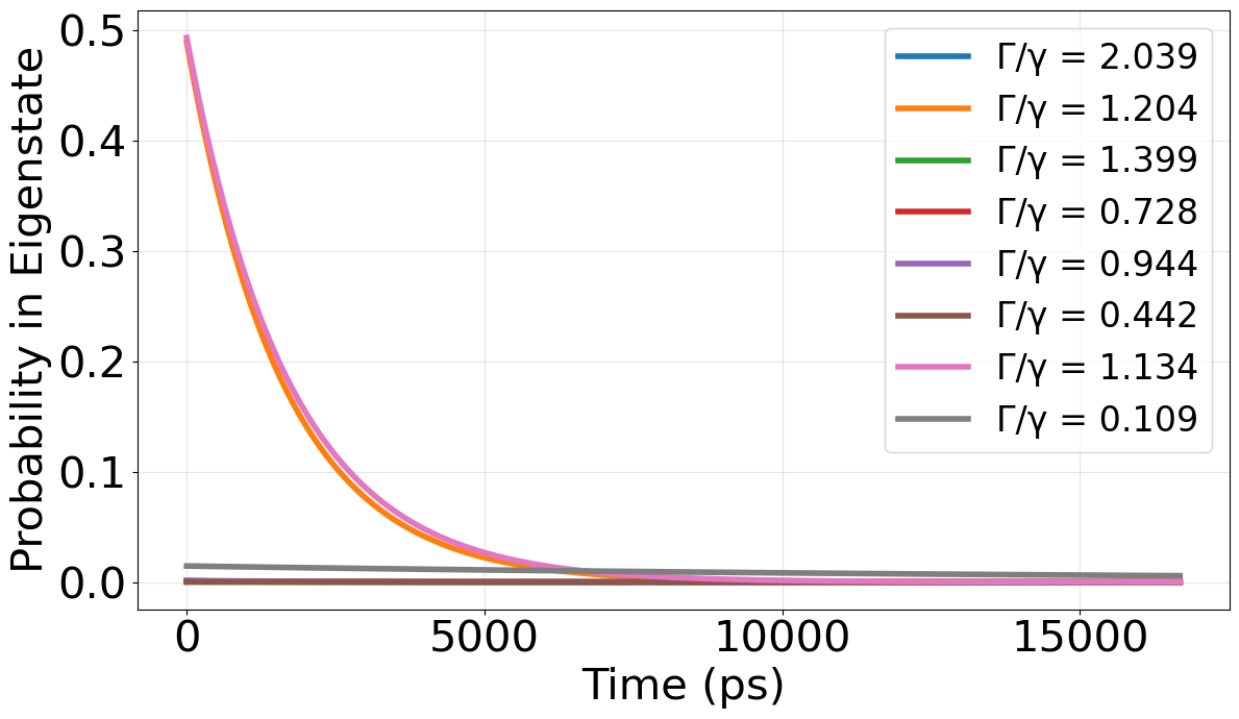} \label{poptrp:f}} \\
    \subfloat[]{\includegraphics[width=0.40\linewidth]{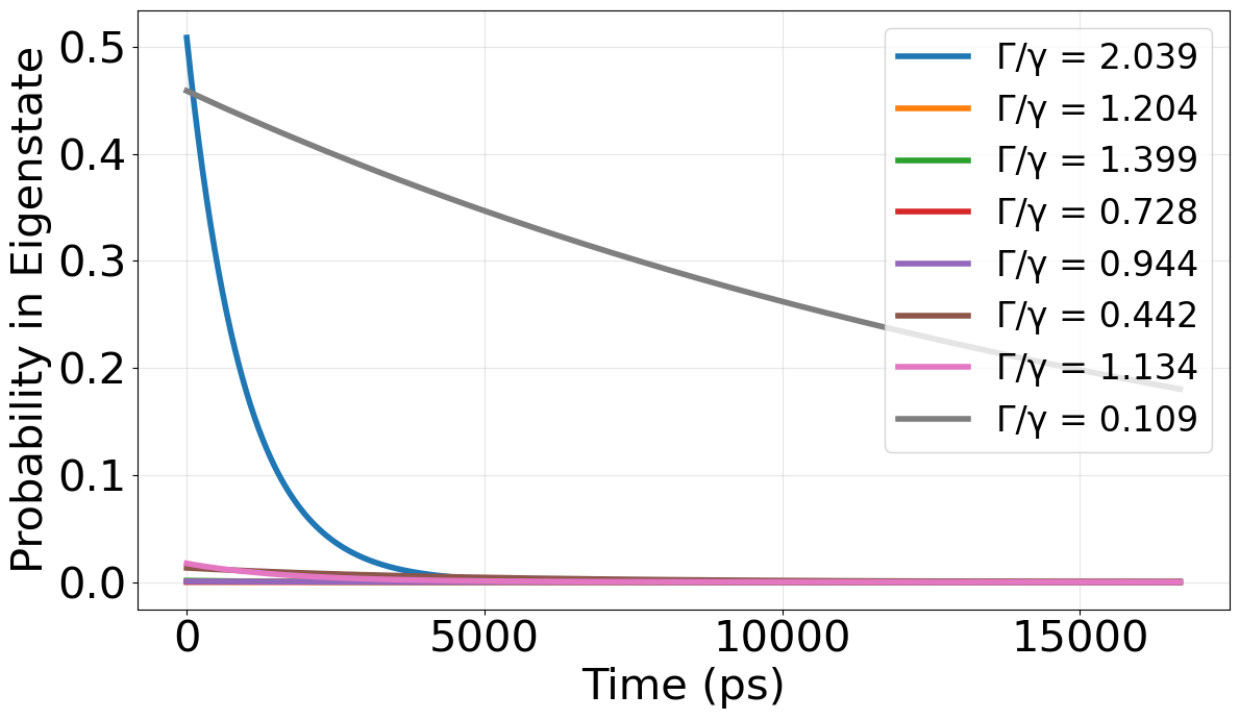} \label{poptrp:g}}
    \quad
    \subfloat[]{\includegraphics[width=0.40\linewidth]{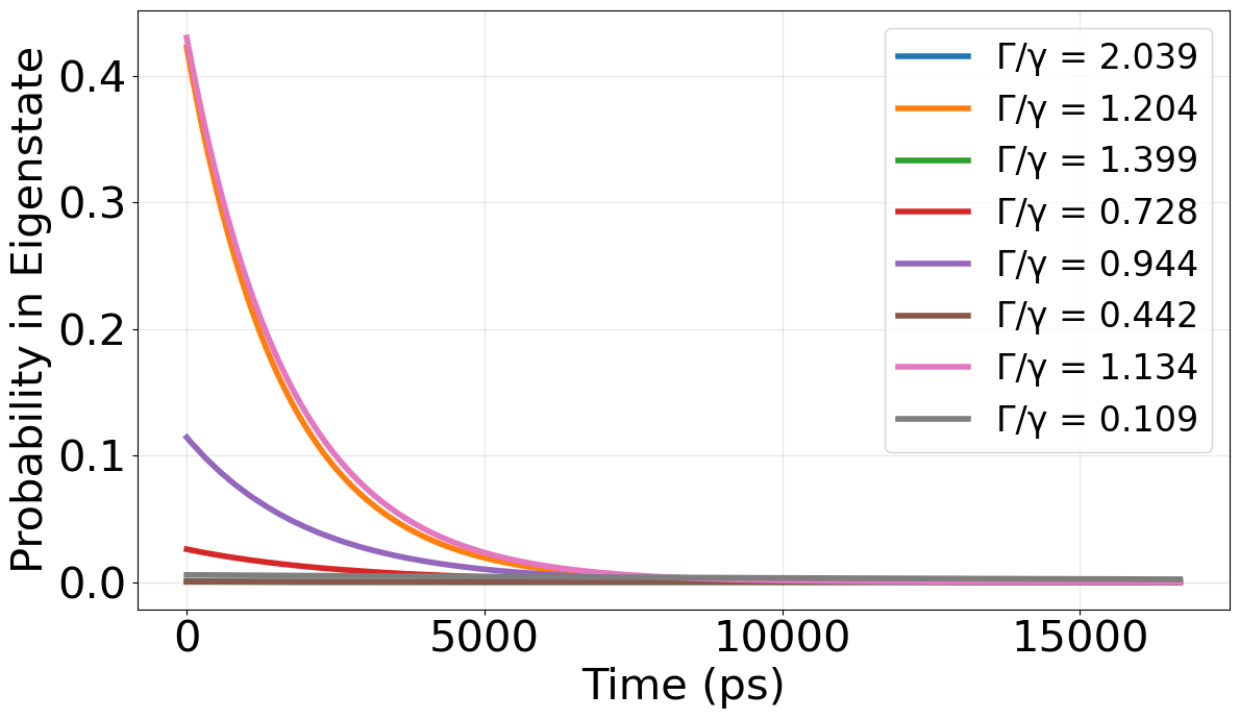} \label{poptrp:h}}
    \caption{Population dynamics for site-localized initial excitations. Each panel shows the site-resolved excitation population over time for an initial excitation localized at a specific tryptophan chromophore: (a) Trp1, (b) Trp2, (c) Trp3, (d) Trp4, (e) Trp5, (f) Trp6, (g) Trp7, (h) Trp8. To interpret the differences in decay/envelope across initial sites, we also analyze the projection onto eigenmodes of the effective non-Hermitian Hamiltonian $H_{\mathrm{eff}}$, classified by their collective radiative rates $\Gamma_j$ relative to the single-site rate $\gamma$ (superradiant/bright: $\Gamma_j/\gamma>1$; subradiant/dark: $\Gamma_j/\gamma<1$); a bright-to-dark crossover can be identified when the total projected weight in modes with $\Gamma_j/\gamma<1$ exceeds that in modes with $\Gamma_j/\gamma>1$. Time is reported in picoseconds (ps) in all panels.}
    \label{Figpoptrp}
\end{figure}

The population traces for all eight preparations are shown in Figure~\ref{Figpoptrp}(a to h). (The time axis in Fig.~\ref{Figpoptrp} spans the same ps--ns window used for the single-dimer simulations, enabling direct comparison of faster vs.\ slower leakage across sites.) The temporal behavior depends strongly on the injection site. Preparations at Trp4 or Trp7 yield markedly slower population decay, while preparations at Trp1 or Trp5 relax more rapidly. This site dependence reveals different couplings to radiative channels. Localized states with larger overlap on subradiant sectors of the non-Hermitian spectrum retain population internally for longer, whereas those aligned with superradiant sectors export excitation quickly.

In terms of information flow, a localized injection sets the initial direction for correlation propagation. Sites such as Trp4 and Trp7 not only slow population leakage but also promote internal redistribution of phase and amplitude information before loss to the environment. In contrast, injections at Trp1 or Trp5 favor outward flow, leaving little time for correlations to circulate within the network. Decomposition of each preparation into radiative and protected components therefore acts as a site-selective router that steers both energy and correlations along distinct pathways.

These observations show that spatial location controls access to long lived correlation preserving subspaces. Consequently, site-specific excitation can select quantum lifetimes and shape the balance between internal correlation transport and external emission, a principle that may be relevant for natural light harvesting and for targeted control strategies in bio-inspired excitonic platforms. More specifically, the injection residue sets the overlap with bright versus dark radiative sectors, biasing the dynamics toward rapid export or transient retention; in vivo this bias could be shaped by oxidation hotspots, local binding environments, and tubulin-state regulation \cite{Ehrenshaft2015,Schoneich2018,Kalra2023,JankeMagiera2020,GoodsonJonasson2018}.

\subsection{Coherence Transfer from One Tubulin to a Spiral}
\label{subsec:one-to-spiral}

\begin{figure}[t]
    \centering
    \includegraphics[width=1\linewidth]{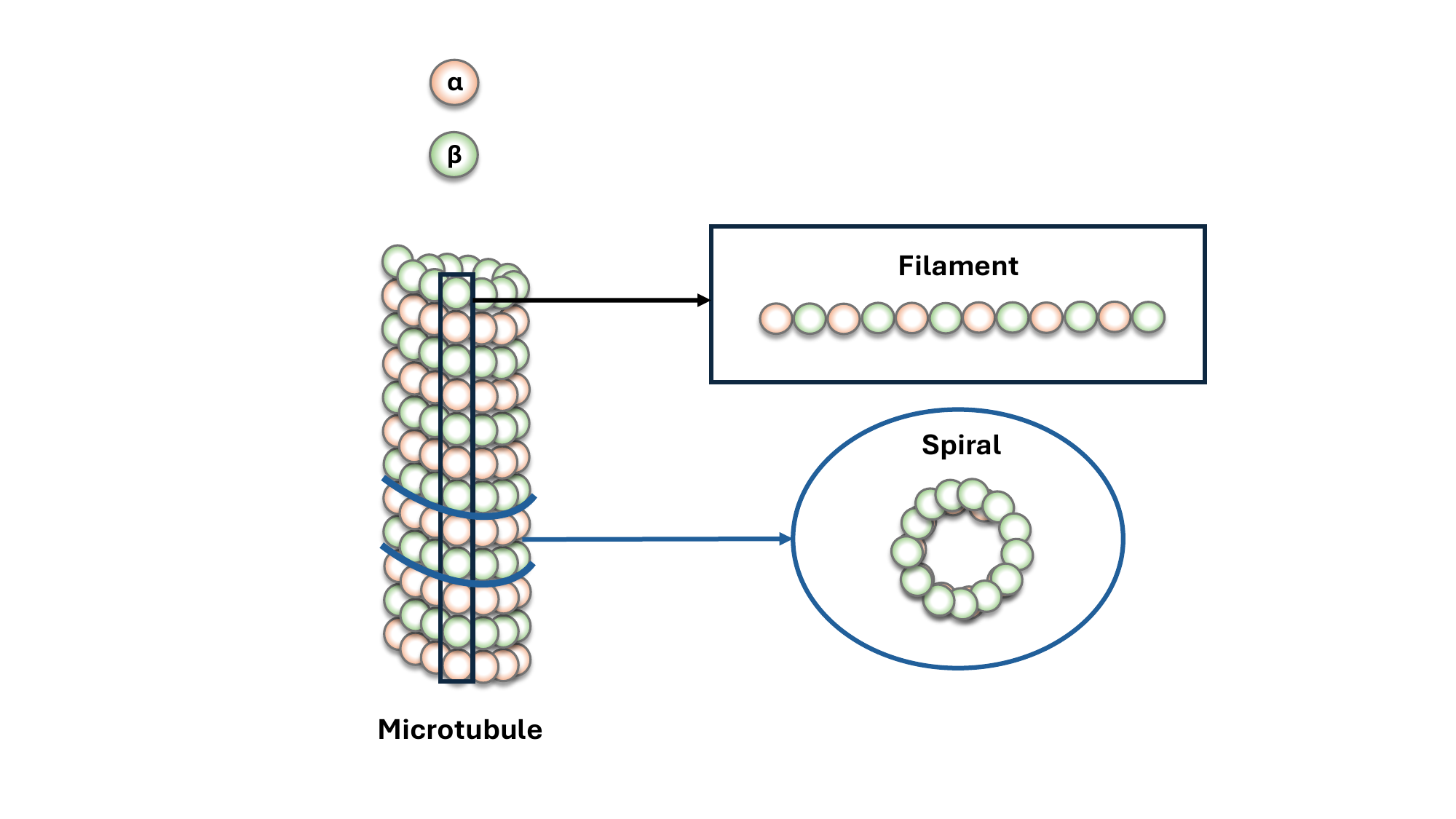}
    \caption{
    Schematic defining the subsystem partitions used in this work. A filament denotes a protofilament-aligned chain of $\alpha\beta$ tubulin dimers along the microtubule longitudinal axis (black rectangle). A spiral denotes one circumferential turn (top/cross-section view) consisting of 13 $\alpha\beta$ dimers around the microtubule cylinder (blue ring). Microtubule segments with $S$ spirals correspond to stacking $S$ such circumferential turns along the longitudinal axis. Orange and green indicate $\alpha$- and $\beta$-tubulin, respectively (not to scale).
    }
    \label{Fig:spiral_filament_schematic}
\end{figure}

In the previous subsection, we analyzed dynamics within a single tubulin dimer. Here, we keep the same focal dimer and the same initial preparations, but progressively change its environment: first by placing it in a two-tubulin system, and then in microtubule segments containing one and two spirals. In our construction, a “spiral” denotes one circumferential turn of the microtubule lattice consisting of 13 dimers around the cylinder; microtubule segments with additional spirals are formed by stacking successive turns along the longitudinal axis (see Fig.~\ref{Fig:spiral_filament_schematic}). For the spiral calculations (systems larger than a single dimer), the time evolution was performed on national high-performance computing systems provided by the Digital Research Alliance of Canada. For each embedding, we track the four largest pairwise $L_1$ coherences within the focal tubulin, labeled by site indices $(i,j)$, to resolve how embedding redirects correlation pathways and alters information flow.

\begin{figure}[t]
  \centering
  % Row 1: single tubulin (a-d)
  \subfloat[]{\includegraphics[width=0.24\linewidth]{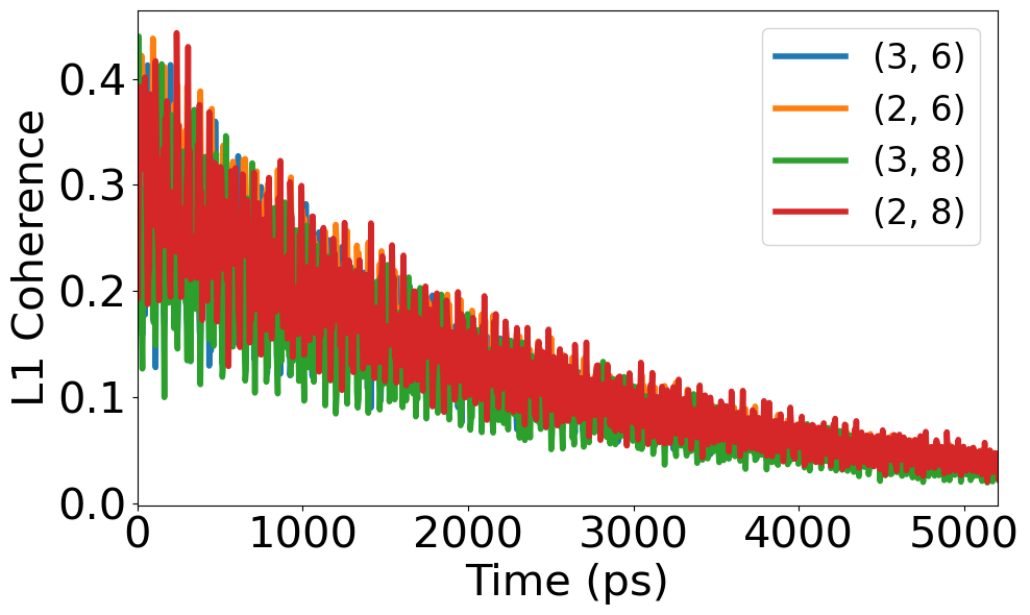}\label{fig:coh:a}}\hfill
  \subfloat[]{\includegraphics[width=0.24\linewidth]{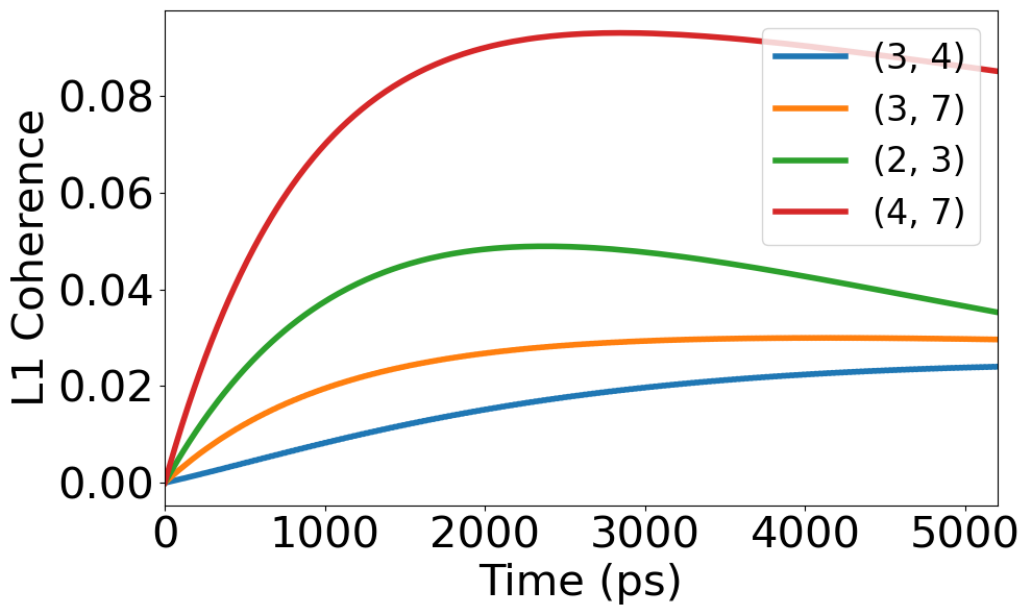}\label{fig:coh:b}}\hfill
  \subfloat[]{\includegraphics[width=0.24\linewidth]{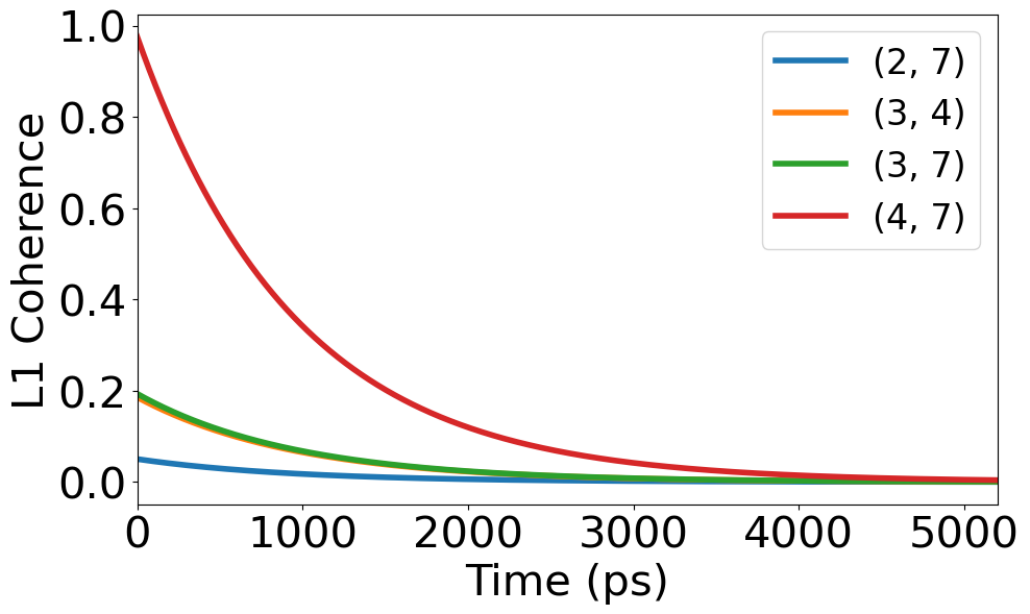}\label{fig:coh:c}}\hfill
  \subfloat[]{\includegraphics[width=0.24\linewidth]{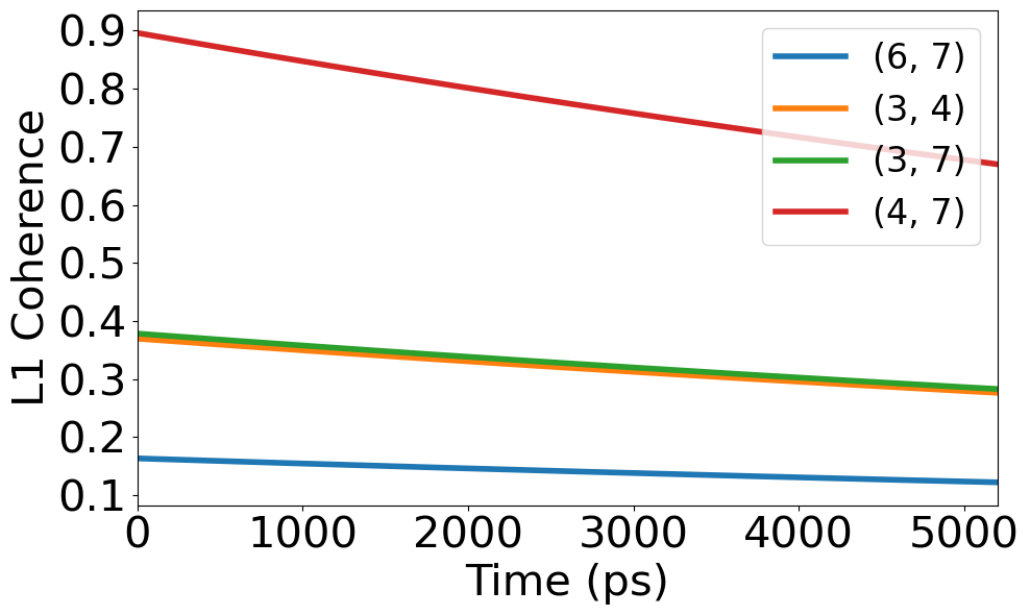}\label{fig:coh:d}}\\[1ex]

  % Row 2: two tubulins (e-h)
  \subfloat[]{\includegraphics[width=0.24\linewidth]{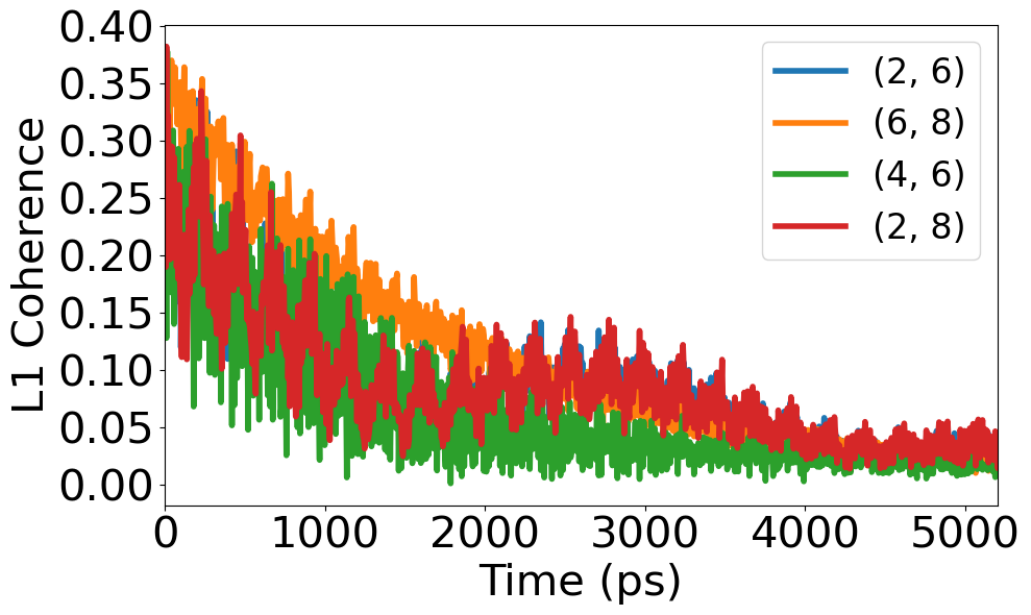}\label{fig:coh:e}}\hfill
  \subfloat[]{\includegraphics[width=0.24\linewidth]{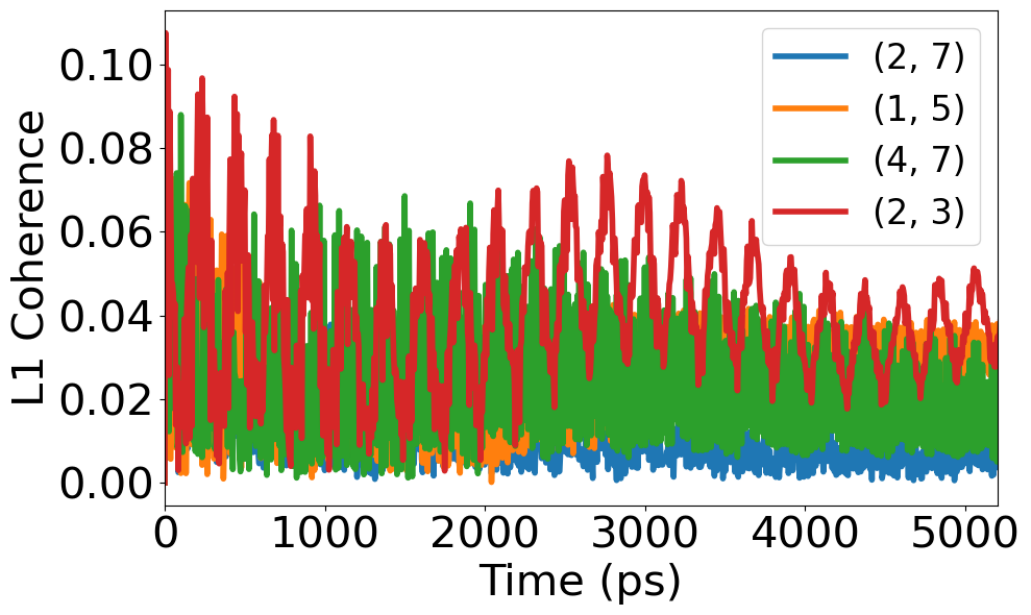}\label{fig:coh:f}}\hfill
  \subfloat[]{\includegraphics[width=0.24\linewidth]{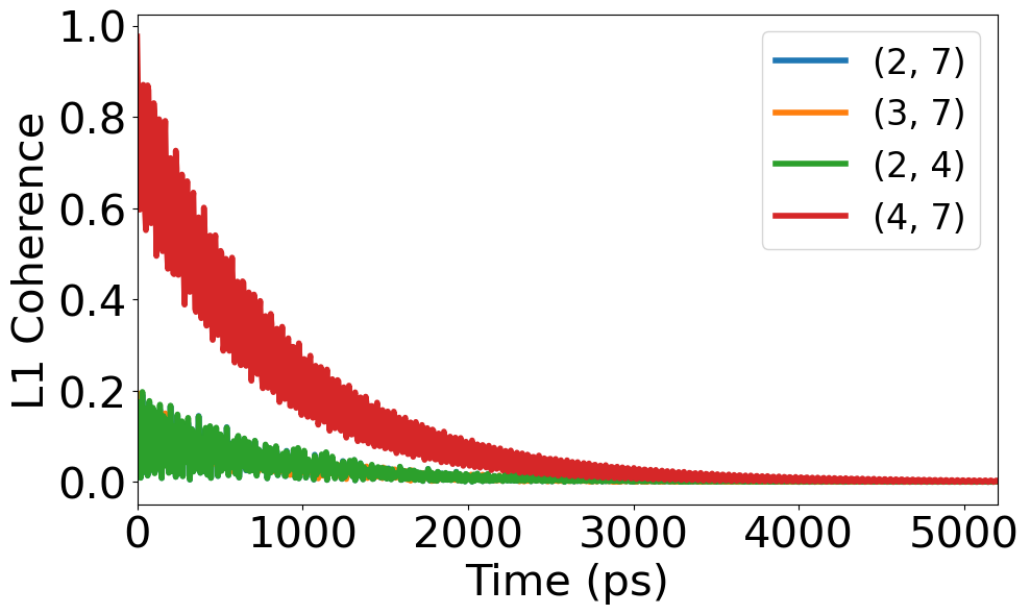}\label{fig:coh:g}}\hfill
  \subfloat[]{\includegraphics[width=0.24\linewidth]{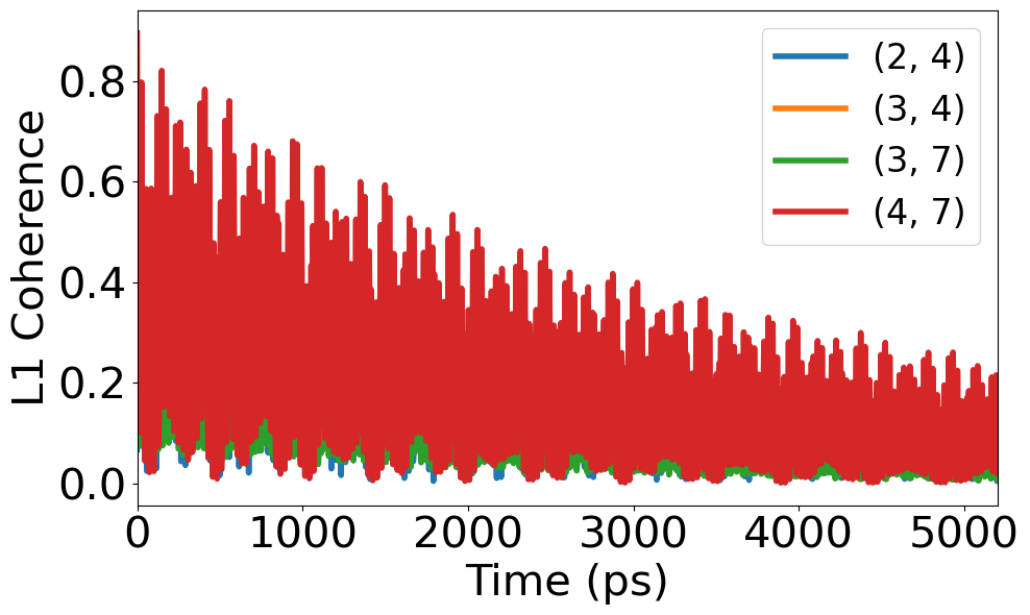}\label{fig:coh:h}}\\[1ex]

  % Row 3: spiral (i-l)
  \subfloat[]{\includegraphics[width=0.24\linewidth]{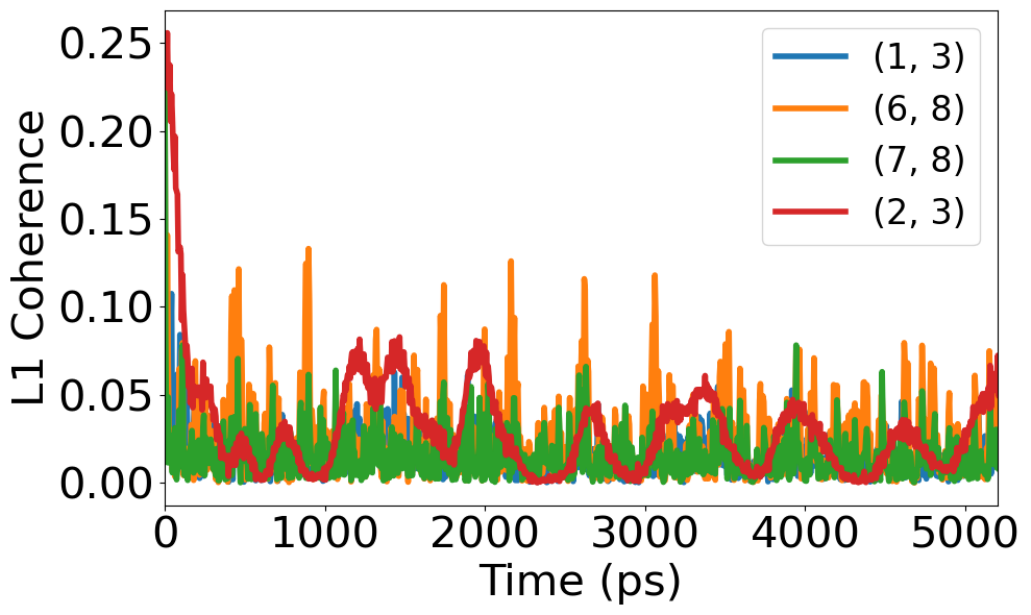}\label{fig:coh:i}}\hfill
  \subfloat[]{\includegraphics[width=0.24\linewidth]{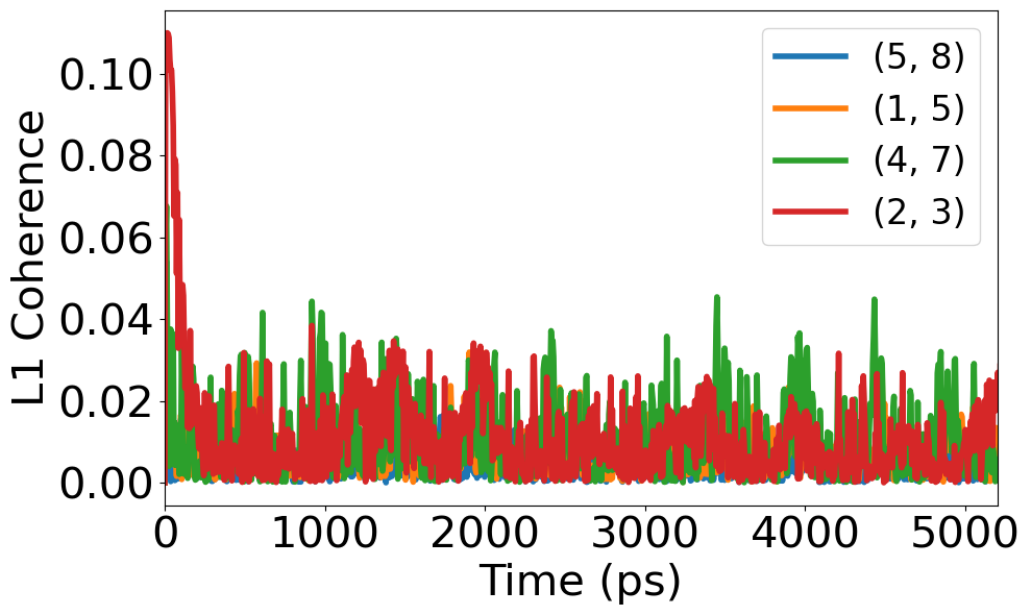}\label{fig:coh:j}}\hfill
  \subfloat[]{\includegraphics[width=0.24\linewidth]{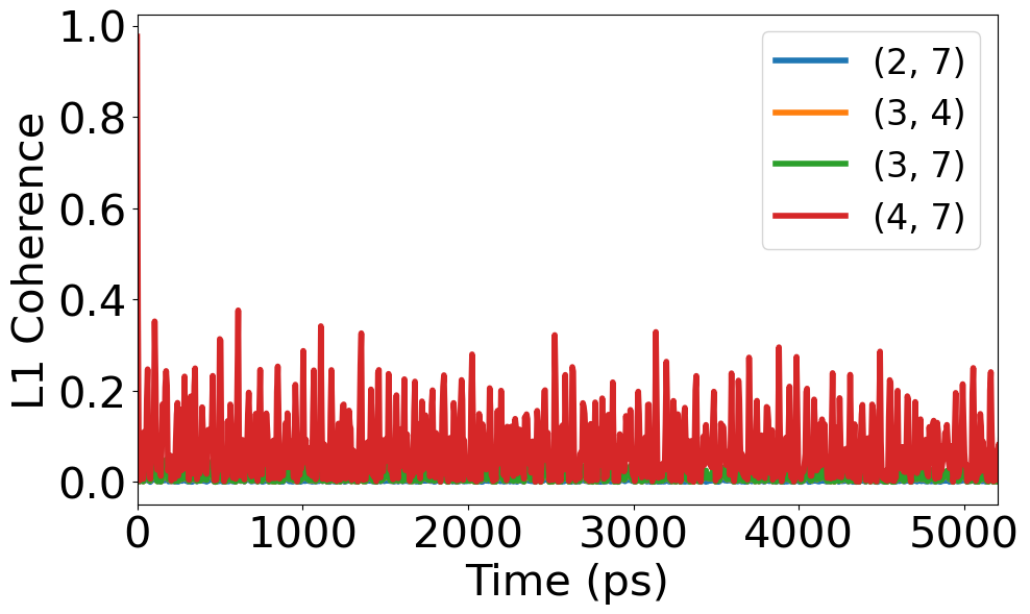}\label{fig:coh:k}}\hfill
  \subfloat[]{\includegraphics[width=0.24\linewidth]{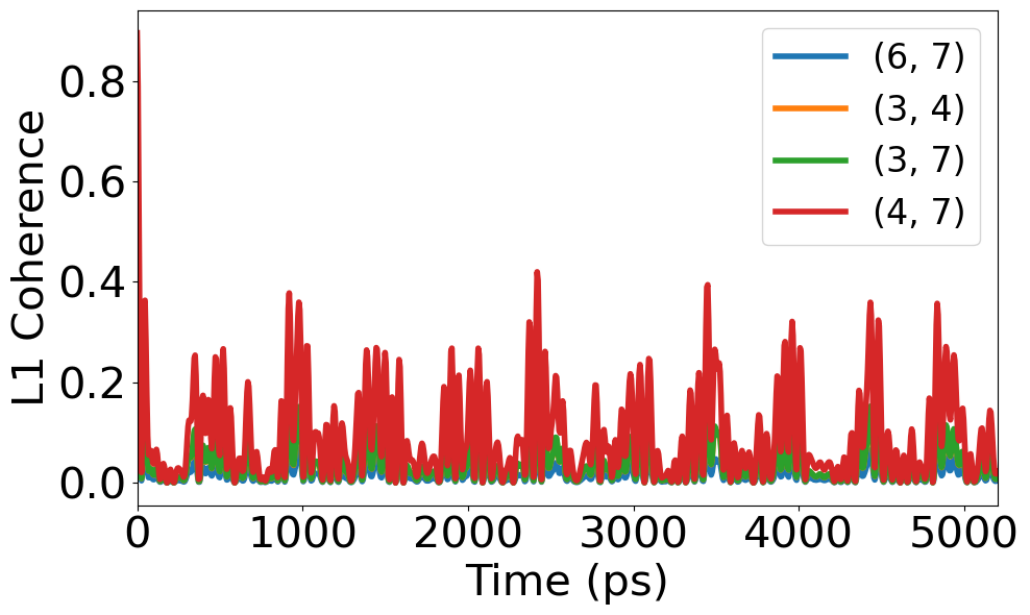}\label{fig:coh:l}}

% Row 4: spiral (m-p)
  \subfloat[]{\includegraphics[width=0.24\linewidth]{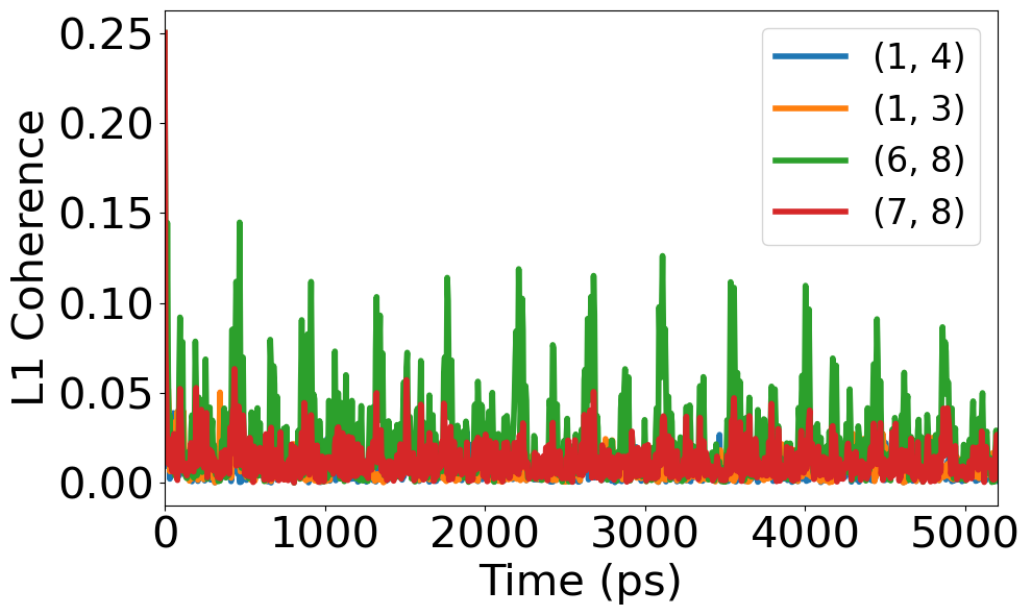}\label{fig:coh:m}}\hfill
  \subfloat[]{\includegraphics[width=0.24\linewidth]{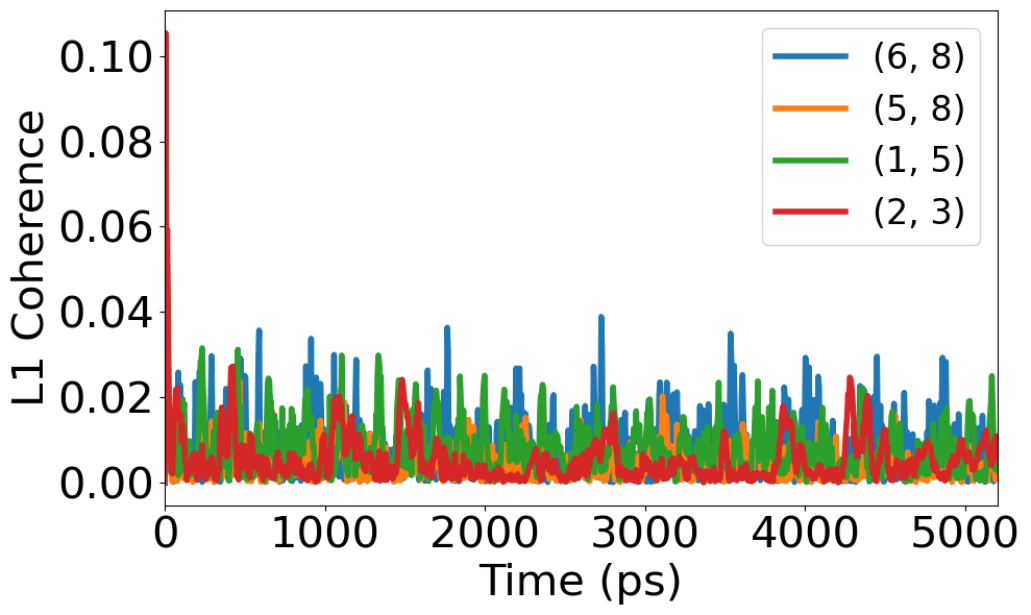}\label{fig:coh:n}}\hfill
  \subfloat[]{\includegraphics[width=0.24\linewidth]{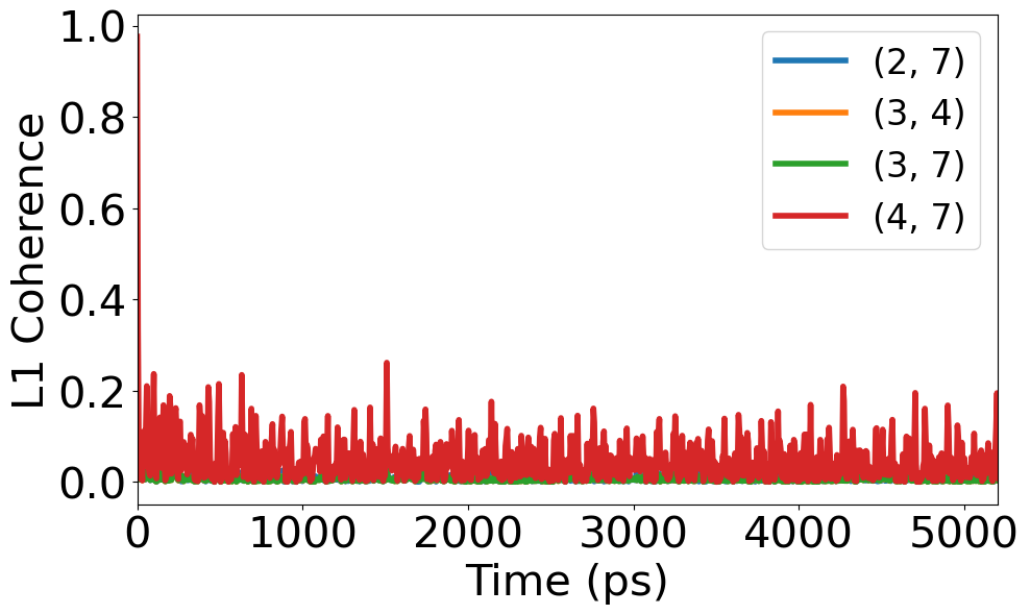}\label{fig:coh:o}}\hfill
  \subfloat[]{\includegraphics[width=0.24\linewidth]{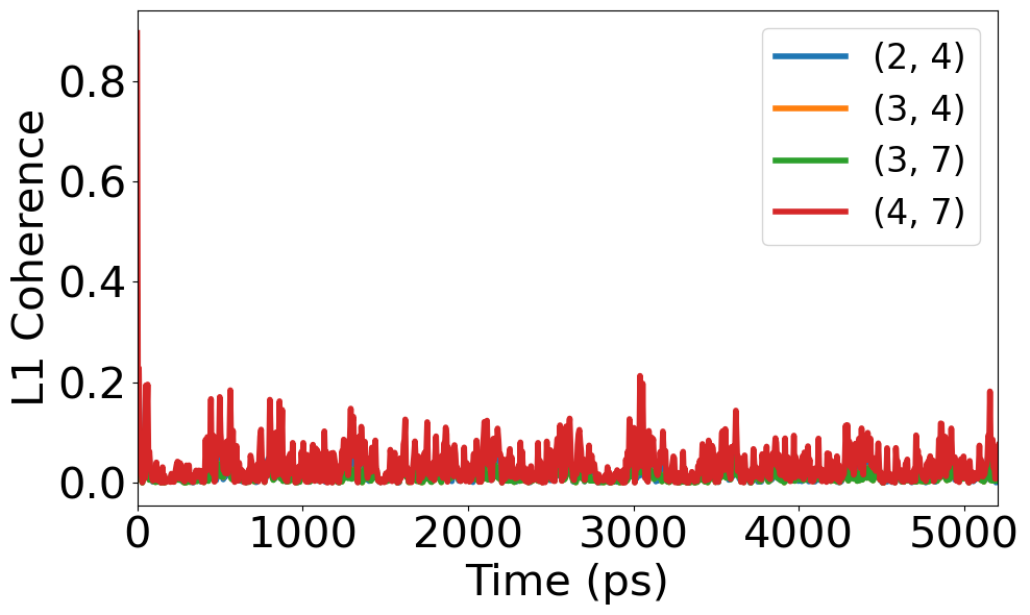}\label{fig:coh:p}}
  \caption{Top four pairwise \(L_1\) coherences across embeddings and initial states.
Columns (all rows): from left to right, maximally coherent, maximally mixed, superradiant, and subradiant initial states. Rows: (a–d) single tubulin; (e–h)
two-tubulin system; (i–l) one spiral; (m–p) two spirals. Each panel shows the
four site pairs with the largest \(L_1\) coherence within the tracked tubulin. Pairs $(i,j)$ denote tryptophan site indices (labels) defined in Fig.~\ref{fig:tub}. Time is reported in picoseconds (ps) in all panels.}
  \label{fig:one-to-spiral:matrix}
\end{figure}

Embedding reshapes the balance between internal circulation of correlations and outward loss. As shown in Figure~\ref{fig:one-to-spiral:matrix}, the identity of the maximally coherent site pairs changes with embedding, indicating that opening the system to a larger environment redistributes correlations across additional channels. As the environment grows (single $\rightarrow$ dimer $\rightarrow$ spiral), coherence amplitudes within the focal tubulin generally decrease and exhibit more pronounced oscillations, reflecting exchange with surrounding degrees of freedom. Small systems retain a strong dependence on the initial preparation, whereas larger embeddings tend to compress these differences and produce more similar oscillatory patterns across preparations.

We observe the same qualitative trends when repeating the analysis with mutual information (see Appendix~\ref{app:sup}, Figure~\ref{fig:one-to-spiral:matrix:mut}).

Evidence for correlation sharing across distinct tubulins is further shown in Appendix~\ref{app:sup}, Figure~\ref{fig:corr_all}, where inter-tubulin coherence correlations are present for each initial state.

\subsection{Time-Resolved Non-Markovian Backflow on Two Tubulin Subsystems (Single Spiral)}
\label{subsec:nonmarkov-two-tubulin}

Building on Sec.~\ref{subsec:one-to-spiral}, where embedding a single tubulin within larger structures redistributed coherence pathways, we now examine whether the surrounding tubulins act as a structured reservoir that can store and then return information to a local subsystem. Here, environment means the other tubulins around the initially prepared tubulin: its partner in a dimer and, in the spiral case, the remaining tubulins. In vivo, tubulin is embedded in an aqueous solvent and hydration shell; such solvent-mediated effects are not included explicitly in the present radiative-loss model and are left for future extensions \citep{DelGiudice1985,DelGiudice1986}.

We consider a single spiral of 13 tubulins, each with eight tryptophans. For every neighbor \(T_k\) of the initially prepared tubulin \(T_1\), we form the two tubulin subsystem \(X_k=T_1\cup T_k\) and study its dynamics within the single excitation manifold while tracing out the other 11 tubulins. Memory effects are quantified via the trace distance backflow of Ref.~\citep{Breuer2009PRL}. Let \(\{|s\rangle\}\) denote single excitation basis states on individual tryptophans, with index sets \(T_1=\{s_1,\ldots,s_8\}\) and \(T_k=\{r_1,\ldots,r_8\}\). All other tubulins start in the global ground state \(|g\rangle\). We use normalized uniform superpositions over a full tubulin,
\[
|10\rangle = \frac{1}{\sqrt{8}}\sum_{s\in T_1} |s\rangle,
\qquad
|01\rangle = \frac{1}{\sqrt{8}}\sum_{r\in T_k} |r\rangle,
\]
which represent single excitations delocalized over the eight tryptophans of \(T_1\) or \(T_k\) and are orthogonal in the single excitation space. Two orthogonal pairs, both confined to \(\mathrm{span}(\{|g\rangle\}\cup T_1 \cup T_k)\), are propagated under the same Liouvillian: a population contrast pair \(|10\rangle\) versus \(|01\rangle\), and a phase contrast pair \((|10\rangle \pm |01\rangle)/\sqrt{2}\). During the evolution, excitations that leave \(X_k\) into the remainder of the spiral are traced out. Non-Markovian information backflow is indicated by revivals of the trace distance
\[
D_k(t)=\tfrac{1}{2}\,\bigl\|\rho^{(1)}_{X_k}(t)-\rho^{(2)}_{X_k}(t)\bigr\|_1,
\]
which quantifies the operational distinguishability of the two reduced states of $X_k$ (larger $D_k$ means they can be more easily told apart by an optimal measurement). For CP divisible (Markovian) reduced dynamics, $D_k(t)$ is contractive and cannot increase; therefore intervals with $\dot D_k(t)>0$ indicate information backflow from the traced-out tubulins into the subsystem~\citep{Breuer2009PRL,Rivas2010PRL}.
The associated scalar non-Markovianity for \(X_k\) is the total positive variation
\[
\mathcal N(k)=\int_{\dot D_k(t)>0}\dot D_k(t)\,\mathrm dt,
\]
reported for both initial pairs.

\begin{figure}[t]
  \centering
  \includegraphics[width=0.32\linewidth]{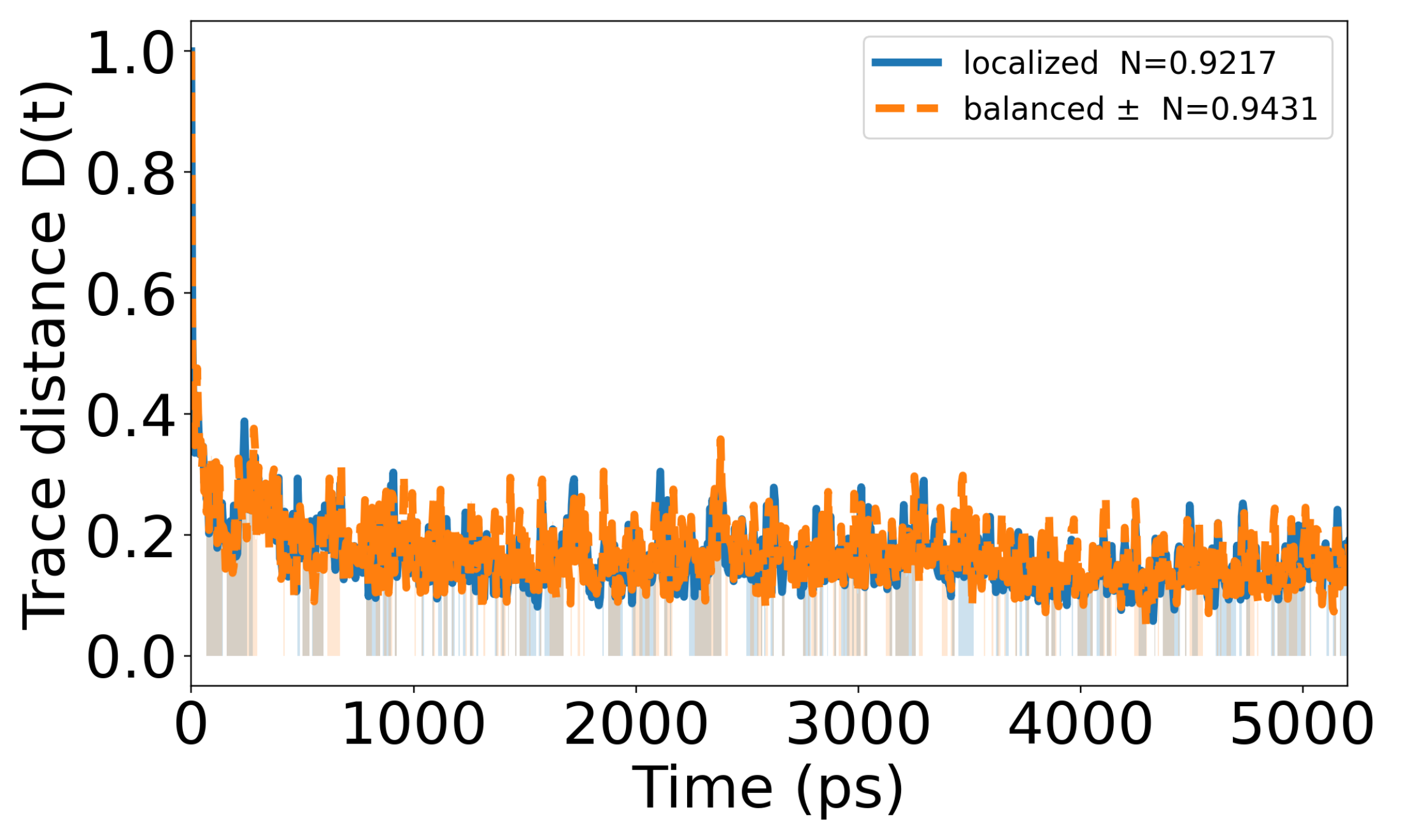}\hfill
  \includegraphics[width=0.32\linewidth]{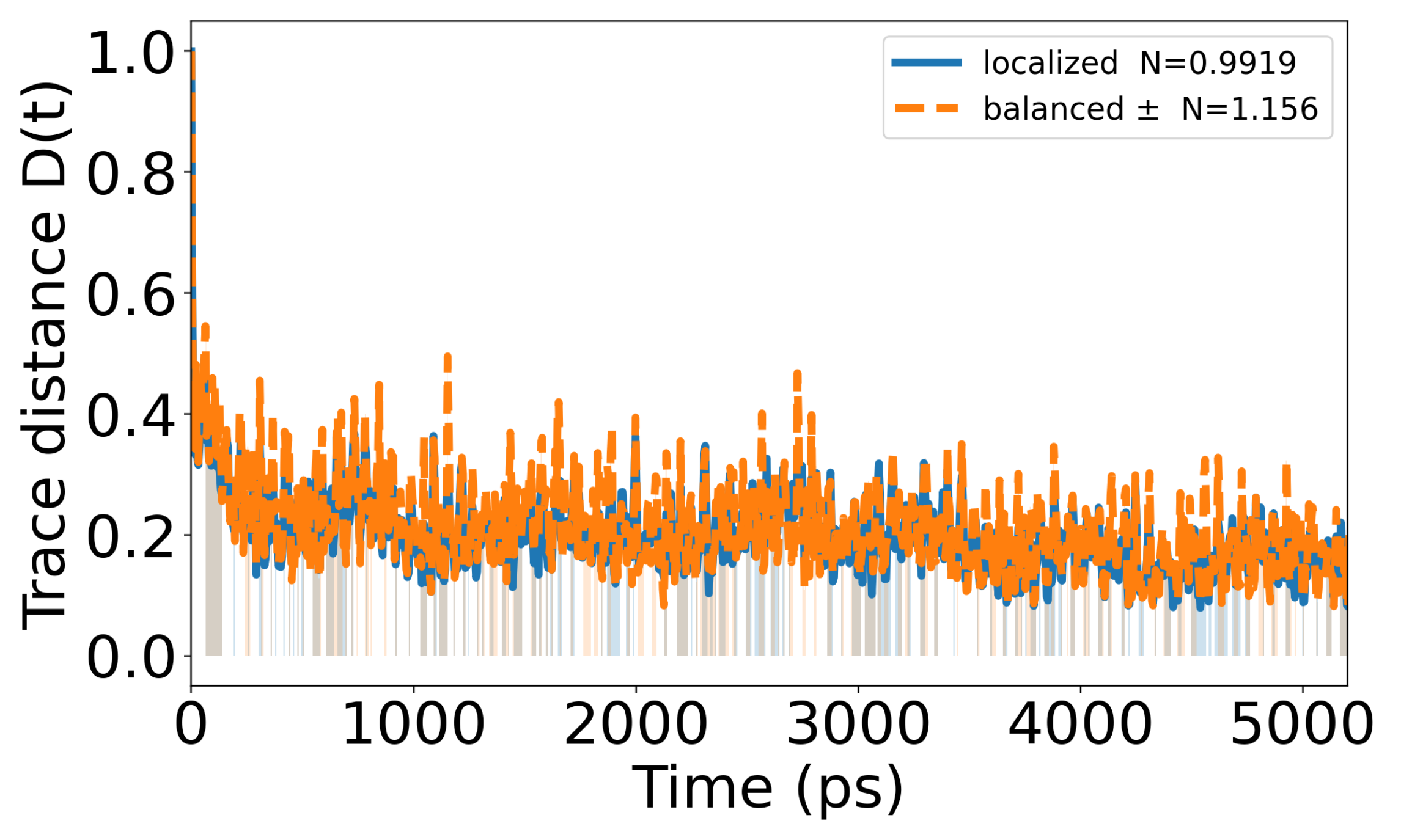}\hfill
  \includegraphics[width=0.32\linewidth]{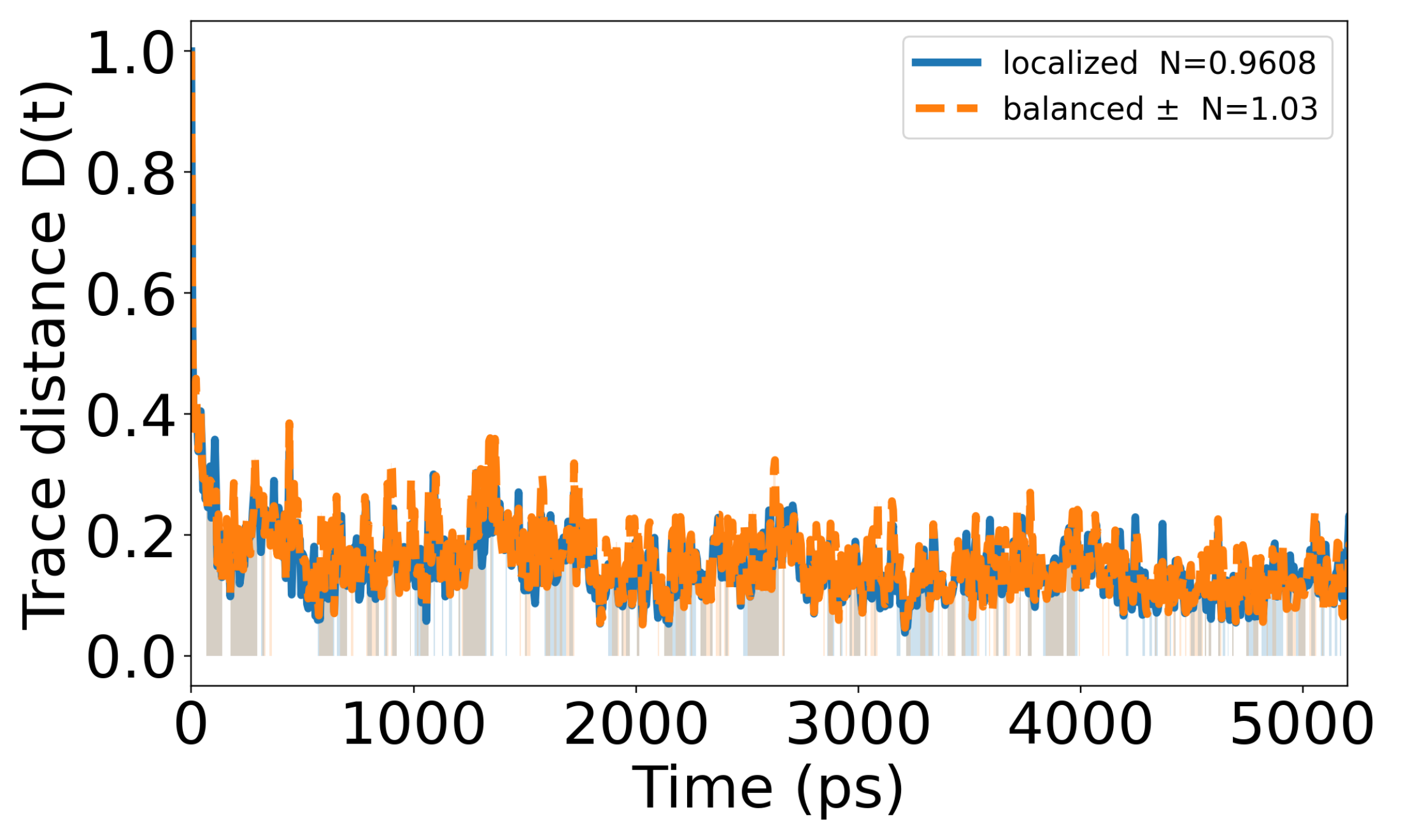}

  \vspace{0.6em}

  \includegraphics[width=0.6\linewidth]{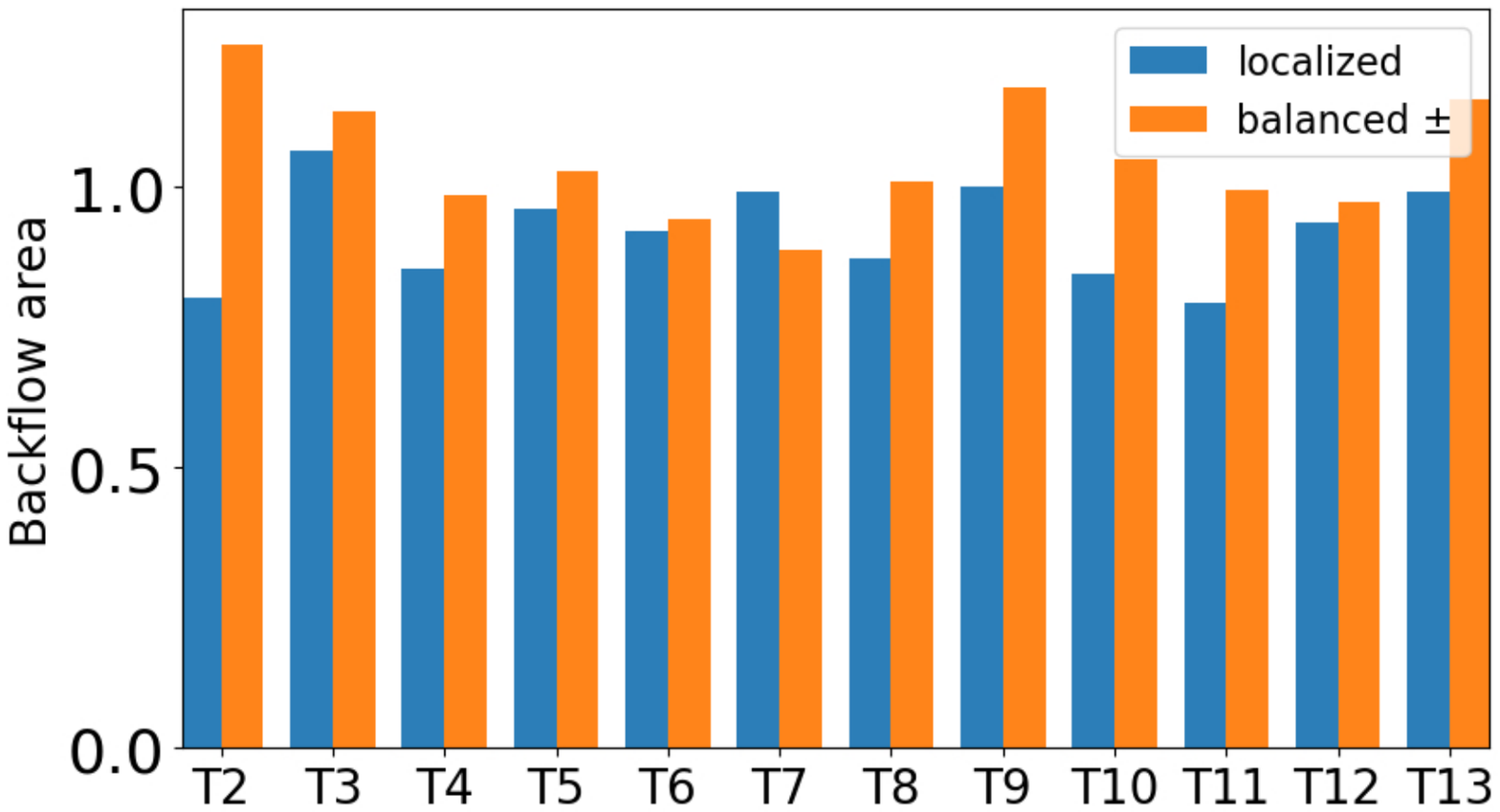}
  \caption{Information backflow on two tubulin subsystems in a single spiral. Top: trace distance dynamics \(D_k(t)\) for \(X_2=(T_1,T_2)\), \(X_9=(T_1,T_9)\), and \(X_{13}=(T_1,T_{13})\); solid curves show the population contrast pair \(|10\rangle\) versus \(|01\rangle\); dashed curves show the phase contrast pair \((|10\rangle\pm|01\rangle)/\sqrt{2}\). Shaded intervals mark \(\dot D_k(t)>0\) (information backflow). Time is reported in picoseconds (ps) in all panels. Bottom: integrated backflow \(\mathcal N(k)\) across \(k=2,\ldots,13\) for both preparations.}
  \label{fig:nonmarkov_tubulin_pairs}
\end{figure}

As shown in Fig.~\ref{fig:nonmarkov_tubulin_pairs}, \(D_k(t)\) exhibits revivals for every neighbor tested, yielding nonzero \(\mathcal N(k)\) across \(k=2,\ldots,13\). Specific neighbors (notably \(X_2\), \(X_9\), \(X_{13}\)) show larger responses, consistent with geometry dependent couplings in the spiral. The phase contrast preparation generally produces stronger backflow than the population contrast pair, which points to a leading role for coherence and phase revivals rather than pure population exchange.

Information backflow is relevant for neuronal microtubules because it shows that correlations are not irreversibly lost to surrounding tubulins; the dimer or spiral can transiently store and return population and coherence to a local subsystem. In ordered bundles inside neurons, such structured reservoir effects could provide mesoscale temporal correlations (i.e., intermediate-time memory effects in the reduced subsystem dynamics due to transient information backflow from the surrounding tubulin network) that could in principle influence intracellular signaling and excitonic energy migration by buffering and synchronizing fluctuations. A testable consequence is that the backflow measure \(\mathcal N(k)\) should vary systematically with local geometry, energetic detuning, temperature or ionic conditions, and the initial phase structure.

\subsection{Transition to Larger Assemblies}

The previous sections examined how embedding a single tubulin within a dimer or a spiral redistributes coherence and can produce information backflow. We now extend the analysis to larger microtubule assemblies arranged as filaments and as ideal spirals with up to 100 spirals, in order to connect local behavior with system level trends. For these sizes, direct simulation with the Lindblad master equation is not practical, so we use the effective non-Hermitian Hamiltonian in Equation (\ref{non_her_eq}) to access eigenmodes and their radiative decay rates. The section proceeds in two steps. First, we quantify pairwise correlated coherence for superradiant and subradiant eigenstates across ordered and disordered structures. Second, we extract radiative lifetimes and analyze how superradiant speedup and subradiant protection scale with size and depend on disorder.

\subsubsection{Correlated Coherence Across Microtubule Structures}

We compute the pairwise correlated coherence (based on the $L_1$ norm) between Trp sites for the superradiant and subradiant eigenstates. This is evaluated across 100 spiral microtubule configurations constructed either from repeated 1JFF structural units or from randomly selected tubulin dimers taken from molecular dynamics simulations (see appendix \ref{sec:geometry_construction}).

\begin{figure}[t]
    \centering
    \subfloat[]{\includegraphics[width=0.35\linewidth]{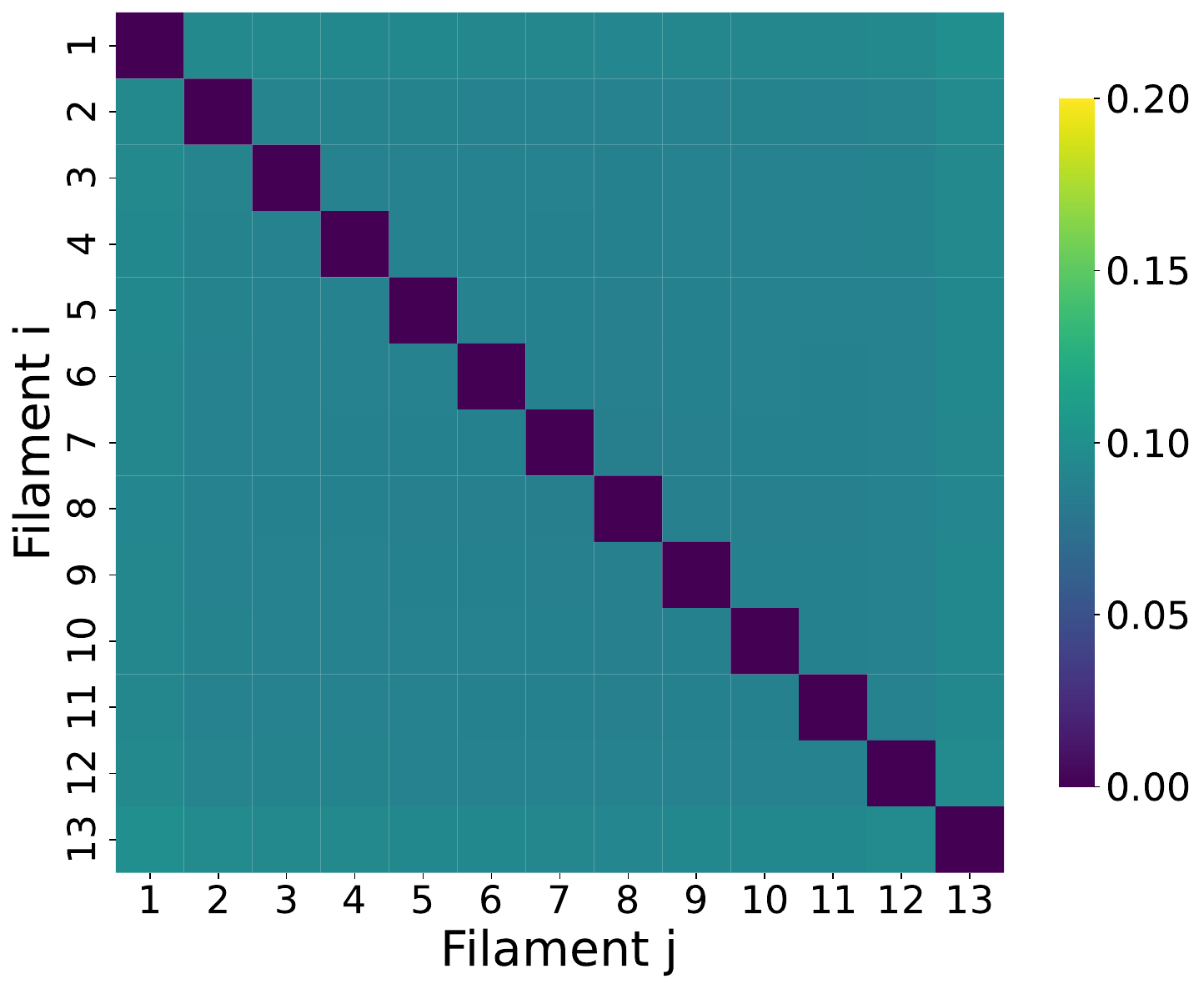}  \label{Fig1:a}}
    \quad
    \subfloat[]{\includegraphics[width=0.35\linewidth]{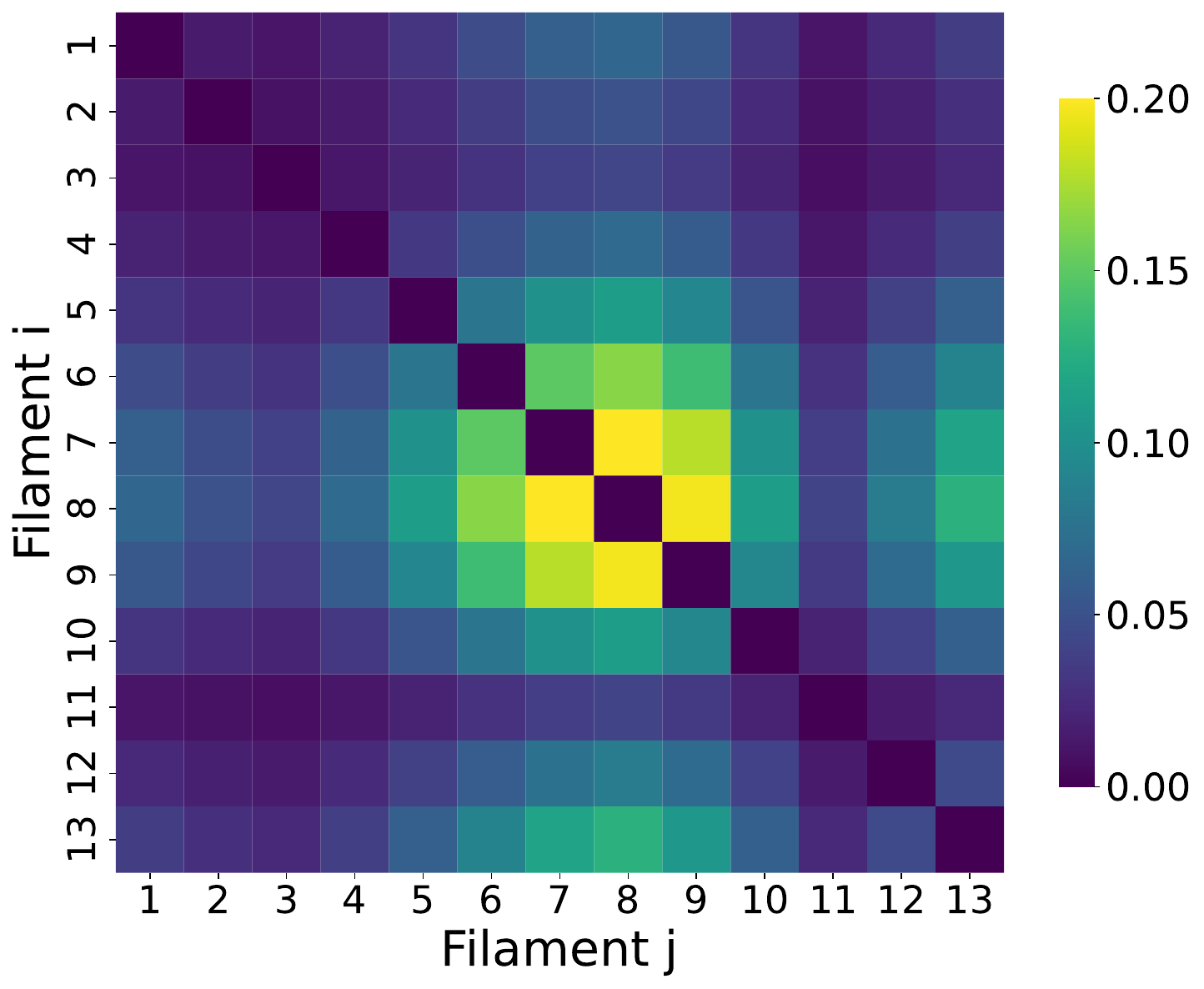}  \label{Fig1:b}} \\
    \subfloat[]{\includegraphics[width=0.35\linewidth]{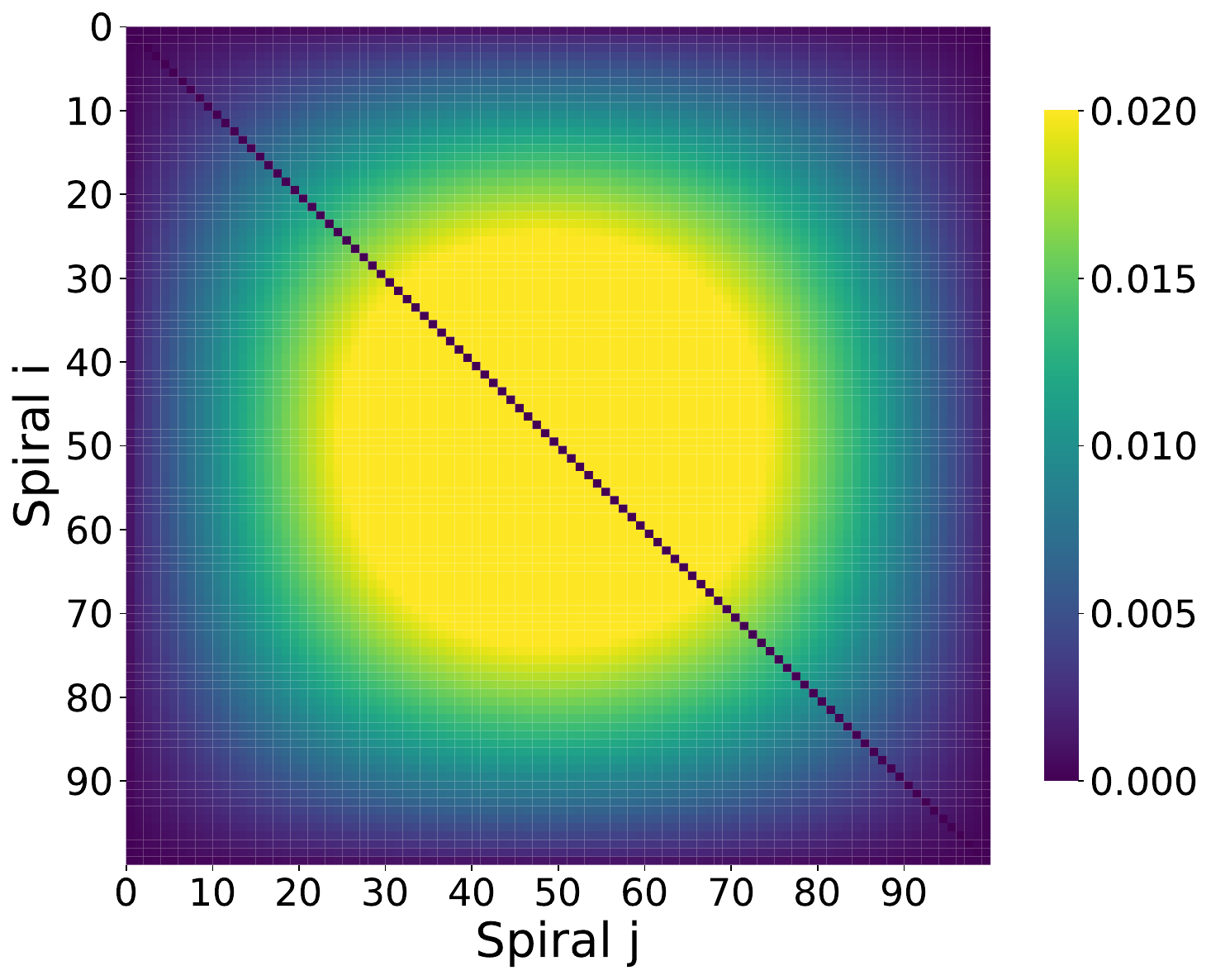}  \label{Fig1:c}}
    \quad
    \subfloat[]{\includegraphics[width=0.35\linewidth]{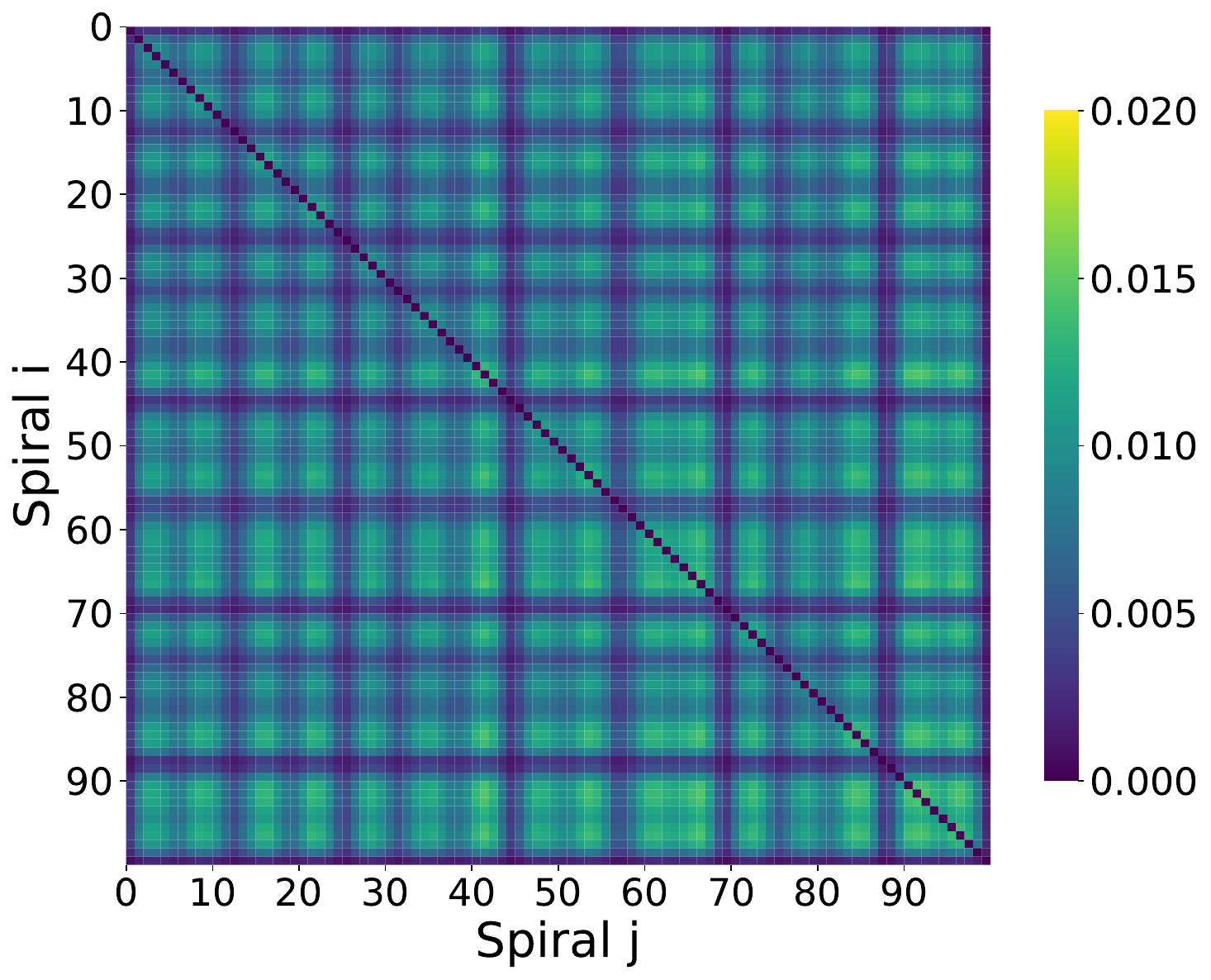}  \label{Fig1:d}}
    \caption{
    Correlated coherence between tubulin dimers in a microtubule constructed from the symmetric 1JFF PDB structure.
    All panels correspond to a microtubule segment with $13$ filaments (protofilaments) and $100$ spirals (circumferential turns); each spiral contains $13$ tubulin dimers (see Fig.~\ref{Fig:spiral_filament_schematic}).
    (a–b) show coherence between filaments for the superradiant and subradiant states, respectively.
    (c–d) show coherence between spirals. In this ordered configuration, superradiant coherence is delocalized across the microtubule, reflecting long-range quantum correlations.
    }
    \label{Fig:1jff_coherence}
\end{figure}

\begin{figure}[t]
%    \centering
    % Random MD snapshots
    \subfloat[]{\includegraphics[width=0.30\linewidth]{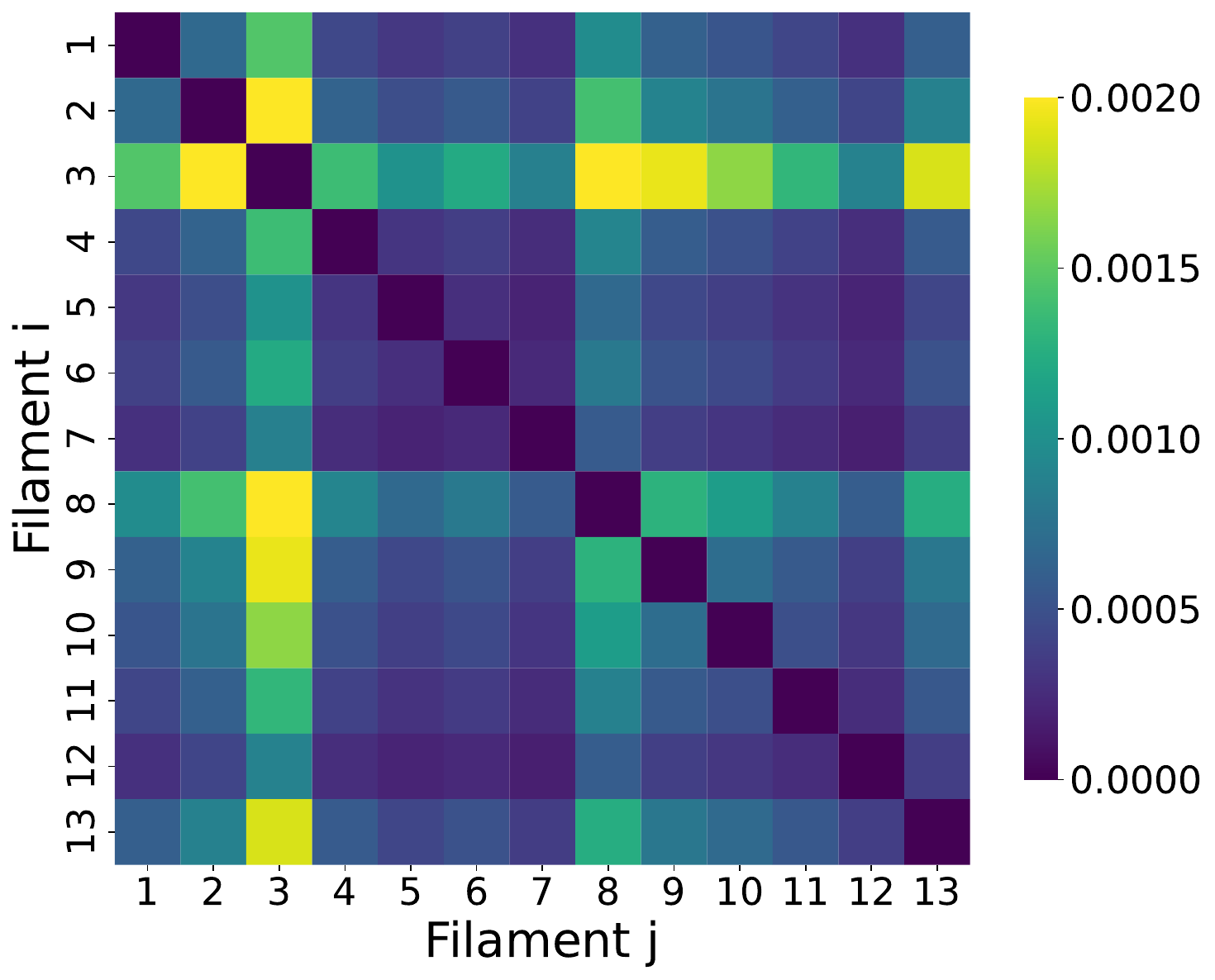}  \label{Fig2:a}}
    \quad
    \subfloat[]{\includegraphics[width=0.30\linewidth]{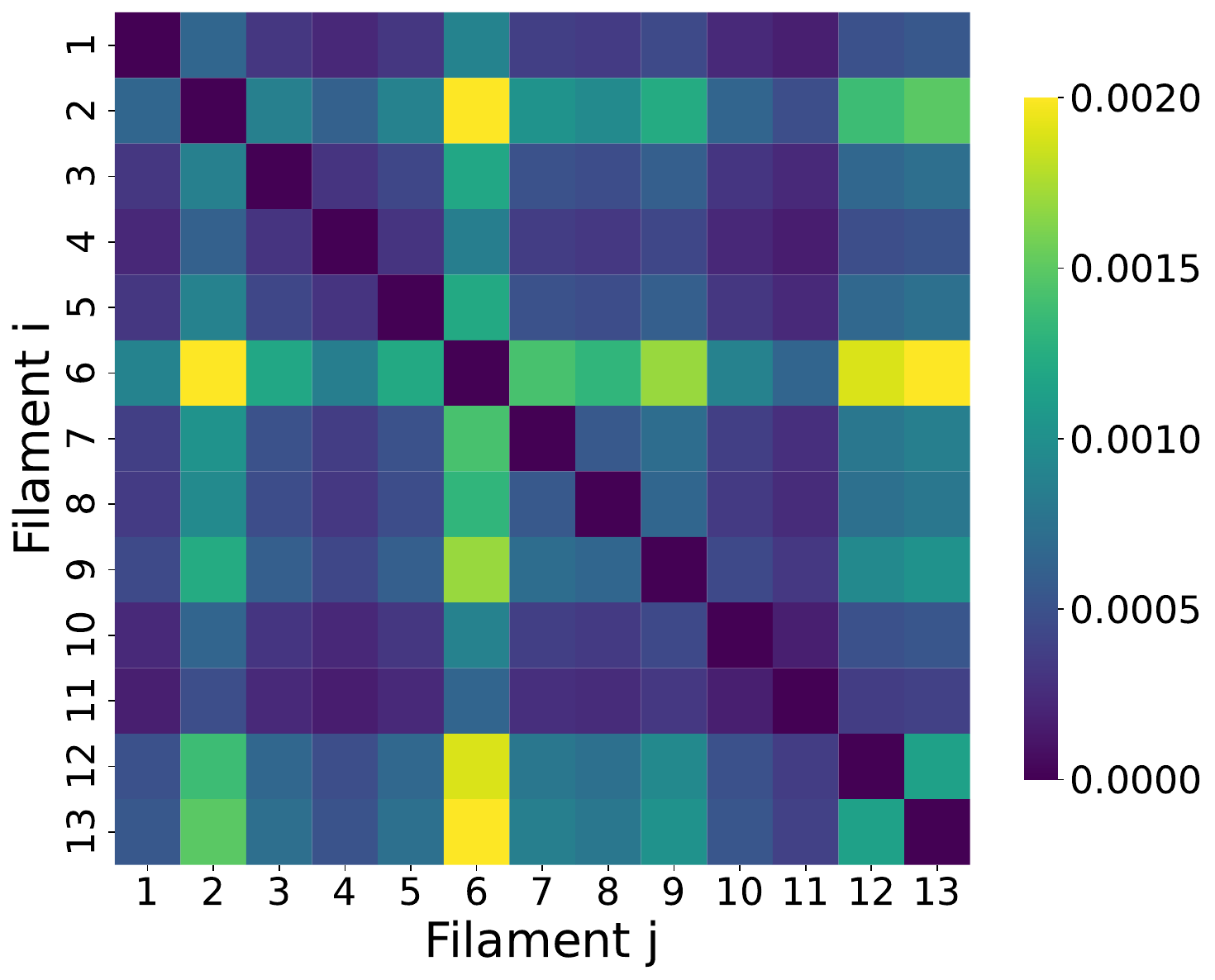}  \label{Fig2:b}} \\
    \subfloat[]{\includegraphics[width=0.30\linewidth]{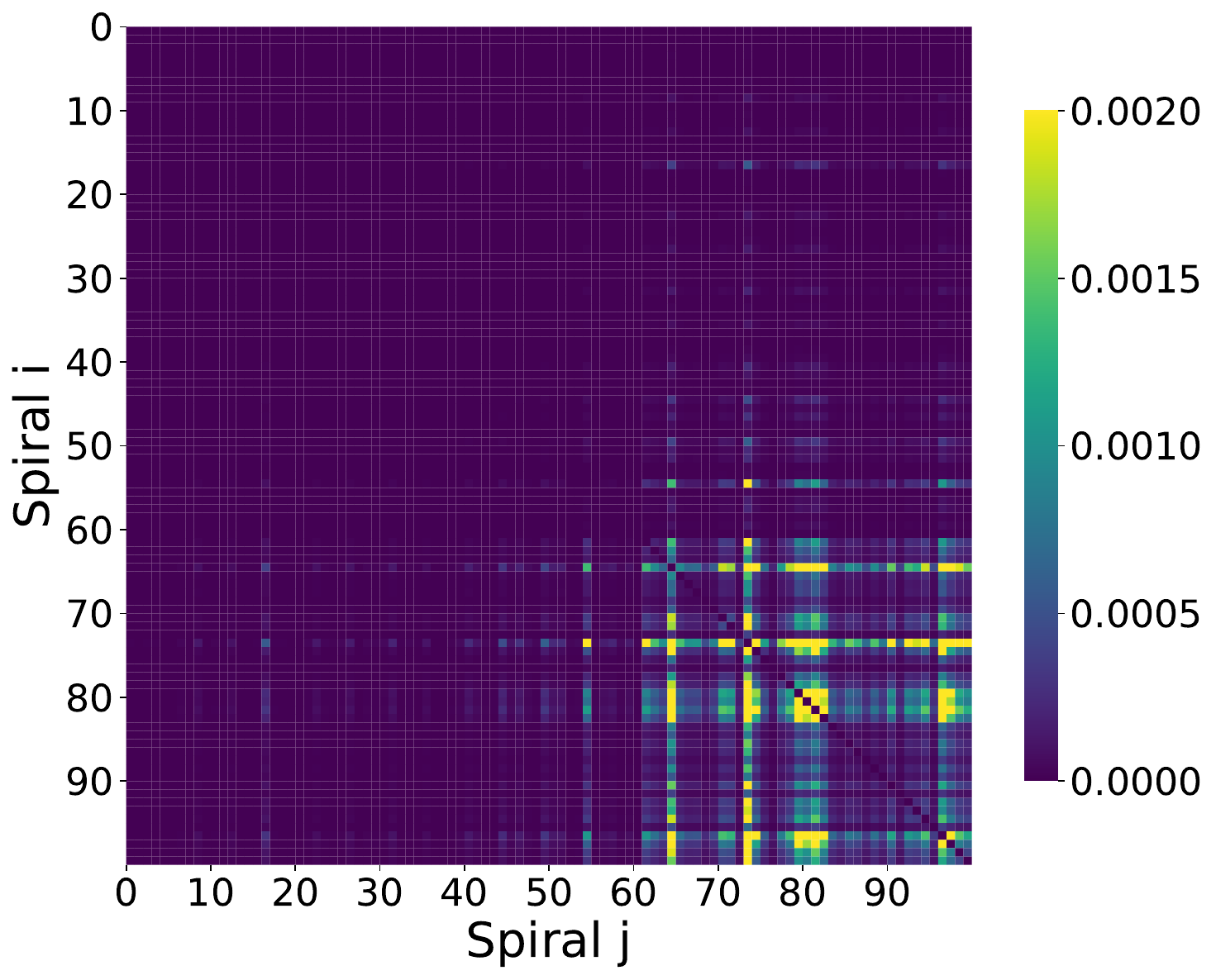}  \label{Fig2:c}}
    \quad
    \subfloat[]{\includegraphics[width=0.30\linewidth]{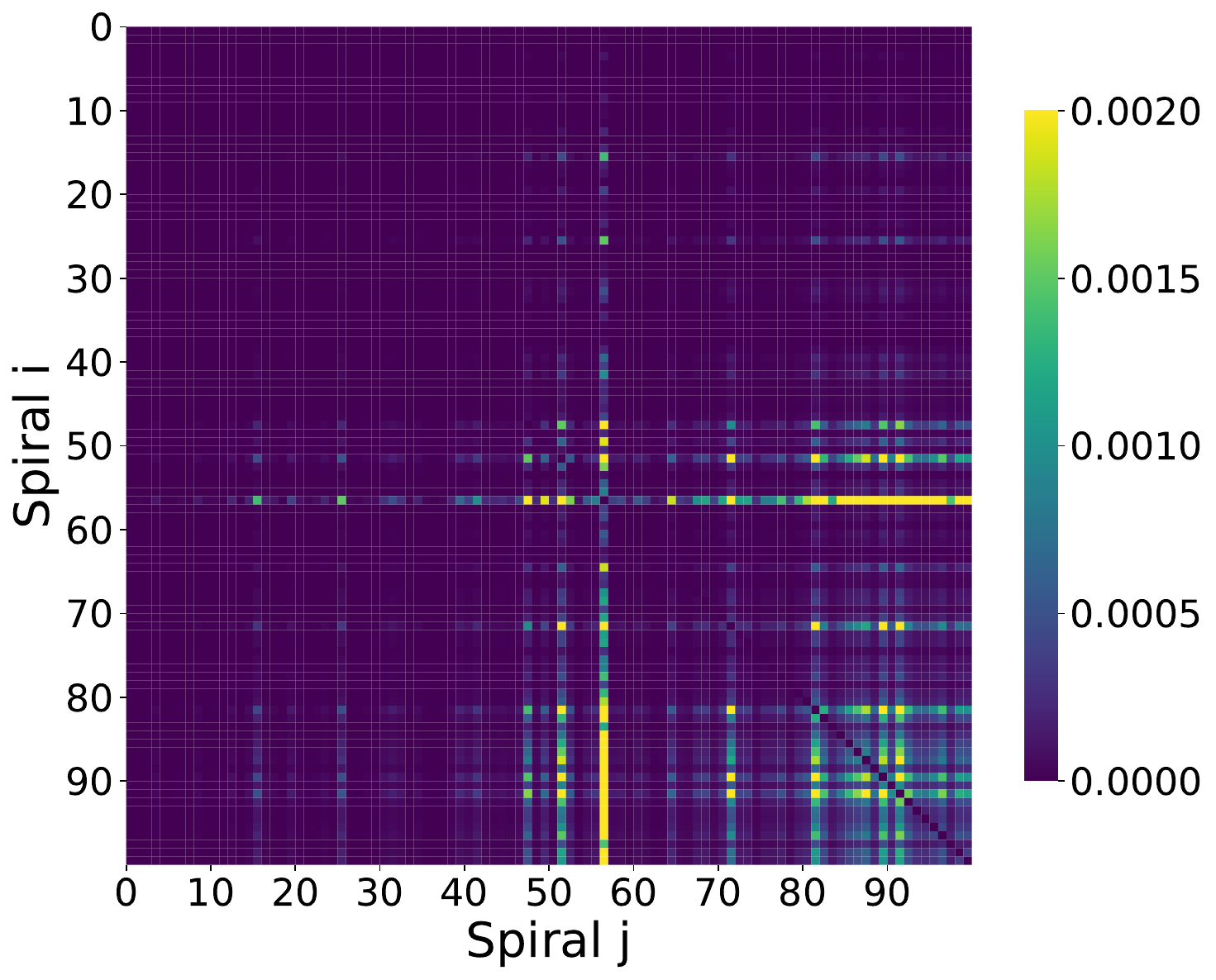}  \label{Fig2:d}} \\
    % Static disorder
    \subfloat[]{\includegraphics[width=0.30\linewidth]{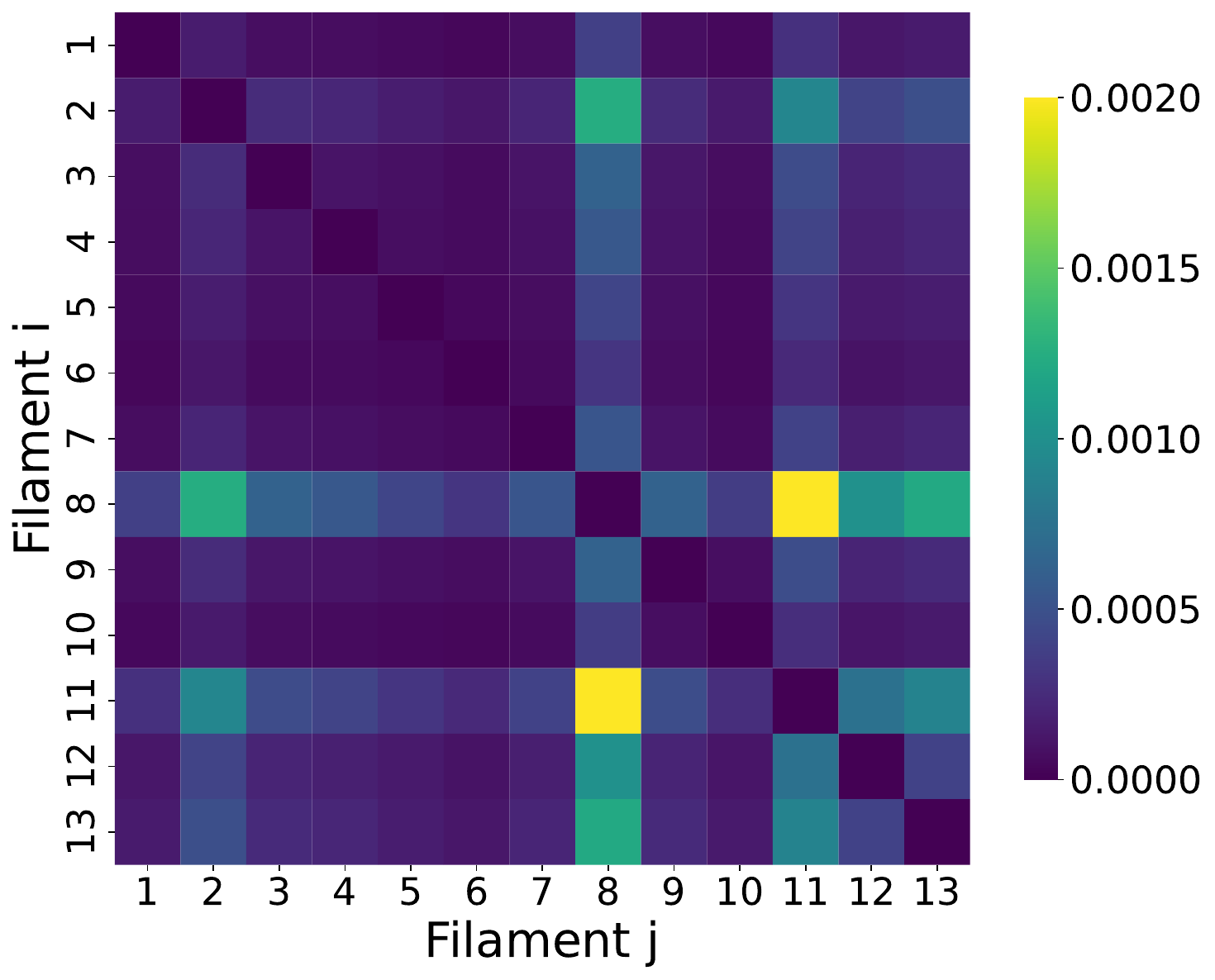}  \label{Fig2:e}}
    \quad
    \subfloat[]{\includegraphics[width=0.30\linewidth]{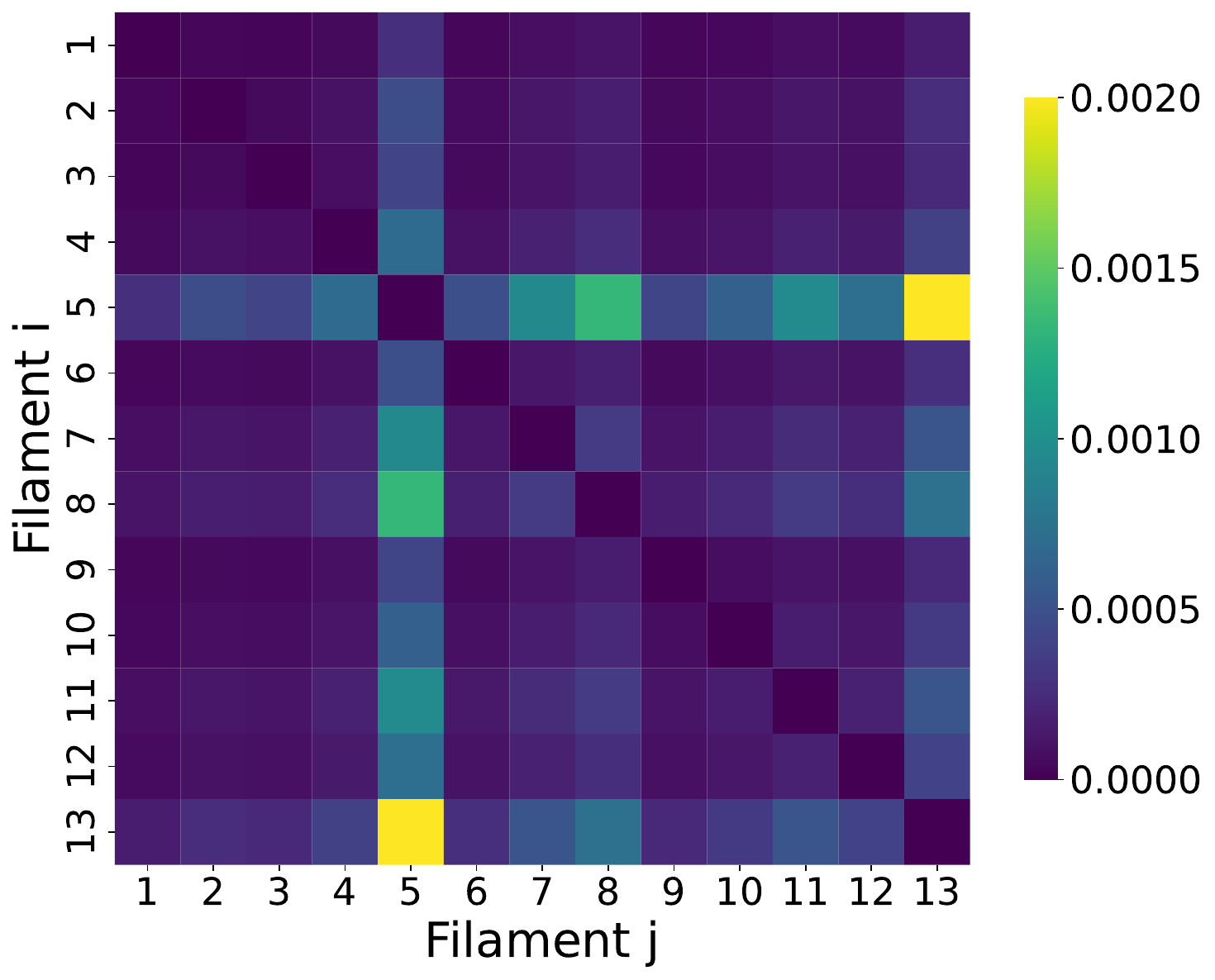}  \label{Fig2:f}} \\
    \subfloat[]{\includegraphics[width=0.30\linewidth]{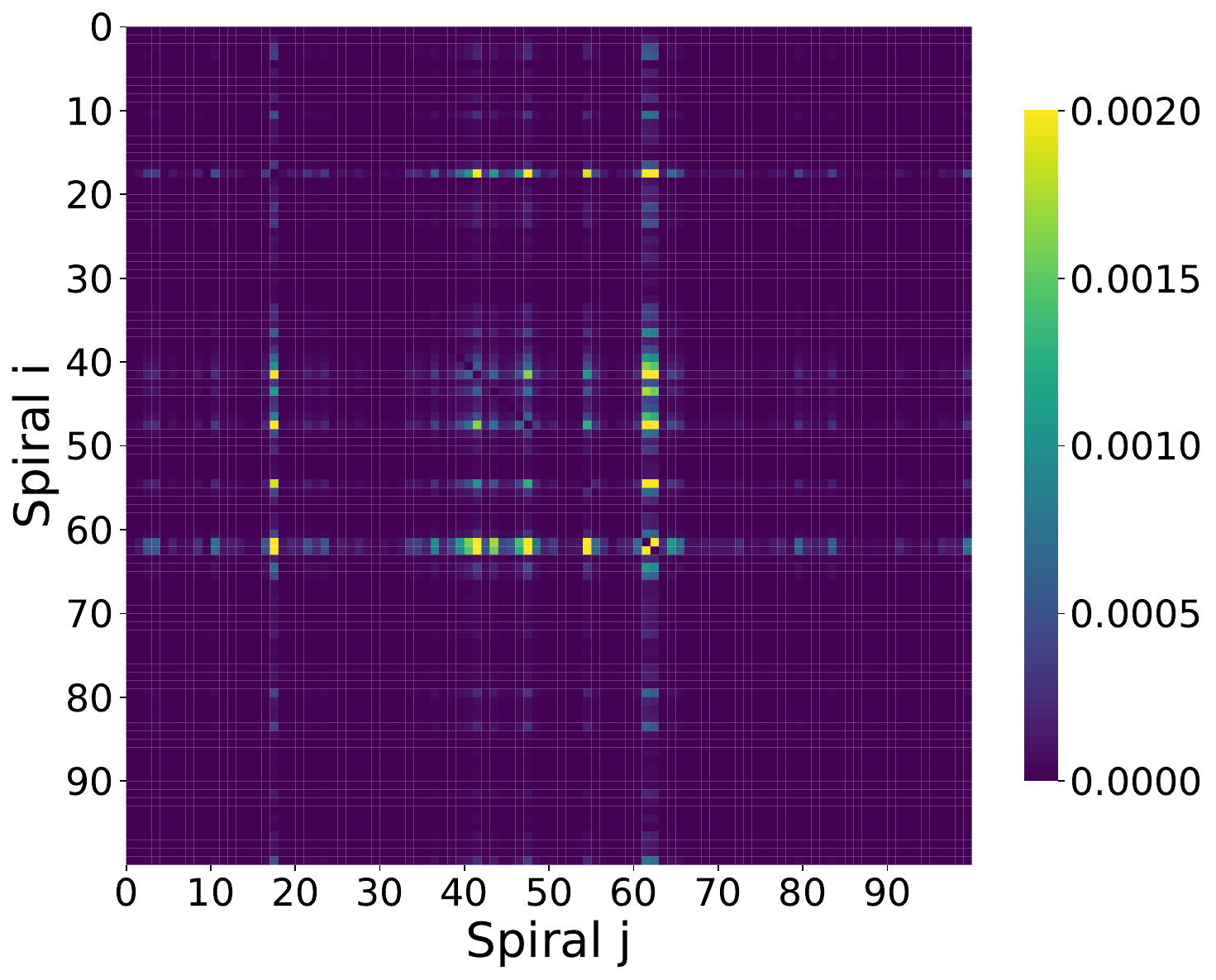}  \label{Fig2:g}}
    \quad
    \subfloat[]{\includegraphics[width=0.30\linewidth]{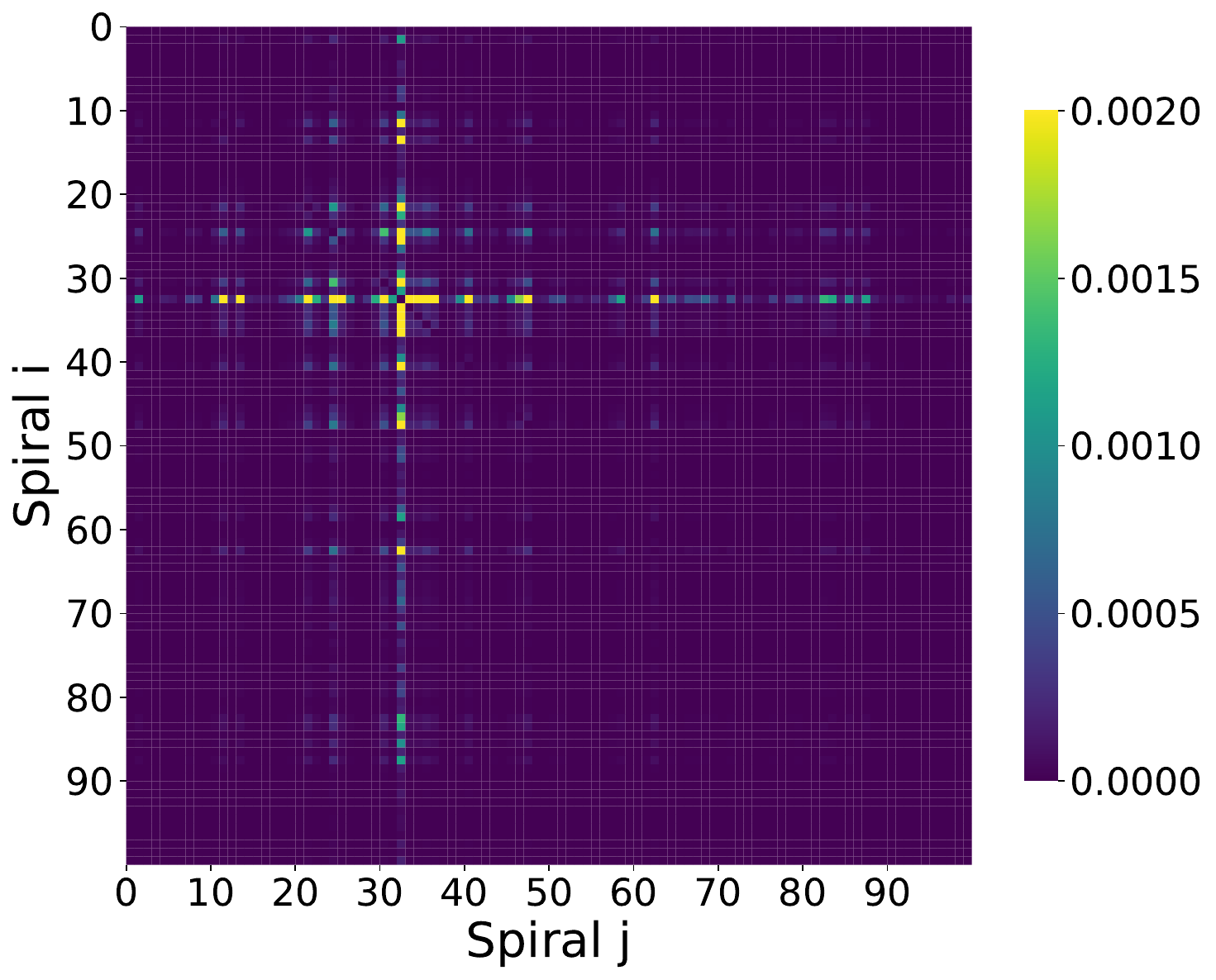}  \label{Fig2:h}}
    \caption{
    Correlated coherence in disordered microtubules. All panels correspond to a microtubule segment with $13$ filaments (protofilaments) and $100$ spirals (circumferential turns); each spiral contains $13$ tubulin dimers (see Fig.~\ref{Fig:spiral_filament_schematic}).
    (a–d) correspond to a structure built from randomly sampled frames of a molecular dynamics trajectory (structural/spatial disorder in the dipole geometry).
    (e–h) show coherence in a symmetric microtubule (based on 1JFF) where static energetic disorder has been added via random diagonal fluctuations in the Hamiltonian. In both cases, coherence in the superradiant state is significantly reduced and becomes localized, indicating that both spatial and static disorder disrupt long-range coherence, in contrast to the ordered case in Fig.~\ref{Fig:1jff_coherence} (here, ``reduced'' means reduced relative to the ordered 1JFF-based microtubule under the same correlated-coherence measure). Note that the color-bar ranges in Fig.~\ref{Fig:1jff_coherence} and the present figure differ.}
    \label{Fig:disordered_coherence}
\end{figure}

For the large microtubule segments, correlated coherence is evaluated between two composite subsystems (each subsystem being a collection of Trp sites) rather than between individual chromophores. Specifically, we consider two geometric partitions: (i) spiral--spiral coherence, where each spiral contains 13 tubulins and therefore 104 Trp sites, and (ii) filament--filament coherence, where each filament subsystem contains 100 tubulins and therefore 800 Trp sites. We then compute correlated coherence between all pairs of such subsystems across the assembly, producing the matrices shown in Figs.~\ref{Fig:1jff_coherence} and~\ref{Fig:disordered_coherence}.
We compute the pairwise correlated coherence (based on the $L_1$ norm) between composite subsystems defined by collections of Trp sites for the superradiant and subradiant eigenstates.

Figures~\ref{Fig:1jff_coherence} and~\ref{Fig:disordered_coherence} highlight the impact of symmetry and disorder on coherence in microtubule systems. In the symmetric microtubule constructed by repeating the 1JFF PDB structure (Fig.~\ref{Fig:1jff_coherence}), coherence in the superradiant state is broadly distributed across the entire lattice, indicating long-range delocalization. In contrast, Fig.~\ref{Fig:disordered_coherence} demonstrates that introducing disorder, either through spatial variability by randomly sampling tubulin dimers from a molecular dynamics simulation (panels a–d) or through static diagonal noise in the Hamiltonian (panels e–h), substantially disrupts this coherence. In both cases, the superradiant coherence becomes localized and fragmented, confirming that both structural and energetic disorder suppress long-range quantum coherence in the system.

% Figure 2: Random MD snapshots + Static disorder

\subsubsection{Lifetimes of Subradiant and Superradiant States}

To characterize the radiative behavior of the system, we compute the physical decay rates $\Gamma_i^{\text{phys}}$ (in cm$^{-1}$) from the imaginary part of the eigenvalues of the non-Hermitian effective Hamiltonian.

The corresponding radiative lifetimes $\tau_i$ (in seconds) are obtained using the spectroscopic relation:
\[
\tau_i = \frac{1}{2\pi c \Gamma_i^{\text{phys}}},
\]
where \( c = 2.99792458 \times 10^{10} \, \text{cm/s} \) is the speed of light.

We identify:
\begin{itemize}
    \item $\tau_{\min}$: the lifetime of the most superradiant state (corresponding to the maximum decay rate $\Gamma_{\max}$),
    \item $\tau_{\max}$: the lifetime of the most subradiant state (corresponding to the minimum decay rate $\Gamma_{\min}$).
\end{itemize}

These lifetimes provide direct insight into the extent of radiative enhancement or suppression in different structural configurations.

We emphasize that the lifetimes reported here are defined within the Markovian radiative-loss model used throughout the paper. Concretely, $\Gamma_i^{\text{phys}}$ is extracted from the imaginary part of the eigenvalues of the effective non-Hermitian generator associated with collective emission, which corresponds to a memoryless (time-independent) coupling to the electromagnetic bath and therefore yields exponential decay envelopes for each collective mode. The non-Markovian behavior discussed in Sec.~\ref{subsec:nonmarkov-two-tubulin} is of a different origin: it appears when we trace out part of a finite, structured tubulin environment and analyze reduced dynamics on a subsystem, which can exhibit information backflow even if the global radiative decay model is Markovian. Under static or structural disorder, the lifetime curves in this section should therefore be interpreted as quantifying how disorder reshapes bright--dark rate separation within the Markovian radiative model (e.g., by reducing collectivity and localizing bright components), while subsystem-level memory effects may still occur due to the structured embedding.

We first consider a microtubule constructed by repeating the 1JFF tubulin unit. We then incorporate static disorder into the model by modifying the site energies in the unperturbed Hamiltonian as follows:
\begin{equation}
H_0 \rightarrow H_0 + \sum_{n=0}^{N-1} \epsilon_n(W) |n\rangle \langle n|,
\end{equation}
where \( W \) is the disorder strength in cm$^{-1}$, and \( \epsilon_n(W) \) is a uniformly distributed random variable within the range
\[
\left[ -\frac{W}{2}, \frac{W}{2} \right].
\]

Finally, we consider structural disorder by constructing a microtubule from molecular dynamics data, where tubulin units are randomly positioned.

\begin{figure}[t]
  %  \centering
    \includegraphics[width=0.6\linewidth]{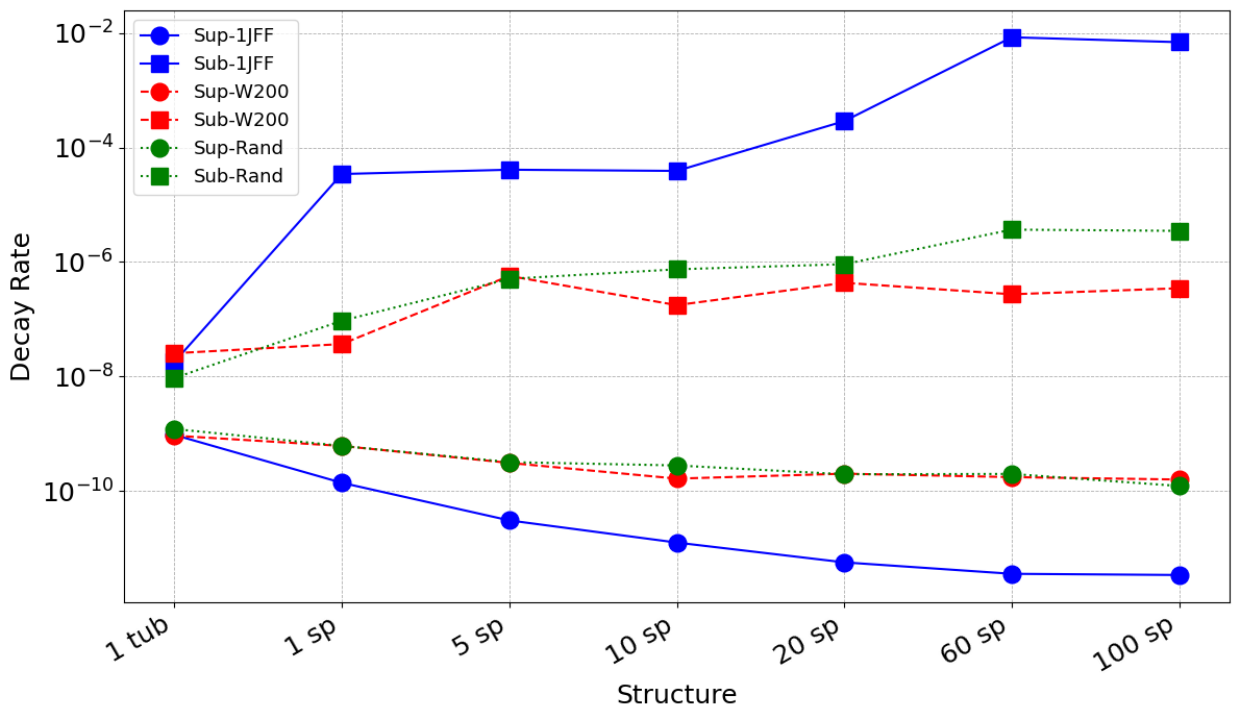}
    \caption{
        Radiative lifetimes of the most superradiant (circles) and subradiant (squares) states as a function of structural size, from a single tubulin dimer up to a microtubule segment containing 100 spirals (100~sp), where one spiral (sp) denotes one circumferential turn consisting of 13 dimers in our construction. The solid blue line corresponds to the ideal periodic structure generated using PDB 1JFF. The red dashed line represents the same structure with static disorder of strength $W = 200$~cm$^{-1}$. The green dotted line corresponds to a structurally disordered microtubule constructed from randomly arranged tubulin units obtained from molecular dynamics simulation.
    }
    \label{fig:decay}
\end{figure}

As seen in Figure~\ref{fig:decay}, increasing the system size in the absence of disorder strongly enhances superradiance (shorter lifetimes) and subradiance (longer lifetimes). For example, the superradiant lifetime decreases significantly while the subradiant one can reach the millisecond range. This shows how cooperative radiative effects intensify with structural size.

However, when either static or structural disorder is introduced, the overall trend remains, but the difference between the smallest and largest structures becomes much smaller, particularly for the superradiant states. For these, the gain in radiative acceleration from one tubulin to 100 spirals becomes marginal, on the order of microseconds. Subradiant states are comparatively more robust, still reaching lifetimes several orders of magnitude longer than superradiant states even under disorder.

This indicates that disorder tends to reduce the contrast between radiative lifetimes and that superradiant states are more sensitive to imperfections than subradiant ones.

\section{Discussion and Conclusions}
\label{secconc}

We have investigated excitation dynamics in tryptophan networks of tubulin using a Lindblad master equation constructed from an effective non-Hermitian generator with radiative loss and explicit site geometry. By comparing five well defined preparations and tracking site resolved populations together with quantum information metrics, we showed how preparation, geometry, and environment size jointly control the fate of correlations. Progressively embedding a single $\alpha\beta$-tubulin into a pair of coupled tubulins, then into a single spiral, and finally into a two-spiral assembly redistributes the leading $L_1$ coherences within the focal tubulin, lowers their amplitudes, and introduces oscillatory exchange with the surrounding tubulins. Within a single spiral, two tubulin subsystems exhibit non-Markovian backflow detected by revivals of the trace distance, with a stronger response for phase contrast preparations. Scaling up to large ordered assemblies with the non-Hermitian eigenmode analysis confirmed the complementary roles of superradiant export and subradiant retention, and revealed a clear sensitivity to symmetry and disorder. Ordered structures sustain long-range correlated coherence and widen the gap between bright and dark lifetimes, whereas static energetic and structural disorder localize the superradiant component, suppress long-range coherence, and reduce lifetime contrast while leaving subradiant protection comparatively robust.

Taken together, these results identify microtubule tryptophan networks as structured reservoirs in which symmetry, spatial arrangement, and initial phase determine whether information is rapidly broadcast through bright channels or preferentially stored in dark subspaces \citep{Ho2013,Li1992Coherence}. The use of quantum information tools provides a compact and operational language for this partition of roles. The $L_1$ norm and correlated coherence resolve where coherence is shared, the trace distance quantifies memory through backflow, and the spectral widths of the effective generator organize radiative lifetimes across scale. This joint dynamical and spectral picture offers a coherent framework for discussing amplification, buffering, and timing in biological excitonic networks.

Biologically, our initial-state comparison does not imply that microtubules must use quantum correlations for neural processing; rather, it shows that a structured tryptophan network can route dynamics differently depending on how an excitation is prepared (localized vs delocalized, and with different relative phases), biasing the evolution toward rapid export (bright channels) or transient retention (dark subspaces). In cells, plausible sources of site-specific preparation include localized absorption events and redox/ROS-driven chemistry on aromatic residues \citep{Ehrenshaft2015,Schoneich2018}. Any downstream coupling, if present, would most plausibly be indirect, through changes in dissipation patterns or local interaction landscapes that regulate established microtubule functions (transport, MAP binding, and tubulin-code-dependent modulation) \citep{GoodsonJonasson2018,Hirokawa2010,JankeMagiera2020}. This motivates future work that links bright-versus-dark routing to specific biochemical readouts and experimentally measurable conditions. Relatedly, the trace-distance revivals reported in Sec.~\ref{subsec:nonmarkov-two-tubulin} indicate short-time memory effects in reduced tubulin subsystems, suggesting that nearby excitation events in a finite spiral need not behave as independent memoryless decays \citep{Breuer2009PRL,Rivas2010PRL}.

Our results also connect naturally to earlier proposals that treat microtubules as quantum information channels. In those works, excitation migration and quantum relaxation in tryptophan chains and dipole networks were shown to support entanglement transfer and biologically relevant coherence times \cite{ShirmovskyChizhov2023,Shirmovsky2024a,Shirmovsky2024b,ShirmovskyShulga2021,ShirmovskyShulga2023,Saenko2025} (here, ``biologically relevant'' is used in a limited operational sense: persistence of population/coherence in protected subspaces over timescales that could, in principle, overlap with downstream molecular degrees of freedom evolving on ns--ms scales). Here we add a complementary radiative perspective, showing that, even in the presence of collective emission, structured tryptophan networks can transiently store and redistribute correlations, with symmetry and preparation determining whether they behave more like broadcasting channels or short-term quantum buffers.

Two timescales should be distinguished. First, coherent redistribution within a local Trp network occurs on the ps--ns scale set by dipole couplings (Secs.~3.1--3.4). Second, the radiative lifetimes of collective eigenmodes can extend to much longer scales in ordered assemblies (Sec.~3.7.2), while disorder tends to localize bright components and compress lifetime contrasts. By ``functional relevance'' we therefore do not claim a specific cellular mechanism; rather, we provide a concrete timescale-matching criterion: whether retention of population/coherence in subradiant (dark) sectors overlaps with the timescales of molecular state variables that could be sensitive to local excitation or electrostatic reorganization (e.g., conformational microstate occupancy, binding/unbinding kinetics of interaction partners, or chemically mediated state changes). This framing can guide experiments aimed at testing whether, and under what structural conditions, optical excitations couple measurably to downstream microtubule-associated dynamics, including in neuronal contexts where microtubules can be comparatively stable.

The present model characterizes relaxation following an impulsive, single-excitation preparation under radiative loss. In living systems, nonequilibrium biochemical inputs (e.g., energy release associated with GTP hydrolysis, redox/ROS chemistry, or enzymatic activity) could act as driving, either indirectly by modulating site energies/couplings or, in a phenomenological description, by maintaining excited-state population out of equilibrium. Within the Lindblad framework, such driving can be included either (i) as incoherent pumping through additional jump operators (e.g., $L^{\mathrm{pump}}_n=\sqrt{p_n}\,\sigma_n^{+}$ with pump rate $p_n$), or (ii) as coherent driving through an added Hamiltonian term (e.g., $H_{\mathrm{drive}}(t)=\sum_n \Omega_n(t)\,(\sigma_n^{+}+\sigma_n^{-})$). Incoherent pumping sustains nonequilibrium populations but typically introduces additional dephasing, so it does not necessarily extend coherence unless the pumping selectively populates (or repeatedly refills) symmetry-protected dark/subradiant subspaces; coherent driving can maintain phase relations but corresponds to a distinct physical regime (e.g., an external optical field). In the limit of rare, localized injection events, the driven dynamics can often be approximated by mixtures over the localized preparations already studied in Sec.~3.4. These extensions provide a clear next step for assessing how continuous energy throughput reshapes the balance between superradiant export and subradiant retention, particularly in the presence of disorder.

There are also clear theoretical next steps. Extending the master equation to include non-radiative decay, structured dephasing, and explicit vibronic coupling would refine the link between geometry and memory. Beyond the single excitation manifold, many-body effects and exciton-exciton interactions could be addressed with stochastic Liouville or time-nonlocal kernels to assess the stability of backflow and subradiant protection under crowding. On the structural side, graph-based analyses of eigenmodes and coherence pathways, together with ensemble averaging over realistic static and dynamic disorder, can connect microscopic organization to emergent timescales. Finally, casting the dynamics in information theoretic terms such as coherence length, channel distinguishability, and resource conversion rates may help compare microtubule networks with other biological light responsive systems within a unified quantitative language.

In summary, master equation dynamics augment static spectral analysis and reveal how geometry, preparation, and environment size regulate the balance between export and retention of quantum information in tubulin-based networks. The combination of non-Hermitian eigenmode structure with quantum information metrics yields a versatile toolbox for biological systems, clarifying when coherence is delocalized, when it is protected, and when memory effects reshape the local flow of information.

%\newpage

%%%%end of paper

%below is left over of template

%\end{document}

%%%%%%%%%%%%%%%%%%%%%%%%%%%%%%%%%%%%%%%%%%

%%%%%%%%%%%%%%%%%%%%%%%%%%%%%%%%%%%%%%%%%%
\vspace{6pt} 

%%%%%%%%%%%%%%%%%%%%%%%%%%%%%%%%%%%%%%%%%%
%% optional
%\supplementary{The following supporting information can be downloaded at:  \linksupplementary{s1}, Figure S1: title; Table S1: title; Video S1: title.}


% Only used for preprtints:
% \supplementary{The following supporting information can be downloaded at the website of this paper posted on \href{https://www.preprints.org/}{Preprints.org}.}

%\begin{tabularx}{\textwidth}{lll}
%\toprule
%\textbf{Name} & \textbf{Type} & \textbf{Description} \\
%\midrule
%S1 & Python script (.py) & Script of python source code used in XX \\
%S2 & Text (.txt) & Script of modelling code used to make Figure X \\
%S3 & Text (.txt) & Raw data from experiment X \\
%S4 & Video (.mp4) & Video demonstrating the hardware in use \\
%\bottomrule
%\end{tabularx}
%}

%%%%%%%%%%%%%%%%%%%%%%%%%%%%%%%%%%%%%%%%%%
\section*{Author Contributions}
Conceptualization: L.G., O.P., and T.J.A.C.; Methodology: L.G., O.P., and T.J.A.C.; Formal analysis: L.G.; Resources: T.J.A.C.; Data curation: L.G.; Writing---original draft preparation: L.G.; Writing---review and editing: L.G., O.P., and T.J.A.C.; Visualization: L.G.; Supervision: T.J.A.C. and O.P.; Project administration: T.J.A.C.; Funding acquisition: T.J.A.C.

\section*{Funding}
This research was undertaken in part thanks to funding to T.J.A.C. from the Canada Research Chairs Program (CRC-2022-00204) and the University of Waterloo. L.G. acknowledges support from the University of Waterloo Provost’s Program for Interdisciplinary Postdoctoral Scholars.

\section*{Data Availability Statement}
Data supporting the findings of this study are available from the corresponding author (L.G.) upon reasonable request.

\section*{Acknowledgments}
During the preparation of this manuscript, the authors used an AI-assisted language tool to help improve the clarity and readability of the text. All AI-assisted content was reviewed and edited by the authors, who take full responsibility for the integrity and accuracy of the final manuscript. This research was enabled in part by support provided by the Digital Research Alliance of Canada (alliancecan.ca) to T.J.A.C.

\section*{Conflicts of Interest}
The authors declare that they have no conflicts of interest.

%%%%%%%%%%%%%%%%%%%%%%%%%%%%%%%%%%%%%%%%%%
%% Optional

%% Only for journal Encyclopedia
%\entrylink{The Link to this entry published on the encyclopedia platform.}

%\abbreviations{Abbreviations}{
%The following abbreviations are used in this manuscript:
%\\
%
%\noindent 
%\begin{tabular}{@{}ll}
%MDPI & Multidisciplinary Digital Publishing Institute\\
%DOAJ & Directory of open access journals\\
%TLA & Three letter acronym\\
%LD & Linear dichroism
%\end{tabular}
%}

%%%%%%%%%%%%%%%%%%%%%%%%%%%%%%%%%%%%%%%%%%
%% Optional

\appendix

\section{Construction of Ordered and Disordered Microtubule Geometries}
\label{sec:geometry_construction}
\subsection{Microtubule Geometry Construction}

Ordered microtubules are assembled by repeating the 1JFF tubulin dimer and arranging dimers with a left-handed spiral geometry following prior work \citep{celardo2019existence,babcock2024ultraviolet} and using the code provided by Patwa et al.~\cite{patwa2024quantum}. The initial dimer orientation places the alpha and beta subunits along the protofilament direction which we define as the $x$ axis. Each dimer is transformed by the following sequence. Rotation by $-55.38^\circ$ about the longitudinal axis. Rotation by $11.7^\circ$ about an axis through the beta tubulin Trp346 CD2 atom. Translation by $11.2$ nm in the $y$ direction and $0.3$ nm in the $z$ direction. To generate the $N$th dimer in a spiral two additional operations are applied. Rotation by $27.69^\circ$ about the $x$ axis. Translation by $0.9$ nm along $x$. One spiral contains $13$ dimers and the resulting diameter is approximately $22.4$ nm. The radius measured from the microtubule longitudinal axis to the center of mass of a dimer is near $11.2$ nm which lies between the outer surface radius near $13.5$ nm and the inner lumen radius near $9.5$ nm. Multiple spirals are concatenated by translations of $8$ nm along the $x$ axis.

\subsection{Structural Disorder from Molecular Dynamics}

Geometric variability is sampled by molecular dynamics of a tubulin dimer and by assembling microtubules from snapshot structures. Simulations were performed using the AMBER molecular dynamics package (AMBER~22, \texttt{pmemd}; 2022) \cite{amber2005} at $310$ K for a production length of $6.2$ ns. The initial template was taken from PDB entry 1TVK with missing residues completed by homology modeling (Swiss-Model) \cite{swissmodel}. Protonation states including histidine assignments were set near neutral pH using a continuum electrostatics based tool (H++ at pH~7, with histidines further refined using PROPKA) \cite{hpp3,propka3}. Tubulin was described with the ff14SB force field \cite{ff14sb}, and the tubulin--nucleotide complexes (GTP- and GDP-bound forms) were assembled in LEaP (ff14SB loaded in LEaP) using standard Amber libraries and any additional Amber-compatible parameter files as needed. The system was neutralized and solvated in LEaP in a TIP3P water box with a buffer of $25$~\AA. After minimization and equilibration a production trajectory was generated and analyzed with cpptraj (AmberTools distributed with AMBER~22) \cite{cpptraj2013}. Trajectory coordinates were saved every 1{,}000 steps (2~ps), yielding $3100$ frames. We extracted these frames and used their coordinates to replace the ordered 1JFF dimer in the construction pipeline. Disordered microtubules were built up to $100$ spirals which corresponds to $1300$ dimers, using high-performance computing resources provided by the Digital Research Alliance of Canada.

\section{Connection Between the Non-Hermitian Hamiltonian 
and the Lindblad Master Equation}
\label{app:NH_to_LME}

We begin with the general Lindblad-type master equation for a density operator $\rho$:
\begin{equation}
\dot{\rho} = -i[H,\rho] + \sum_\mu \gamma_\mu 
\left( L_\mu \rho L_\mu^\dagger - \tfrac{1}{2} \{ L_\mu^\dagger L_\mu, \rho \} \right),
\label{eq:lindblad}
\end{equation}
where $H$ is the system Hamiltonian, $L_\mu$ are collapse (jump) operators, and $\gamma_\mu$ the associated rates.

\subsection{Effective Non-Hermitian Hamiltonian}

We define the effective non-Hermitian Hamiltonian as
\begin{equation}
H_{\mathrm{eff}} = H - \frac{i}{2} \sum_\mu \gamma_\mu L_\mu^\dagger L_\mu.
\end{equation}
Using this, the master equation \eqref{eq:lindblad} can be rewritten as
\begin{equation}
\dot{\rho} = -i \left( H_{\mathrm{eff}} \rho - \rho H_{\mathrm{eff}}^\dagger \right) 
+ \sum_\mu \gamma_\mu L_\mu \rho L_\mu^\dagger.
\label{eq:master_split}
\end{equation}
If the last ``quantum jump'' term is neglected, the dynamics reduce to purely non-Hermitian Hamiltonian evolution. In the quantum-trajectory (unraveling) picture, this corresponds to the conditional no-jump evolution (i.e., evolution conditioned on no emission event), while the jump term restores trace preservation on average \cite{breuer2002theory,WisemanMilburn2009,Carmichael1993}.

\subsection{Effective Hamiltonian in Dipole Networks}

In models of radiatively coupled dipoles, the effective Hamiltonian typically takes the form
\begin{equation}
H_{\mathrm{eff}} = \sum_{n=1}^N \left( \hbar \omega_0 - \tfrac{i}{2}\gamma \right) |n\rangle\langle n| 
+ \sum_{\substack{m,n=1 \\ m\neq n}}^N 
\left( \Omega_{mn} - \tfrac{i}{2}\Gamma_{mn} \right) |m\rangle\langle n|,
\end{equation}
where $|n\rangle$ denotes a localized excitation at site $n$, $\Omega_{mn}$ is the coherent dipole--dipole coupling, and $\Gamma_{mn}$ are the collective radiative decay terms (In the main text we denote the collective decay matrix by $G$; here $\Gamma$ plays the same role).

\subsection{Lindblad Formulation in the Single-Excitation Plus Ground-State Space}

To recover the full open-system dynamics, we return to Eq.~\eqref{eq:master_split}. 
Writing the system Hamiltonian in the site basis,
\begin{equation}
H = \sum_{n=1}^N \hbar \omega_0 |n\rangle\langle n| 
+ \sum_{m\neq n} \Omega_{mn} |m\rangle\langle n|,
\end{equation}
the radiative Lindblad equation can be written as
\begin{equation}
\dot{\rho} = -i[H,\rho] 
+ \sum_{m,n=1}^N \Gamma_{mn} 
\left( |0\rangle\langle m| \, \rho \, |n\rangle\langle 0| 
- \tfrac{1}{2}\{ |n\rangle\langle m|, \rho \} \right),
\label{eq:lme_final}
\end{equation}
where $|0\rangle$ denotes the ground state (absence of excitation).

It is often helpful to define local excitation annihilation and excitation creation operators,
\begin{equation}
\sigma_n^- = |0\rangle\langle n|,\qquad \sigma_n^+ = |n\rangle\langle 0|.
\end{equation}
With this notation, the dissipator becomes
\begin{equation}
\sum_{m,n=1}^N \Gamma_{mn}\left(\sigma_m^- \rho \sigma_n^+ - \tfrac{1}{2}\{\sigma_n^+\sigma_m^-,\rho\}\right),
\end{equation}
which makes explicit that the radiative channels are determined by the Hermitian positive semidefinite matrix $\Gamma$. The standard construction of collective collapse operators is obtained by diagonalizing $\Gamma$, as described next.

\subsection{Collective Collapse Operators from the Decay Matrix}

Since $\Gamma$ is Hermitian and positive semidefinite, it admits an eigendecomposition
\begin{equation}
\Gamma = V \Lambda V^\dagger,\qquad \Lambda = \mathrm{diag}(\gamma_1,\ldots,\gamma_N),
\end{equation}
where the columns $v^{(j)}$ of $V$ are orthonormal eigenvectors and $\gamma_j\ge 0$ are eigenvalues. Substituting this into the dissipator yields a sum of independent Lindblad channels with collective jump operators
\begin{equation}
L_j = \sqrt{\gamma_j}\sum_{n=1}^N v_n^{(j)}\,\sigma_n^- 
= \sqrt{\gamma_j}\sum_{n=1}^N v_n^{(j)}\,|0\rangle\langle n|.
\end{equation}
This makes transparent how superradiant and subradiant channels arise: eigenmodes with large $\gamma_j$ correspond to bright (enhanced-decay) channels, while those with small $\gamma_j$ correspond to dark (suppressed-decay) channels \cite{breuer2002theory,WisemanMilburn2009,Carmichael1993}.

\subsection{Single Excitation Subspace and Conditional Evolution}
\label{app:single_excitation}

When the dynamics are projected onto the single excitation manifold without including the ground state \(\ket{0}\), radiative quantum jumps that take population to \(\ket{0}\) are excluded by construction. In this reduced description, the evolution is the conditional no-jump dynamics generated by the effective non-Hermitian Hamiltonian,
\begin{equation}
\dot{\rho}(t) = -i\!\left(H_{\mathrm{eff}}\rho(t) - \rho(t)H_{\mathrm{eff}}^{\dagger}\right),
\end{equation}
which is not trace preserving and should be interpreted as the evolution conditioned on the absence of emission events (no-jump evolution in a quantum-trajectory unraveling) \cite{breuer2002theory,WisemanMilburn2009,Carmichael1993}.

In the trace preserving, completely positive Lindblad model used in our main simulations, we do include the ground state \(\ket{0}\) and the associated collective jump operators obtained from the eigendecomposition \(G = V\Lambda V^{\dagger}\):
\begin{equation}
L_j = \sqrt{\gamma_j}\sum_{n} v^{(j)}_{n}\,\ket{0}\!\bra{n}, 
\qquad 
\Lambda = \mathrm{diag}(\gamma_1,\ldots,\gamma_N).
\end{equation}
The full master equation then reads
\begin{equation}
\dot{\rho}(t) = -i[H,\rho(t)] + \sum_{j}\left(L_j\,\rho(t)\,L_j^{\dagger}
-\tfrac{1}{2}\{L_j^{\dagger}L_j,\rho(t)\}\right),
\end{equation}
which conserves \(\mathrm{Tr}\,\rho(t)\) while allowing population to leave the single excitation sector into \(\ket{0}\) through the collective emission channels.

The non-Hermitian Hamiltonian formalism arises as an approximation to the full Lindblad equation when quantum jumps into the ground state are neglected (e.g., in the restricted single-excitation subspace). The Lindblad master equation generalizes this approach by incorporating both coherent dynamics and irreversible radiative decay in a trace-preserving manner.

\section{Quantum Information Measures}
\label{app:coh}

\subsection{Quantum Coherence}

To quantify quantum coherence within the tryptophan (Trp) network and between its substructures (e.g., spiral and filament regions), we employ the $l_1$-norm of coherence as introduced in the resource theory of quantum coherence. For a density matrix $\rho$ in a fixed reference basis ${ \ket{i} }$, the $l_1$-norm of coherence is defined as:
\begin{equation}
C_{l_1}(\rho) = \sum_{i \neq j} |\rho_{ij}|,
\end{equation}
where $\rho_{ij}$ are the off-diagonal elements of $\rho$. This measure captures the total amount of coherence present in the system relative to the chosen basis.

To investigate how coherence is distributed and shared between different regions, we calculate the correlated coherence between subsystems $A$ and $B$ (e.g., spiral vs. filament domains). The correlated coherence $C_{\text{corr}}$ is defined as:
\begin{equation}
C_{\text{corr}} = C_{l_1}(\rho_{AB}) - C_{l_1}(\rho_A) - C_{l_1}(\rho_B),
\end{equation}
where $\rho_{AB}$ is the density matrix of the full system, $\rho_A = \mathrm{Tr}B[\rho{AB}]$ and $\rho_B = \mathrm{Tr}A[\rho{AB}]$ are the reduced density matrices of subsystems $A$ and $B$, respectively.

This quantity isolates the coherence that is genuinely shared between the two subsystems and is not attributable to local coherences alone.

\subsection{Logarithmic Negativity}

The negativity, denoted as $N(\rho)$, quantifies entanglement in bipartite quantum systems, $\rho_{AB}$ \citep{benabdallah2021dynamics}. It is given by summing over the absolute of the negative eigenvalues, $\lambda_i$, of~the partially transposed density matrix, $\rho_A^{T}$,
\begin{equation}
N(\rho_{AB}) = \sum_i |\lambda_i|.
\end{equation}
 The logarithmic negativity, $E_N$, is related to the negativity and serves as a good indicator of the degree of entanglement. It is defined as
\begin{equation}
\label{logN}
 E_N(\rho_{AB}) = \log_2(2N(\rho_{AB}) + 1).
\end{equation}

\subsection{Mutual Information}

We quantify total correlations by the quantum mutual information
\[
I(A{:}B,t)=S\!\left(\rho_A(t)\right)+S\!\left(\rho_B(t)\right)-S\!\left(\rho_{AB}(t)\right),
\quad
S(\rho)=-\mathrm{Tr}\,\rho\log_2\rho,
\]
where \(\rho_{AB}(t)\) is the reduced density operator on a chosen bipartition \(A\otimes B\), and \(\rho_A=\mathrm{Tr}_B\rho_{AB}\), \(\rho_B=\mathrm{Tr}_A\rho_{AB}\). All reductions respect the single-excitation model: populations outside the chosen subsystem are traced into the local ground state \(|0\rangle\).

\section{Supplementary Results}
\label{app:sup}
\subsection{Mutual Information}

\begin{figure}[t]
  \centering
  % Row 1: single tubulin (a-d)
  \subfloat[]{\includegraphics[width=0.24\linewidth]{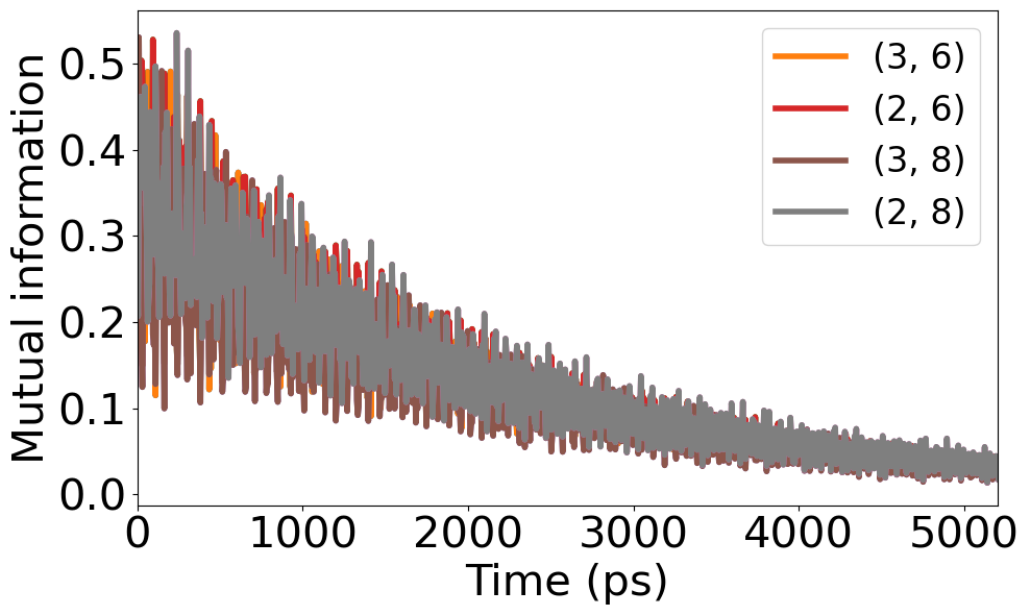}\label{fig:coh:a2}}\hfill
  \subfloat[]{\includegraphics[width=0.24\linewidth]{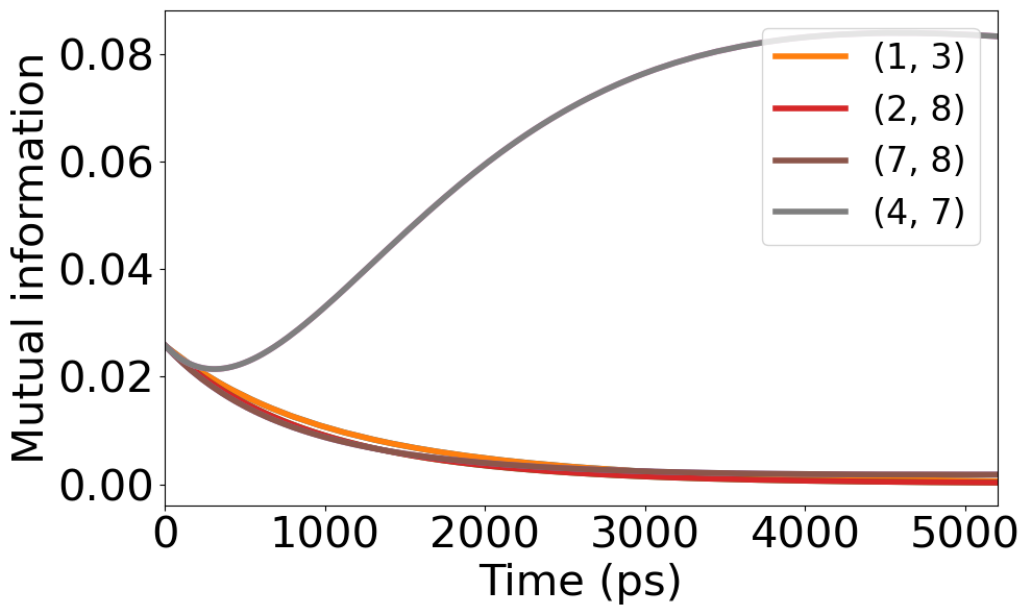}\label{fig:coh:b2}}\hfill
  \subfloat[]{\includegraphics[width=0.24\linewidth]{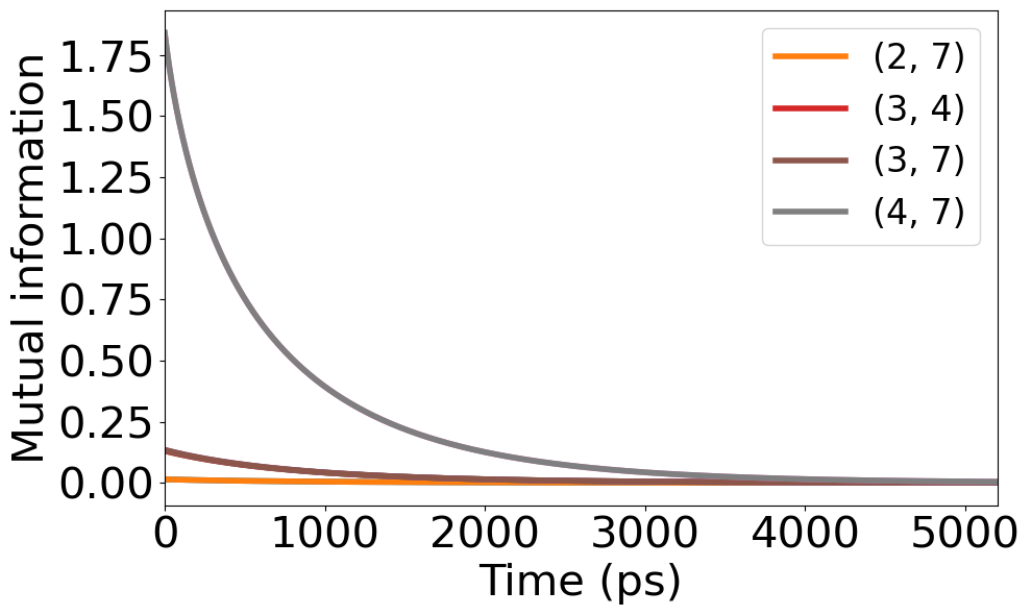}\label{fig:coh:c2}}\hfill
  \subfloat[]{\includegraphics[width=0.24\linewidth]{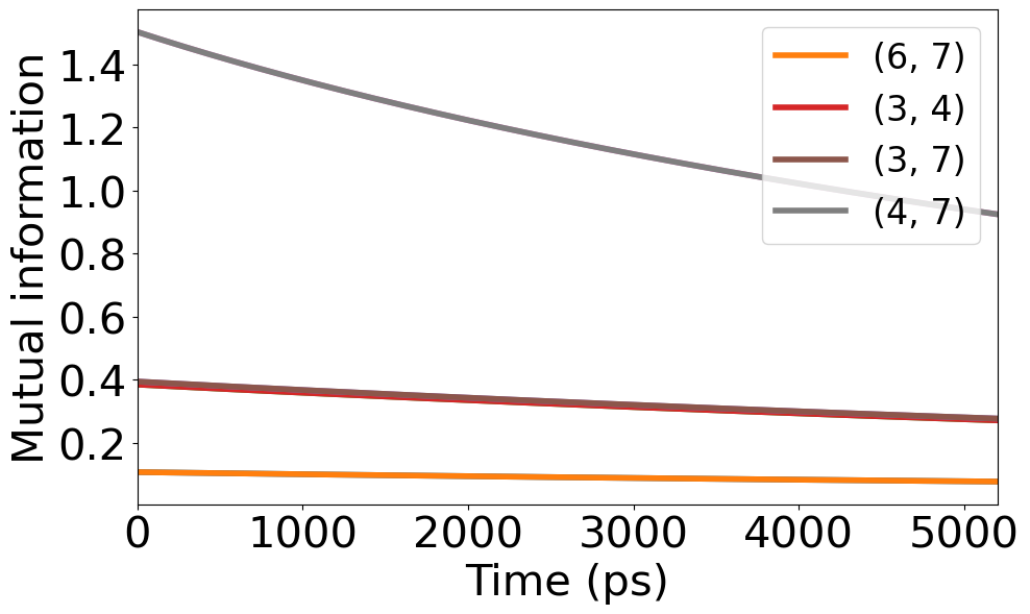}\label{fig:coh:d2}}\\[1ex]

  % Row 2: two tubulins (e-h)
  \subfloat[]{\includegraphics[width=0.24\linewidth]{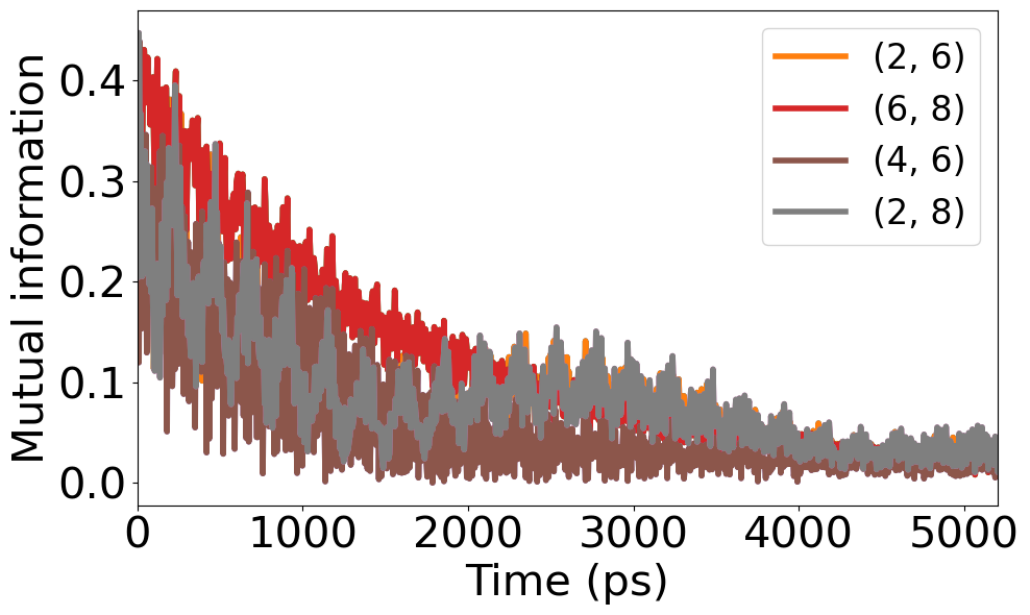}\label{fig:coh:e2}}\hfill
  \subfloat[]{\includegraphics[width=0.24\linewidth]{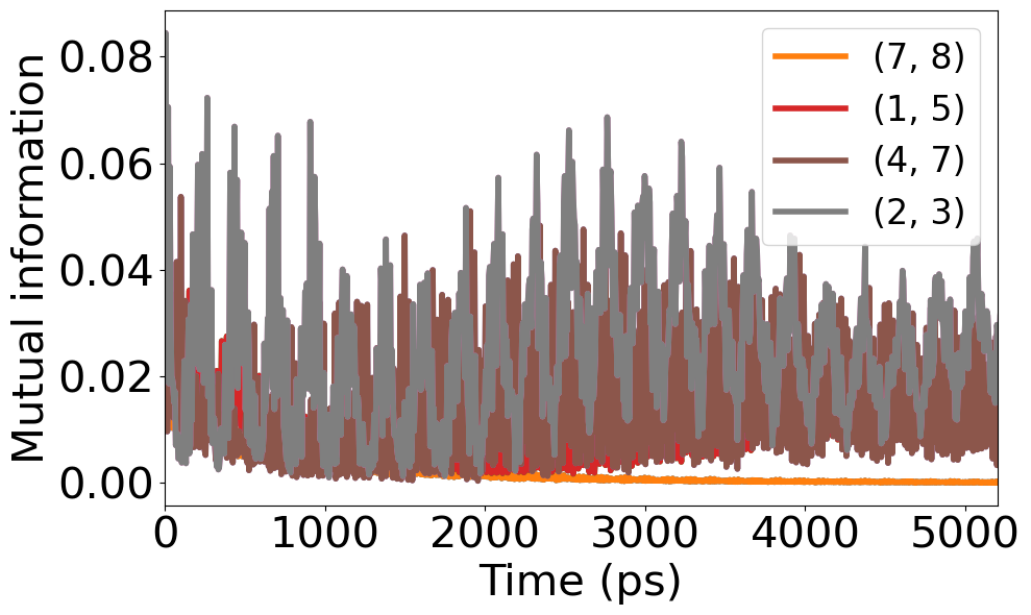}\label{fig:coh:f2}}\hfill
  \subfloat[]{\includegraphics[width=0.24\linewidth]{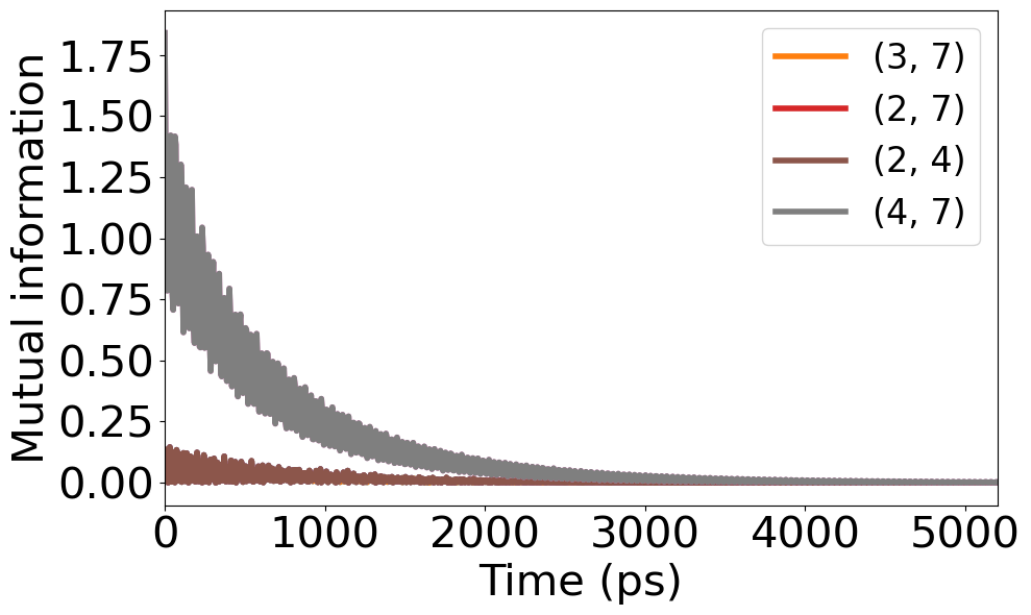}\label{fig:coh:g2}}\hfill
  \subfloat[]{\includegraphics[width=0.24\linewidth]{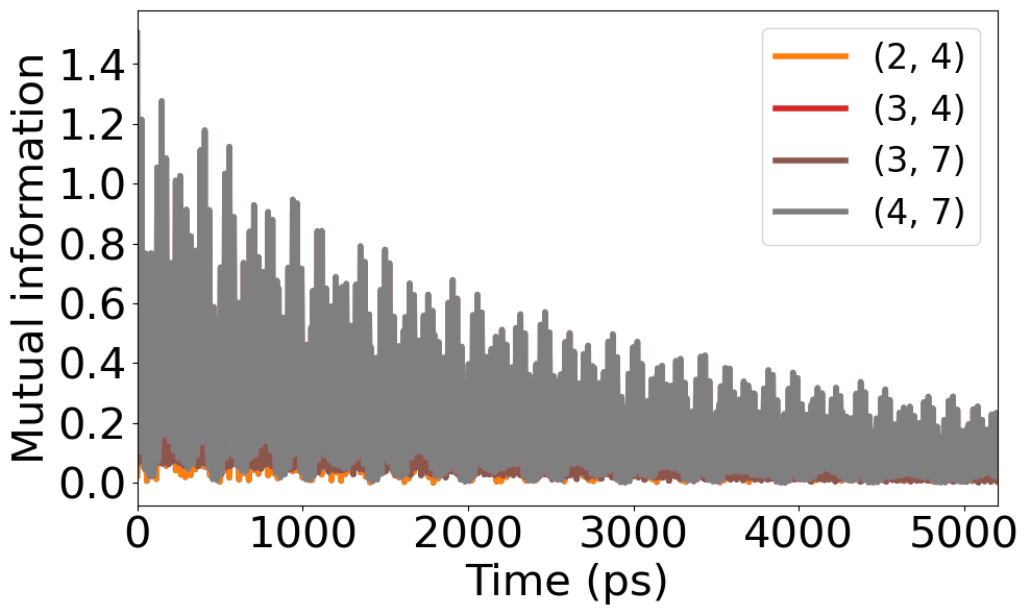}\label{fig:coh:h2}}\\[1ex]

  % Row 3: spiral (i-l)
  \subfloat[]{\includegraphics[width=0.24\linewidth]{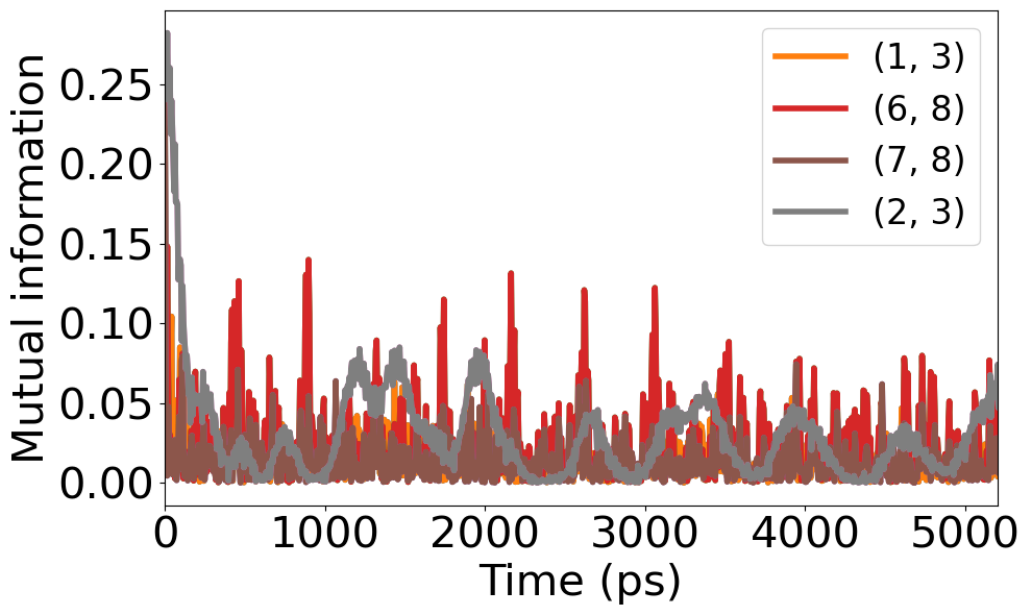}\label{fig:coh:i2}}\hfill
  \subfloat[]{\includegraphics[width=0.24\linewidth]{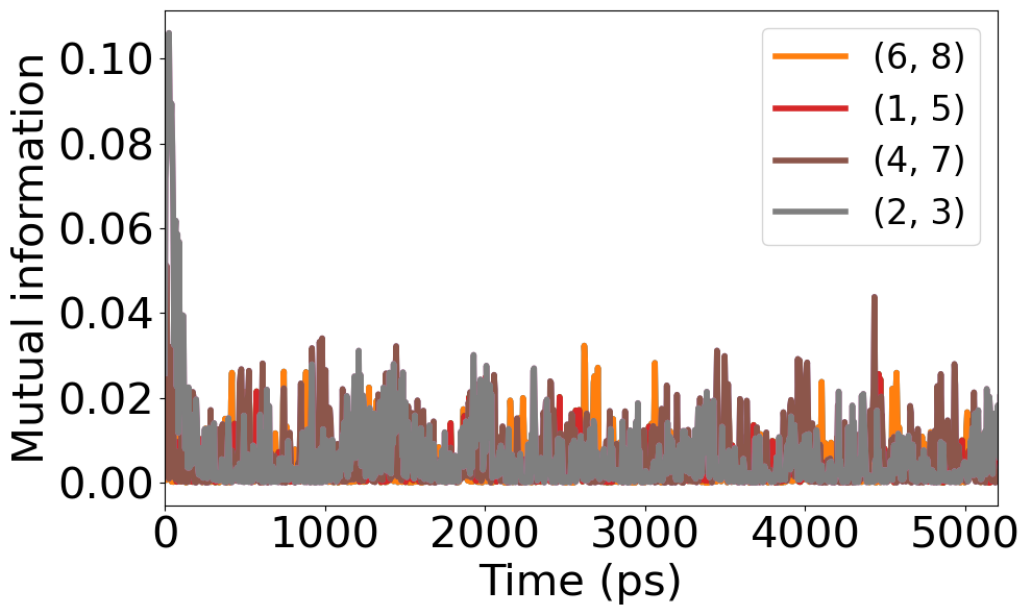}\label{fig:coh:j2}}\hfill
  \subfloat[]{\includegraphics[width=0.24\linewidth]{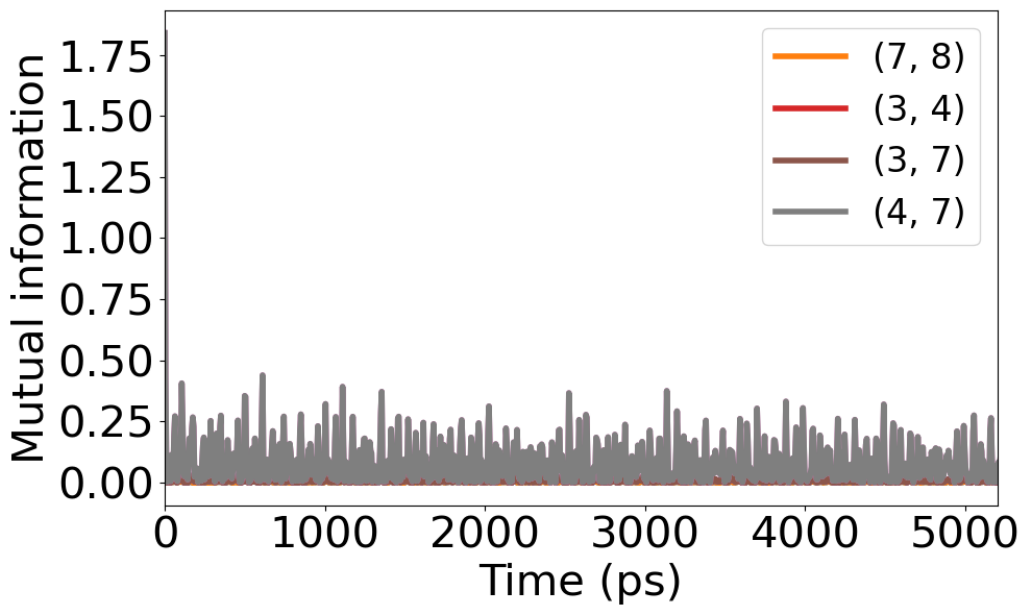}\label{fig:coh:k2}}\hfill
  \subfloat[]{\includegraphics[width=0.24\linewidth]{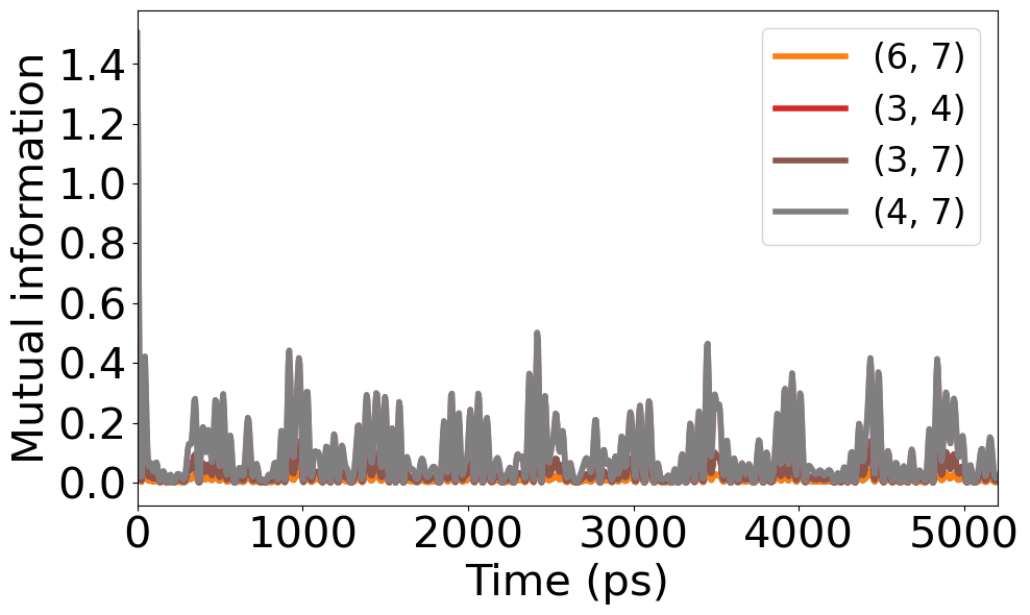}\label{fig:coh:l2}}

% Row 4: spiral (m-p)
  \subfloat[]{\includegraphics[width=0.24\linewidth]{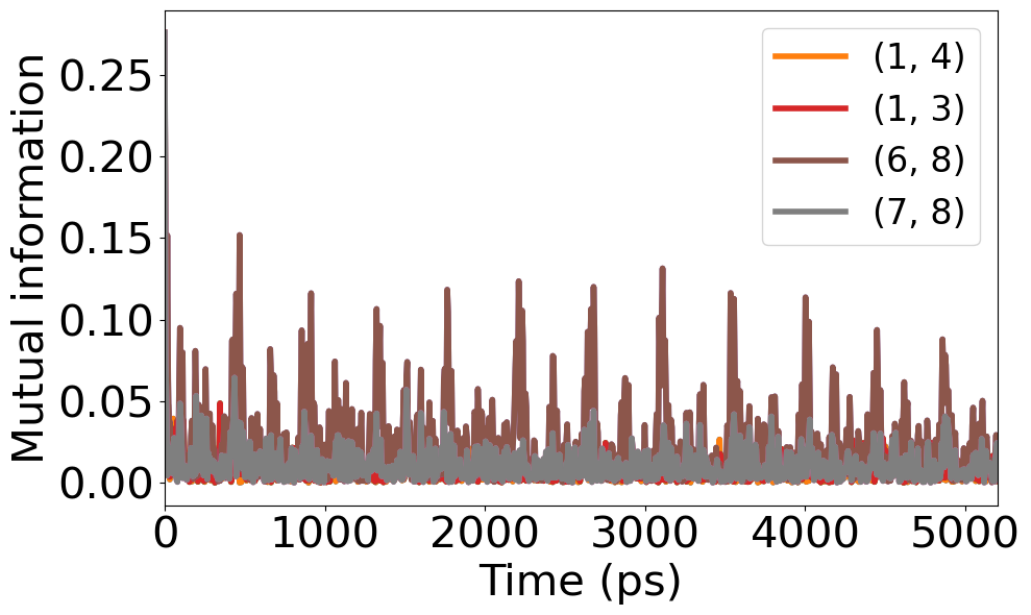}\label{fig:coh:m2}}\hfill
  \subfloat[]{\includegraphics[width=0.24\linewidth]{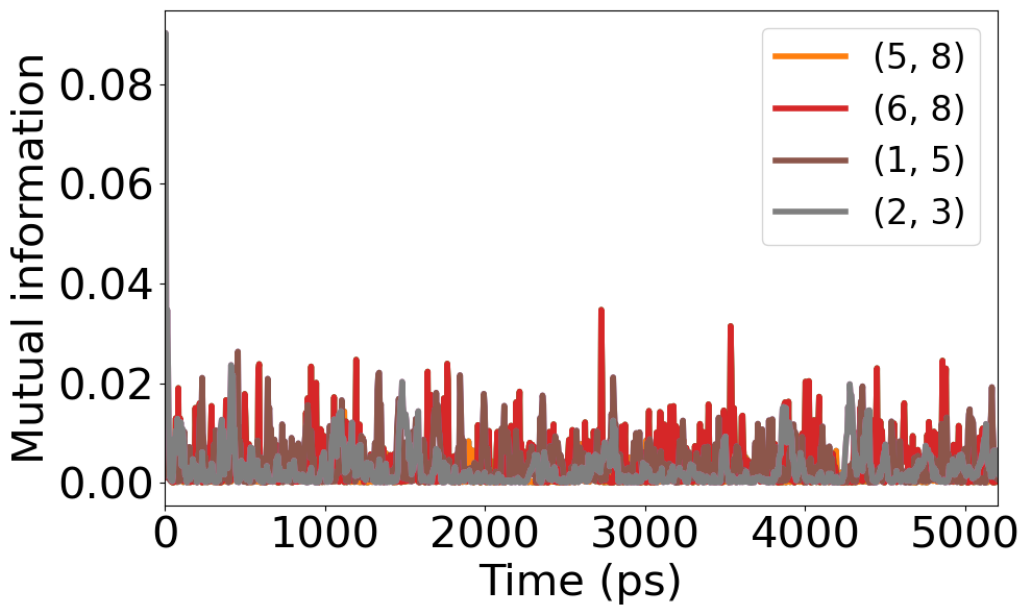}\label{fig:coh:n2}}\hfill
  \subfloat[]{\includegraphics[width=0.24\linewidth]{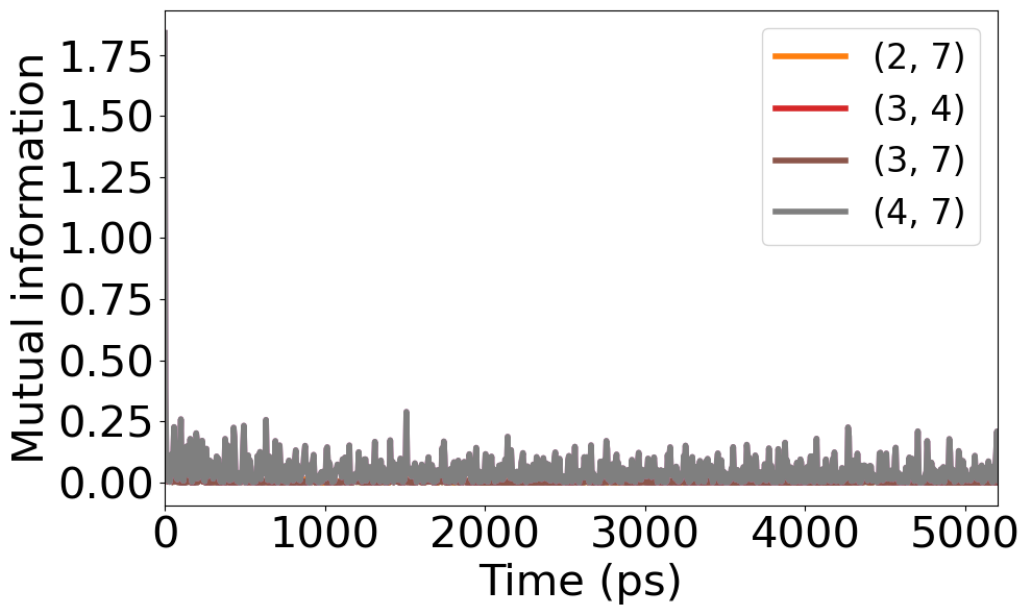}\label{fig:coh:o2}}\hfill
  \subfloat[]{\includegraphics[width=0.24\linewidth]{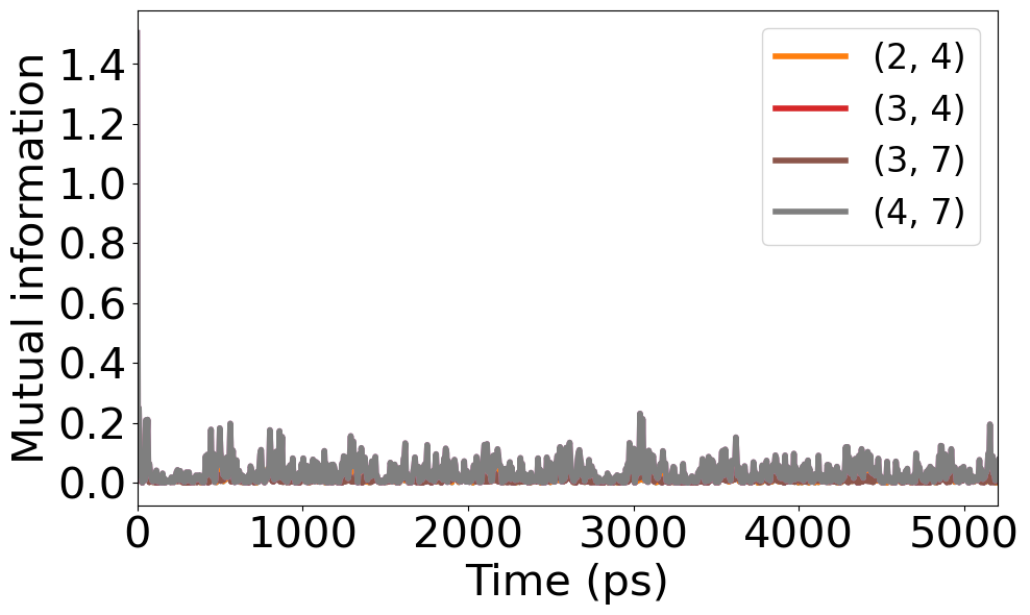}\label{fig:coh:p2}}
  \caption{Top four pairwise mutual information values across embeddings and initial states.
Columns (all rows): from left to right, maximally coherent, maximally mixed, superradiant,
and subradiant initial states. Rows: (a–d) single tubulin; (e–h) two-tubulin system; (i–l) one spiral; (m–p) two spirals. Each panel shows the four site pairs with the largest mutual
information within the tracked tubulin. Pairs $(i,j)$ denote tryptophan site indices (labels) defined in Fig.~\ref{fig:tub}. Time is reported in picoseconds (ps) in all panels.}
  \label{fig:one-to-spiral:matrix:mut}
\end{figure}

In Fig.~\ref{fig:one-to-spiral:matrix:mut} we summarize, for each embedding and initial state, the four site pairs with the largest time-averaged pair mutual information, highlighting how the pattern of information sharing evolves from localized (single tubulin) to increasingly delocalized (spirals).

\subsection{Correlated Coherence Across One Spiral}
\begin{figure}[t]
  \centering
  \includegraphics[width=0.48\textwidth]{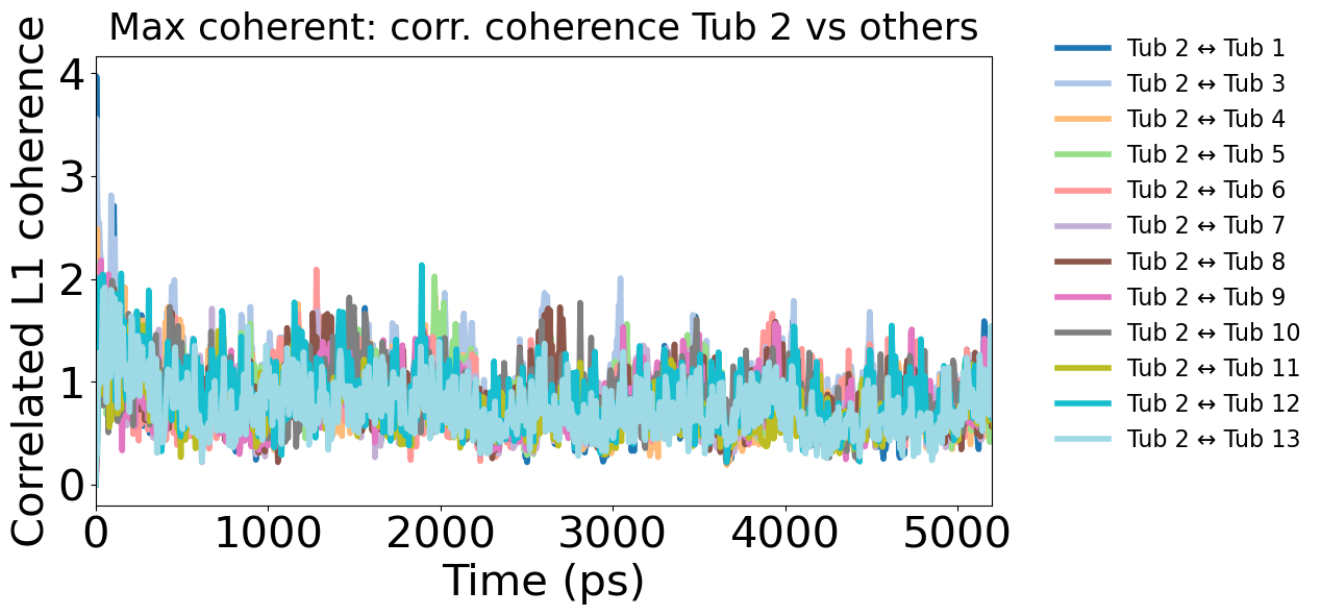}\hfill
  \includegraphics[width=0.48\textwidth]{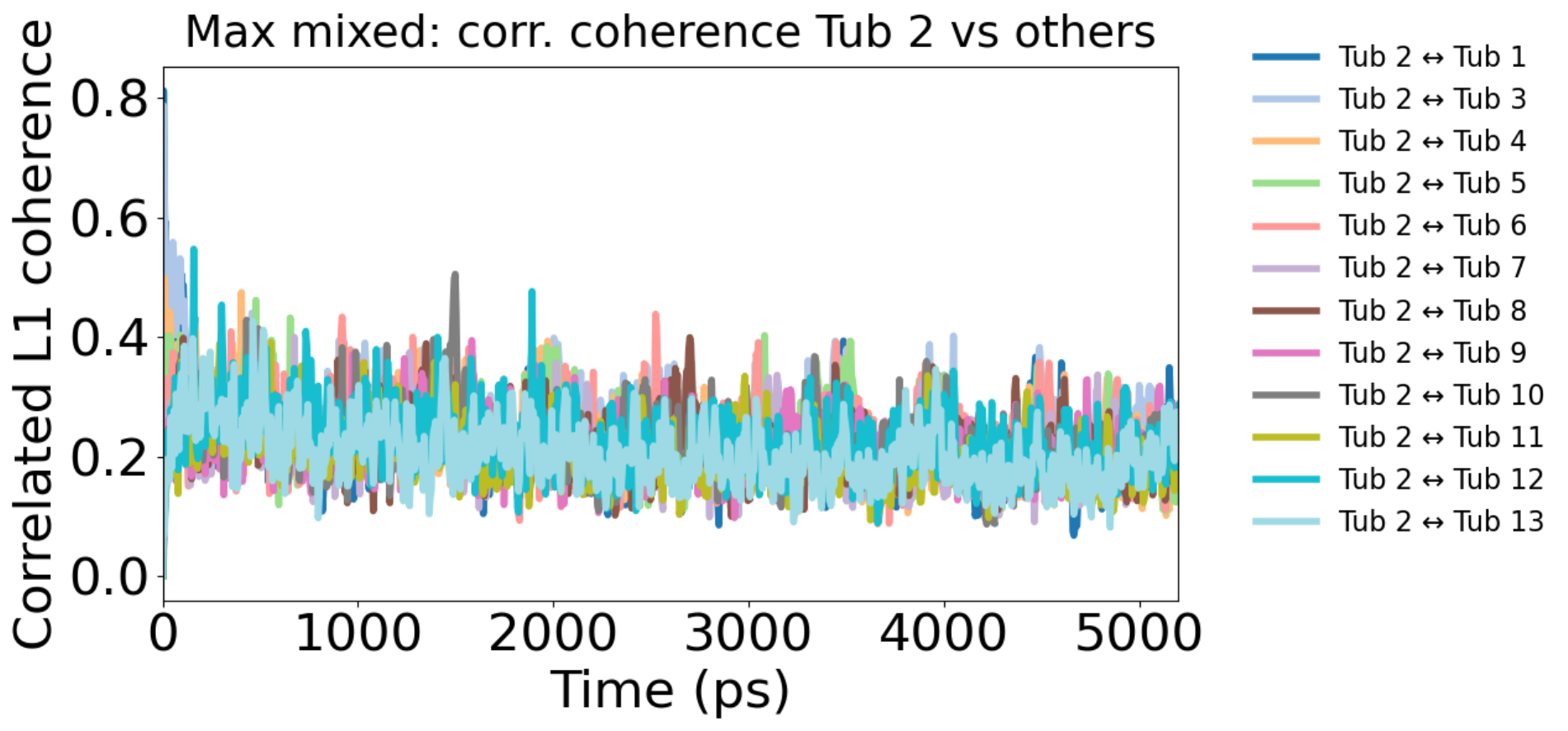}\\[0.5em]
  \includegraphics[width=0.48\textwidth]{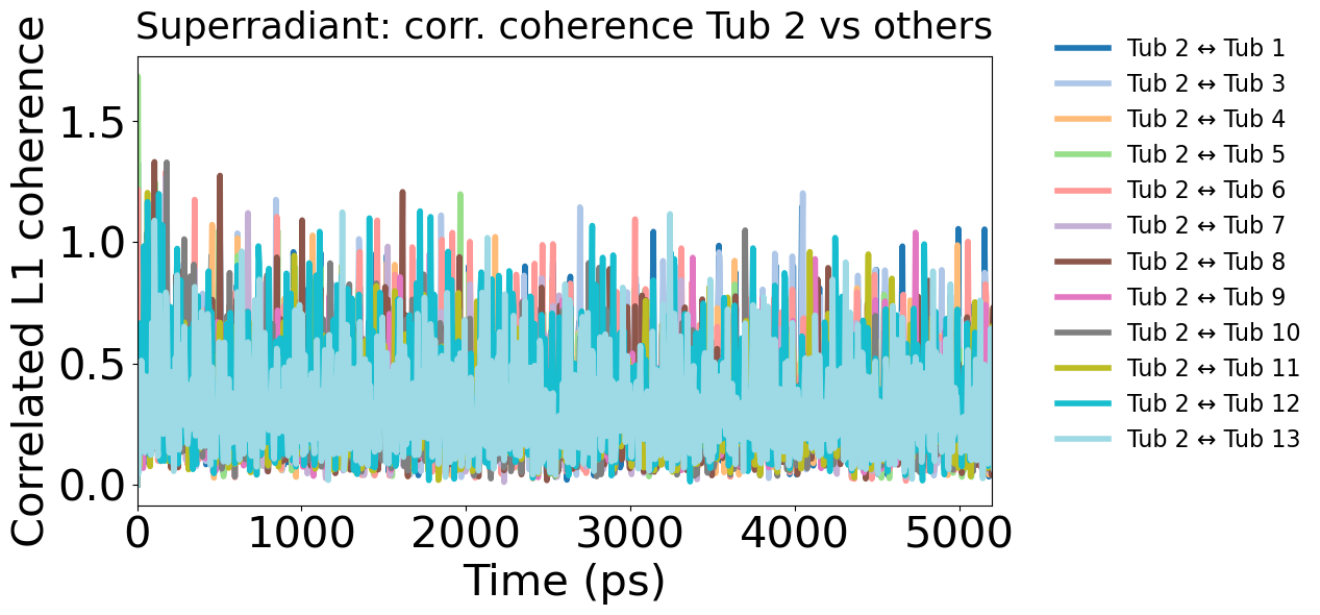}\hfill
  \includegraphics[width=0.48\textwidth]{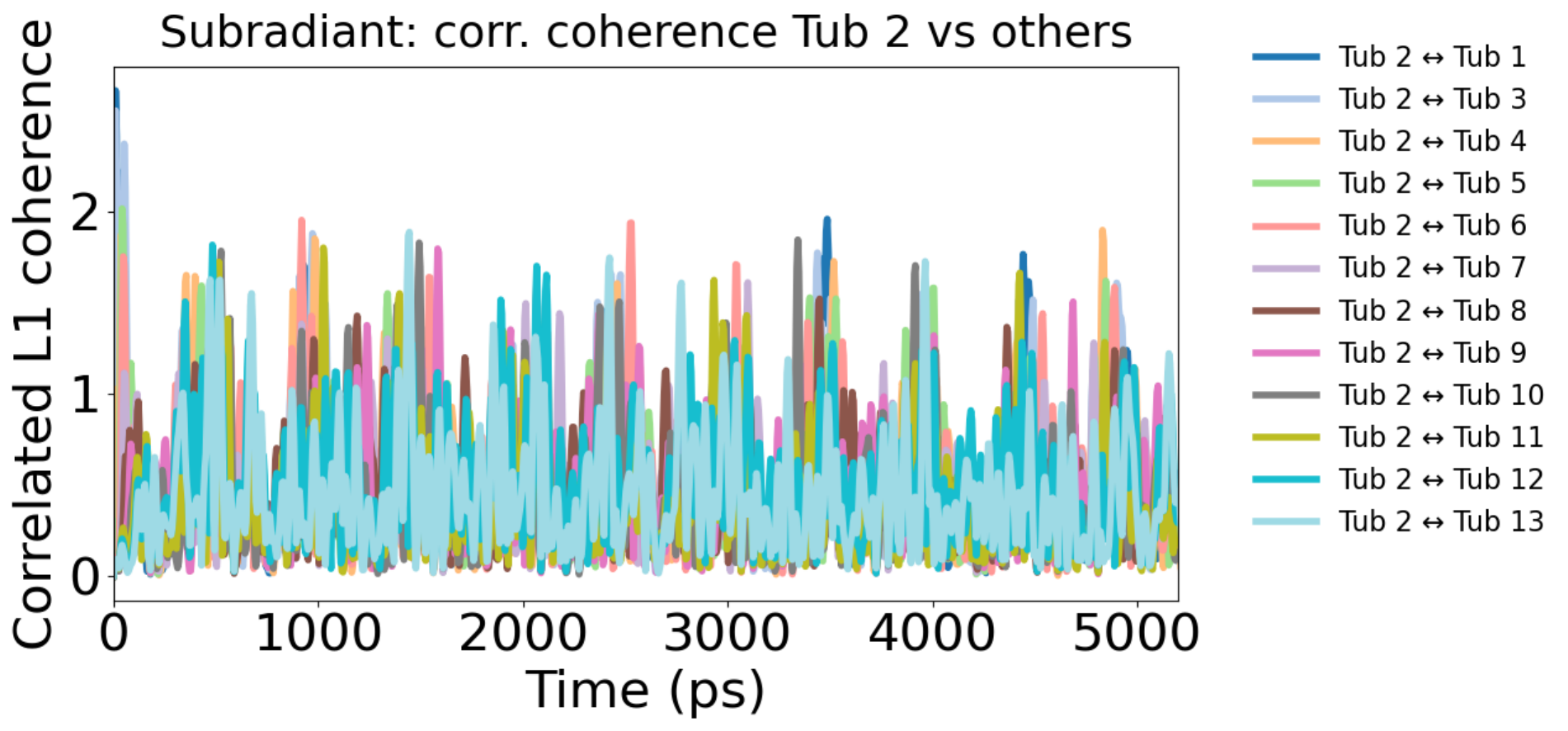}
  \caption{Correlated coherence between the initially excited tubulin (no.~2) and all other tubulins in the spiral, shown for four distinct initial conditions. Each line corresponds to a pair $(2,k)$ with $k\in\{1,3,\dots,13\}$. The correlated coherence quantifies the portion of $\ell_1$-coherence shared between subsystems beyond their local contributions. Time is reported in picoseconds (ps) in all panels.}
  \label{fig:corr_all}
\end{figure}

We evaluate how quantum coherence is shared between the excited tubulin (no.~2) and the remaining tubulins in a single spiral (Fig.~\ref{fig:corr_all}).  
For each pair $(2,k)$, the correlated coherence is computed.
For every initial condition, maximally coherent, maximally mixed, superradiant, and subradiant, we plot the time evolution of $C_{\mathrm{corr}}(2{:}k)$ for all partner tubulins $k\neq2$.

\end{document}